\documentclass[useAMS,usenatbib]{mn2e}

\usepackage{graphicx}
\usepackage{epsfig}
\usepackage{color}

\usepackage{longtable}
\usepackage{colortbl}
\usepackage{hhline}

\usepackage{natbib}
\bibpunct{(}{)}{;}{a}{}{,} 

\newcommand{\eps}{eps}

\newcommand{\amiga}{\texttt{AMIGA}}
\newcommand{\ahf}{\texttt{AHF}}

\newcommand{\mlapm}{\texttt{MLAPM}}
\newcommand{\mhf}{\texttt{MHF}}

\def\LCDM{$\Lambda$CDM}
\def\hkpc{$h^{-1}{\ }{\rm kpc}$}
\def\hMpc{$h^{-1}{\ }{\rm Mpc}$}
\def\hMsun{$h^{-1}{\ }{\rm M_{\odot}}$}
\def\nbody{$N$-body}
\def\Rvir{$R_{\rm vir}$}
\def\Riso{$R_{\rm iso}$}

\def\zform{$z_{\rm form}$}

\def\lesssim{\mathrel{\hbox{\rlap{\hbox{\lower4pt\hbox{$\sim$}}}\hbox{$<$}}}}
\def\gtrsim{\mathrel{\hbox{\rlap{\hbox{\lower4pt\hbox{$\sim$}}}\hbox{$>$}}}}

\newcommand{\Sec}[1]{Section~\ref{#1}}
\newcommand{\Eq}[1]{Eq.~(\ref{#1})}
\newcommand{\Fig}[1]{Figure~\ref{#1}}

\newcommand{\bq}{\begin{equation}}
\newcommand{\eq}{\end{equation}}
\newcommand{\bqa}{\begin{eqnarray}}
\newcommand{\eqa}{\end{eqnarray}}

\newcommand{\apj}{ApJ}
\newcommand{\apjs}{ApJS}
\newcommand{\mnras}{MNRAS}

\newcommand{\prd}{PhRD}
\newcommand{\apjl}{ApJL}
\newcommand{\nat}{Nature}

\usepackage{ifthen}

\newcommand{\simus}[1]{
	\ifthenelse{\equal{#1}{01-01}}{C1\ (#1)}{}%
	\ifthenelse{\equal{#1}{04-03}}{C2\ (#1)}{}%
	\ifthenelse{\equal{#1}{01-07}}{C3\ (#1)}{}%
	\ifthenelse{\equal{#1}{02-07}}{C4\ (#1)}{}%
	\ifthenelse{\equal{#1}{03-05}}{C5\ (#1)}{}%
	\ifthenelse{\equal{#1}{04-04}}{C6\ (#1)}{}%
	\ifthenelse{\equal{#1}{01-02}}{C7\ (#1)}{}%
	\ifthenelse{\equal{#1}{01-10}}{C8\ (#1)}{}%
	\ifthenelse{\equal{#1}{Box20b}}{G1\ (#1)}{}%
}
\newcommand{\simu}[2]{#1}	

\newcommand{\AHF}{\texttt{AHF}}

\newcommand*{\DOT}{.}

\newcommand{\s}{\hphantom{0}}

\renewcommand{\vec}[1]{\textit{\textbf{#1}}}

\definecolor{lightgrey}{gray}{0.9}
\definecolor{lightred}{rgb}{1.0,0.85,0.85}
\definecolor{lightgreen}{rgb}{0.85,1.0,0.85}

\begin{document}

  \title[Tidal Streams of Subhaloes]
	{The Tidal Streams of Disrupting Subhaloes in Cosmological Dark Matter Haloes}

  \author[Warnick, Knebe \& Power]
	 {Kristin Warnick\thanks{E-mail: kwarnick@aip.de}$^{1}$, 
	   Alexander Knebe$^{1}$ \& Chris Power$^{2}$ \\
	   $^1$ Astrophysikalisches Institut Potsdam, An der Sternwarte
	   16, 14482 Potsdam, Germany\\
	   $^2$ Centre for Astrophysics \& Supercomputing, Swinburne 
	   University of Technology, PO Box 218, Hawthorn, 
	   Victoria 3122, Australia}

\date{Received / Accepted}

\pagerange{\pageref{firstpage}--\pageref{lastpage}} \pubyear{2007}
\maketitle

\label{firstpage}

\begin{abstract}
We present a detailed analysis of the properties of tidally stripped
material from disrupting substructure haloes or \emph{subhaloes} in 
a sample of high resolution cosmological $N$-body host haloes ranging from 
galaxy- to cluster-mass scales. We focus on devising methods to recover the
infall mass and infall eccentricity of subhaloes from the properties
of their tidally stripped material (i.e. tidal streams). Our analysis
reveals that there is a relation between the scatter of stream particles
about the best-fit debris plane and the infall mass of the
progenitor subhalo. This allows us to reconstruct the infall mass from
the spread of its tidal debris in space. We also find that the
spread in radial velocities of the debris material (as measured by
an observer located at the centre of the host) correlates with the infall
eccentricity of the subhalo, which allows us to reconstruct its orbital 
parameters.
We devise an automated method to identify leading and trailing arms that can,
in principle at least, be applied to observations of stellar streams from 
satellite galaxies. This method is based on the energy distribution of material 
in the tidal stream. Using this method, we show that the mass associated with 
leading and trailing arms differ.
While our analysis indicates that tidal streams can be used to recover certain 
properties of their progenitor subhaloes (and consequently satellites), we do 
not find strong correlations between host halo properties and stream properties. 
This likely reflects the complicated relationship between the stream and the host,
 which in a cosmological context is characterised by a complex mass accretion 
history, an asymmetric mass distribution and the abundance of substructure. 
Finally, we confirm that the so-called ``backsplash'' subhalo
population is present not only in galaxy cluster haloes but also in
galaxy haloes. The orbits of backsplash subhaloes brought them inside
the virial radius of their host at some earlier time, but they now
reside in its outskirts at the present-day, beyond the virial
radius. Both backsplash and bound subhaloes experience similar mass
loss, but the contribution of the backsplash subhaloes to the overall
tidal debris field is negligible.

\end{abstract}


\section{Introduction}

The currently favoured model for cosmological structure formation is the 
$\Lambda$CDM model. This model assumes that we live in a spatially flat 
Universe whose matter content is dominated by non-baryonic Cold Dark 
Matter (CDM) and whose present-day expansion rate is accelerating, driven 
by Dark Energy ($\Lambda$). In the context of this model, the formation 
and evolution of cosmic structure proceeds in a ``bottom-up'' or 
hierarchical manner. Low mass gravitationally bound structures merge to 
form progressively more massive structures, and it is through such a 
merging hierarchy that galaxies, groups and clusters form 
\cite[e.g.][]{Davis.etal.85, Springel.Frenk.White.06}.

The tidal disruption of satellite galaxies around our Galaxy and others is 
characteristic of the hierarchical merging scenario. As a satellite galaxy 
orbits within the gravitational potential of its more massive host, it is 
subject to a tidal field that may vary in both space and time. The 
gravitational force acting on the satellite strips a stream of tidal 
debris from it, and in some cases tidal forces may be sufficient to lead 
to the complete disruption of the system. This tidal debris tends to form 
two distinct arms -- a \emph{leading} arm ahead of the satellite and a 
\emph{trailing} arm following the satellite.

The process by which a satellite galaxy suffers mass loss and the
subsequent formation of a stream of tidal debris (hereafter referred
to as a \emph{tidal stream}) has been studied extensively using
numerical simulations \cite[e.g.][]{Johnston.98, Ibata.etal.01,
  Helmi.04}. A number of these studies have followed the orbital
evolution of an individual satellite galaxy with a known mass profile,
realised with many particles, in a static analytic and
non-cosmological\footnote{By ``non-cosmological'' we mean that the
  systems were evolved in isolation, in the absence of the background
  expansion of the Universe, without infall and accretion, and without
  a population of substructures.} host potential \citep[e.g.][and many 
others]{Johnston.98, Ibata.etal.01, Helmi.04}. This approach allows for 
well-defined test cases to be investigated and for a systematic survey of 
orbital and structural parameters to be carried out. For example, 
\cite{Helmi.04} showed that the width of the tidal debris in the plane 
perpendicular to its orbit increases when the host halo potential is 
flattened rather than spherical.

In this regard, it is important to note that stream properties are 
sensitive to both the internal properties of its progenitor satellite and 
the details of the orbit that it follows.  For example, we expect the 
length and width of a stream to tightly correlate with the age of the 
stream and the mass/internal velocity dispersion of its progenitor 
\cite[e.g.][]{Johnston.Hernquist.Bolte.96,
  Johnston.98,Johnston.Sackett.Bullock.01,Penarrubia.etal.06}.
Understanding what role the progenitor satellite and its orbit plays
in shaping stream properties is therefore vital if we wish to use
streams to address more fundamental problems, such as the nature of
the dark matter.\\

While previous studies have provided important insights into the
formation and evolution of tidal streams and their dependence on the
properties of both satellites and host dark matter haloes, none have
considered tidal streams in cosmological host haloes\footnote{We note
  the work of \citet[][]{Penarrubia.etal.06}, which employs a
  semi-analytical approach to study tidal streams in an evolving host
  potential whose time variation is based on results of cosmological
  simulations.}. Cosmological haloes have complex mass accretion
histories and mass distributions that can at best be crudely
approximated by fitting formulae, and the interplay between a halo's
asphericity, its clumpiness, and how its mass grows in time may be
non-trivial. How this complexity may affect the results of previous
studies, which were based on analytic potentials and approximations to
halo growth, is difficult to say. For example, interactions between
the stream and substructures lead to the dynamical heating, causing an
increase in internal velocity dispersion and a subsequent broadening
of the stream \citep{Moore.etal.99, Ibata.etal.02,
  Johnston.Spergel.Haydn.02}.  Therefore, while studying tidal streams
in cosmological haloes is a challenging problem, such an approach is
necessary if our results are to reflect the complexity of nature
(assuming the validity of the $\Lambda$CDM model, of course!).

This paper is the first of in a series that will examine the formation
and evolution of tidal streams in \emph{cosmological} dark matter
haloes, and determine how the properties of these tidal streams depend
on the properties of the satellite galaxies from which they originate
and on the host dark matter haloes in which they orbit. We use high
resolution cosmological simulations of individual galaxy- and
cluster-mass dark matter haloes -- the \emph{host haloes} -- that
allow us to track \emph{in detail} the evolution of a large population
of their substructures or \emph{satellite galaxies} over multiple
orbits\footnote{For the purposes of our study, we treat substructure
  haloes and satellite galaxies as interchangeable, insofar as both
  suffer mass loss and the tidally stripped material is a dynamical
  tracer of the host potential. However, we note that the
  correspondence between dark matter substructures and luminous
  satellite galaxies is not a straightforward one
  \citep[][]{Gao.DeLucia.etal.04}.}. We seek to answer the questions
we consider to be key to the study of tidal streams;

\begin{itemize}
\item \emph{Can observations of tidal streams be used to infer the properties
  of their parent satellite?}
\item \emph{Can tidal streams reveal properties of the underlying
  dark matter distribution of the host halo?} 
\end{itemize}

The present paper primarily focuses on the first question and can be 
summarised as follows. In \Sec{sec:simulations} we describe the fully 
self-consistent cosmological simulations of structure formation in a 
\LCDM\ universe that we have used in our study. The suite of host haloes 
(consisting of eight cluster and one Milky Way sized dark matter halo) are 
presented in \Sec{sec:hosts}. In \Sec{sec:subhaloes} we shift attention to 
the formation and evolution of debris fields. In this regard, 
\Sec{sec:subhaloes} deals with mass loss from subhaloes and the 
contribution of so-called ``backsplash'' satellites. These subhaloes are 
found outside the virial region of the host at the present day, but their 
orbits took them inside the virial radius at earlier times. We show that 
they exist not only in cluster-sized haloes as found by e.g. 
\cite{Gill.etal.05.3} and \cite{Moore.Diemand.Stadel.04}, but also in our 
galactic halo. In \Sec{sec:streams_sat} we examine how well satellite 
properties such as its original mass and the eccentricity of its orbit can 
be reproduced using the whole debris field of the corresponding satellite. 
Furthermore, we present in \Sec{sec:trailing/leading} a method to 
distinguish between trailing and leading arms in an automated fashion and 
we investigate properties of the streams such as their mass, shape and 
velocity dispersion. We pay particular interest to the energy distribution 
of debris, since it can be used to distinguish between leading and 
trailing arm observationally\footnote{It is important to remember that our 
tidal streams consist of ``particles'' of dark matter, whereas an observer 
detects stellar streams. However, the physical processes involved in the 
formation of both dark matter and stellar streams are identical, and so we 
can think of them interchangeably.}. Finally, we summarise our results in 
\Sec{sec:summary} and comment on their significance in the context of 
future papers, which will try to establish the missing link between the 
properties of tidal streams and the host haloes only briefly touched upon 
in this study.

\section{Simulations \& Analysis Methods}  
\label{sec:simulations}

\subsection{The Simulations}

Our analysis is based on a set of nine high-resolution cosmological
\nbody\ simulations that were used in the study of
\citet[][]{Warnick.06}. These simulations focus on the formation
and evolution of a sample of galaxy- and cluster-mass dark matter
haloes forming in a spatially flat \LCDM\ cosmology with
$\Omega_0$=0.3, $\Omega_\Lambda$=0.7, $\Omega_b$=0.04, $h$ = 0.7, and
$\sigma_8$ = 0.9. Each halo contains at least $10^6$ particles at
$z$=0 and each simulation was run with an effective spatial resolution
of $\leq0.01 R_{\rm vir}$.

Eight of the haloes -- the cluster-mass systems, C1-C8 -- were
simulated using the publicly available adaptive mesh refinement code
\mlapm\ \cite[][]{Knebe.Green.Binney.01}. We first created a set of
four independent initial conditions at redshift $z=45$ in a standard
\LCDM\ cosmology ($\Omega_0 = 0.3,\Omega_\lambda = 0.7, \Omega_b =
0.04, h = 0.7, \sigma_8 = 0.9$). $512^{3}$ particles were placed in a
box of side length 64\hMpc\ giving a mass resolution of $m_p = 1.6
\times 10^{8}$\hMsun.  For each of these initial conditions
we iteratively collapsed eight adjacent particles to one particle
reducing our particle number to 128$^3$ particles. These lower mass
resolution initial conditions were then evolved until $z=0$. At $z=0$,
eight clusters (labeled C1-C8) from our simulation suite were
selected, with masses in the range 1--3$\times 10^{14}$\hMsun. Then,
as described by \citet{Tormen.97}, for each cluster the particles
within five times the virial radius were tracked back to their
Lagrangian positions at the initial redshift ($z=45$). Those particles
were then regenerated to their original mass resolution and positions,
with the next layer of surrounding large particles regenerated only to
one level (i.e. 8 times the original mass resolution), and the
remaining particles were left 64 times more massive than the particles
resident with the host cluster. This conservative criterion was
selected in order to minimise contamination of the final
high-resolution haloes with massive particles.

The ninth (re-)simulation was performed using the same (technical)
approach but with the ART code
\citep[][]{Kravtsov.Klypin.Khokhlov.97}. Moreover, this particular run
(labeled G1) describes the formation of a Milky Way type dark matter
halo in a box of side 20\hMpc. It is the same simulation as ``Box20''
presented in \citet{Prada.etal.06} and for more details we refer the
reader to that study. The final object consists of about two million
particles at a mass resolution of $6 \times 10^{5}$\hMsun\ per
particle and a spatial force resolution of 0.15\hkpc\ has been
reached. All nine simulations have the required temporal resolution
(i.e. $\Delta t \approx$ 200 Myrs for the clusters and $\Delta t
\approx$ 20 Myrs for the galactic halo, respectively) to accurately
follow the formation and evolution of subhaloes within their respective
hosts and hence are well suited for the study presented here. The high
mass and force resolution of our simulations is sufficient to resolve
an abundance of substructure (the ``satellites'' in our study) within
the virial radius of each host.

The physical properties of the haloes (based upon the halo finding
technique explained below) are summarised in Table~\ref{t:haloprop}
alongside some additional indicators quantifying the mass assembly
histories (cf. \Sec{sec:hosts}).

\subsection{Halo Finding}

Both the haloes and their subhaloes are identified using
\ahf\footnote{{\small \textbf{A}MIGA}'s-{\small
    \textbf{H}}alo-{\small\textbf{F}}inder; \ahf\ can be downloaded
  from \texttt{http://www.aip.de/People/aknebe/AMIGA}. \amiga\ is the
  successor to \mlapm.}, an MPI parallelised modification of the
\mhf\footnote{\mlapm's-\texttt{H}alo-\texttt{F}inder} algorithm
presented in \cite{Gill.etal.04.1}. \ahf~utilises the adaptive grid
hierarchy of \mlapm\ to locate haloes (subhaloes) as peaks in an
adaptively smoothed density field. Local potential minima are computed
for each peak and the set of particles that are gravitationally bound
to the peak are returned. If the peak contains in excess of 20
particles, then it is considered a halo (subhalo) and retained for
further analysis.

For each halo (subhalo) we calculate a suite of canonical properties
from particles within the virial/truncation radius. We define the
virial radius $R_{\rm vir}$ as the point at which the density profile
(measured in terms of the cosmological background density $\rho_b$)
drops below the virial overdensity $\Delta_{\rm vir}$, i.e. $M(<R_{\rm
  vir})/(4\pi R_{\rm vir}^3/3) = \Delta_{\rm vir} \rho_b$. Following
convention, we assume the cosmology- and redshift-dependent definition
of $\Delta_{\rm vir}$; for a distinct (i.e. host) halo in a \LCDM\
cosmology with the cosmological parameters that we have adopted,
$\Delta_{\rm vir}=340$ at $z=0$. This prescription is not appropriate
for subhaloes in the dense environs of their host halo, where the
local density exceeds $\Delta_{\rm vir} \rho_b$, and so the density
profile will show a characteristic upturn at a radius $R \lesssim
R_{\rm vir}$. In this case we use the radius at which the density
profile shows this upturn to define the truncation radius for the
subhalo. Further details of this approach
can be found in \cite{Gill.etal.04.1}.\\


\subsection{Subhalo Tracking}\label{sec:HaloTracker}

In the present study we are primarily interested in the temporal
evolution and disruption of subhaloes and hence a simple halo finder
will not suffice. Therefore we base our analysis on the halo tracking
method outlined in \cite{Gill.etal.04.1}\footnote{The tracking utility
  \texttt{HaloTracker} is part of the publicly available
  \ahf\ distribution.}. At the formation time of the host halo
we identify all subhaloes in its immediate environment using
\texttt{AHF} and track their subsequent evolution using the set of
particles identified at the initial time. Using the 25 innermost
particles at time $t_1$ we calculate the new subhalo centre at the
subsequent time $t_2$. We then check which of the initial set of
particles remain bound to the subhalo; all unbound particles
constitute ``debris''. Particles are unbound if their speed exceeds
the escape velocity by a factor of 1.5.  This tolerance factor is
introduced to allow for any inaccuracy arising from the assumption of
spherical symmetry or uncertainties in the extent of the subhalo. All
particles, bound as well as unbound ones, are tracked forward to the
present day, which allows us to look at the field of debris particles
for each subhalo.

The tracker does not take into account accretion of mass, and so we may 
slightly underestimate subhalo masses. However, it is very unlikely that 
subhaloes will accrete material as they orbit within the host; rather, 
they will suffer continuous mass loss. In contrast to a simple halo 
finder, the tracker is able to detect ``flickering'' subhaloes that 
temporarily disappear in the dense inner regions of the host (e.g. Power 
\& Knebe 2007). The tracker is in the exceptional position of knowing that 
there should be a subhalo and thus can track it as it orbits within the 
densest innermost parts of the host \cite[see][]{Gill.etal.04.1}.

However, there is one draw-back with this otherwise useful feature: a 
subhalo may completely merge with the host, but its particles still appear 
to constitute a bound structure. Such 'fake' subhaloes appear to lie close 
to the centre of the corresponding host. Thus, we remove these objects by 
neglecting every subhalo whose centre lies within the central 2.5\% of the 
virial radius of the host and which remains within this radius for several 
consecutive time-steps.

\subsection{Orbit analysis}\label{sec:orbitanalysis}

For studying subhaloes and their properties, knowing their individual 
orbits around the host centre is important. We focus on subhalo orbits 
in the rest frame of the host (which encompasses all (bound) matter inside 
its virial region). Because of inherent uncertainties in the accuracy with 
which we can determine a halo's centre, the path of the host halo appears 
to follow a ``zigzag'' trajectory. Therefore, as an initial step we smooth 
over the host's orbit before translating subhalo orbits into the host 
frame. This ensures that subhalo orbits do not artificially ``wiggle'' 
when translated to the host frame because of uncertainties in the host's 
precise centre.

In addition, we need to calculate orbital properties such as the number of
orbits around the host centre and eccentricity. We base our
eccentricity determination on finding the apo- and pericentre of a
subhalo's orbit.  For each apo- ($r_a$) and pericentre ($r_p$) pair,
the corresponding eccentricity can then be defined as: 
\bq
\label{eq:eccentricity}
e = \frac{r_a-r_p}{r_a+r_p} \quad .
\eq

However, noisy density determinations (and the influence of other
subhaloes) may introduce additional minima and maxima in the orbit,
thus leading to artificially low eccentricities mimicking nearly
circular orbits. We get rid of these minima and maxima by applying a
``sanity" test to each supposed apo-pericentre pair with an
eccentricity below 0.1: we estimate the time the subhalo
would need for one circular orbit around the host using its mean
distance $\langle r\rangle$ and speed $\langle v\rangle$ based on the
values at the corresponding apo- and pericentre: $T = 2\pi \langle
r\rangle / \langle v\rangle$. If the time between the suspected apo-
and pericentre is less than 30\% of half of the estimated orbital
time, then this minima/maxima pair is most probably not related to the
motion around the host centre and thus is removed. In this way, we
achieve a much more reliable determination of apo- and pericentres for
each subhalo orbit.

This also affects the number of orbits that we estimate, since it relies 
on the number of found apo- and pericentres. Additionally, partial orbits 
at the beginning and end of the studied time interval are added to the 
number of full orbits derived from the extrema of the orbit leading to 
non-integer orbits.

\section{The Host Haloes}
\label{sec:hosts}

\begin{table*}
\begin{center}
\begin{tabular}{lccccccccccccccccc}
\hline
  Host	
& $M_{\rm vir}$
& $R_{\rm vir}$
& $V_{\rm max}$
& $R_{\rm max}$
& $\sigma_{v}$
& age
& $z_{\rm form}$
& $C_{1/5}$
& $\lambda$
& $\left< \frac{\Delta M}{\Delta t M}\right>$
& $\sigma_{\Delta M/M}$
& $T$
& $c/a$
& $Q$
\\ 
\hline
C1 & 2.9   & 1355 & 1141 & 346 & 1161 & 7.9 & 1.05 & 7.33 & 0.0157 & 0.128 &  0.125  & 0.69 & 0.83  & 0.94  \\
C2 & 1.4   & 1067 &  909 & 338 &  933 & 6.9 & 0.80 & 7.21 & 0.0091 & 0.122 &  0.156  & 0.82 & 0.35  & 1.02 \\
C3 & 1.1   &  973 &  828 & 236 &  831 & 6.9 & 0.80 & 7.34 & 0.0125 & 0.100 &  0.117  & 0.88 & 0.49  & 0.96  \\
C4 & 1.4   & 1061 &  922 & 165 &  916 & 6.6 & 0.75 & 6.66 & 0.0402 & 0.127 &  0.207  & 0.82 & 0.86  & 0.91 \\
C5 & 1.2   & 1008 &  841 & 187 &  848 & 6.0 & 0.64 & 6.37 & 0.0093 & 0.129 &  0.141  & 0.79 & 0.95  & 1.03 \\
C6 & 1.4   & 1065 &  870 & 216 &  886 & 5.5 & 0.57 & 6.98 & 0.0359 & 0.147 &  0.153  & 0.69 & 0.83  & 0.94 \\
C7 & 2.9   & 1347 & 1089 & 508 & 1182 & 4.6 & 0.44 & 6.57 & 0.0317 & 0.844 &  1.068  & 0.81 & 0.62  & 0.93 \\
C8 & 3.1   & 1379 & 1053 & 859 & 1091 & 2.8 & 0.24 & 4.09 & 0.0231 & 0.250 &  0.225  & 0.86 & 0.52  & 1.12 \\
G1 & 0.012 & 214  &  210 &  44 &  202 & 8.5 & 1.23 & 8.42 & 0.0229 & 0.073 &  0.107  & 0.80 & 0.69  & 0.97 \\
\hline

\end{tabular}
\caption{Properties of the host haloes in our simulations. Masses are
  measured in 10$^{14}$\hMsun, velocities in km/s, distances in \hkpc,
  and the age in Gyr. $M_{\rm vir}$ is the virial mass, $R_{\rm vir}$
  is the virial radius, $V_{\rm max}$ is the maximum of the rotation
  curve, $R_{\rm max}$ is the position of this maximum, age is the
  time since half of the present day mass was first assembled, $z_{\rm
    form}$ is the corresponding formation redshift, $C_{1/5}=R_{\rm
    vir}/R_{1/5}$ is a measure of the concentration (where $R_{1/5}$
  is the radius containing 20\% of the virial mass), and $\lambda$ is
  the (classical) spin parameter. The following two columns quantify
  the mass assembly of the hosts (in Gyr$^{-1}$ and Gyr$^{-2}$,
  respectively) and we refer the reader to the corresponding
  \Sec{sec:MAH} for details. The last three columns measure the the
  shape of the hosts based upon our novel technique introduced in
  Appendix~\ref{app:shapemethod} and applied to the potential of the
  hosts in \Sec{sec:shapes} as well as the dynamical state $Q$ defined
  in \Sec{sec:dynamics}.}

\label{t:haloprop}
\end{center}
\end{table*}

Some properties of the hosts used in this paper have already been 
presented in a previous study \citep[][]{Warnick.06}. However, for 
completeness we list relevant properties in Table~\ref{t:haloprop}; the 
meaning of listed quantities are explained in this section. Note that the 
shape measurements we present differ from those in \citet{Warnick.06} 
because we employed a new method for shape estimation (cf. 
\Sec{app:shapemethod}).

\subsection{Formation Times}
\label{sec:formation_times}

The standard definition of the formation time of a dark matter halo is 
based on the simple criterion of \citet[e.g.][]{Lacey.Cole.93}, which 
determines formation time to be the time at which a halo's most massive 
progenitor first contains in excess of half its present day mass. If we 
apply this ``half-mass'' definition to our sample, we find that our haloes 
formed between $3 \lesssim \Delta t_{\rm form} \lesssim 8$ Gyrs ago. 

However, we can define an alternative formation time that is based 
explicitly on a halo's merging history. If we determine the time of the 
halo's last \emph{major} merger, then we can use the age of the Universe 
at this time to define the halo's formation time. When applying this 
criterion we note that the ``half-mass'' formation time actually provides 
a lower limit on formation time; for all host haloes no merger with a mass 
ratio greater than 1:3 occurred since the half-mass formation redshift. We 
decided to use the conservative age determination based upon the half-mass 
criterion.

\subsection{Mass Accretion} 
\label{sec:MAH} 

As already introduced in \citet[][]{Warnick.06} we compute the
dispersion of the fractional mass change rate

\begin{equation}
 \sigma_{\Delta M/M}^2 = \frac{1}{N_{\rm out}-1} \sum_{i=1}^{N_{\rm out}-1}
                        \left(\frac{\Delta M_i}{\Delta t_i M_i} - 
                         \left< \frac{\Delta M}{\Delta t M}\right> \right)^2 \ ,
\end{equation}

\noindent
where $N_{\rm out}$ is the number of available outputs from formation
\zform\ to redshift $z=0$, $\Delta M_i = |M(z_i)-M(z_{i+1})|$ the
change in the mass of the host halo, and $\Delta t_i$ is the
respective change in time. The mean growth rate for a given halo is
calculated as follows

\begin{equation} \label{MassGrowthRate}
 \left< \frac{\Delta M}{\Delta t M}\right> = 
 \frac{1}{N_{\rm out}-1} \sum_{i=1}^{N_{\rm out}-1} \frac{\Delta M_i}{\Delta t_i M_i} \ .
\end{equation}

\noindent
A large dispersion $\sigma_{\Delta M/M}$ is indicative of a violent
formation history whereas a small dispersion reflects a more quiescent
history. Our definition for formation time implies that the host
halo's mass has doubled its mass between the formation time and $z$=0,
and as such the inverse of the age of the host provides an estimate
for the mean growth rate, which can be compared with the rate
calculated from \Eq{MassGrowthRate}.

\subsection{Shape of the Gravitational Potential} 
\label{sec:shapes}

A generic prediction of the CDM model is that dark matter haloes are
triaxial systems, that can be reasonably well approximated as
ellipsoids \citep[][etc.]{Frenk.etal.88, Warren.etal.92, Allgood.etal.06}.

Here we apply the method detailed in Appendix~\ref{app:shapemethod} to
our suite of host haloes and define the triaxiality in the usual way
\citep[e.g., ][]{Franx.Illingworth.deZeeuw.91}

\begin{equation}
 T = \frac{a^2-b^2}{a^2-c^2}
\end{equation}

\noindent where the ellipsoidal axes $a>b>c$ are closely related to
the eigenvalues of the inertia tensor (cf. \Eq{eq:ellipsoid_axes}).

Note that the values for $T$ listed in Table~\ref{t:haloprop} differ from 
those in \cite{Warnick.06}. In the previous study we used all particles 
\emph{within the virial radius} (cf. dashed circle in, for instance, 
\Fig{f:hostshape.01-01.NoSub.adweight}) whereas the present values are 
based on the novel method outlined in Appendix~\ref{app:shapemethod}, 
which uses only particles \textit{within an ellipsoidal equipotential 
shell} with a ``mean'' distance (cf. $R_{\rm iso}$ defined in 
Appendix~\ref{app:isoequi}) from the centre of $\approx R_{\rm vir}$ (cf. 
solid lines in the upper panels of \Fig{f:hostshape.01-01.NoSub.adweight}).

\subsection{Dynamical State}
\label{sec:dynamics}

We use the virial theorem in order to quantify the dynamical state of
our host haloes. For a self-gravitating system of $N$ (collisionless)
particles it is given by

\bq
\label{eq:VirialTheorem}
\frac{1}{2} \frac{d^2I}{dt^2} = 2E_{\rm kin} + E_{\rm pot} \ ,
\eq

\noindent 
where $I$ is the inertia tensor. The potential $E_{\rm pot}$ and
kinetic $E_{\rm kin}$ energies are given by

\bqa 
\label{eq:EkinEpot}
E_{\rm kin} &=& \frac{1}{2}\sum_{i} (m_i v_i^2)\\
E_{\rm pot} &=& \frac{1}{2}\sum_{i} (m_i \Phi_i)	\ ,
\eqa

\noindent
where the summations are over all $N$ particles. If we assume that the
inertia tensor does not depend on time, then
expression~(\ref{eq:VirialTheorem}) reduces to the more familiar

\bq \label{eq:UsualVirial}
Q = 2E_{\rm kin}/|E_{\rm pot}| = 1	\ ,
\eq

\noindent which we refer to as the ``virial ratio''. The summation to
obtain both the kinetic and potential energy is performed over all
particles inside an equipotential surfaces. We chose the one whose
``mean'' distance (cf. $R_{\rm iso}$ defined in
Appendix~\ref{app:isoequi}) to the host's centre is $R_{\rm vir}$,
just as for the shape determination in \Sec{sec:shapes}.

\subsection{Summary -- Host Haloes} 
\label{sec:host_summary}

All our host haloes appear to be relaxed and in dynamical equilibrium
with a range of formation times and triaxialities.  Hence we have an
interesting sample of hosts to study the formation and evolution of
tidal debris from disrupting subhaloes orbiting within and about
them. However, visual inspection of the temporal evolution of the
hosts reveals that C8 is experiencing an ongoing triple merger, which
explains why it has the largest deviation of $Q$ from unity amongst
all haloes in our sample.

\section{The Subhaloes}\label{sec:subhaloes}

Strong tidal forces induced by the host halo will lead to a
substantial mass loss from the orbiting subhaloes. In this section we
examine this mass loss in detail and investigate how it relates to the
orbital parameters of the subhaloes. In particular we consider two
distinct subhalo populations, ``trapped'' and ``backsplash''
subhaloes. Trapped subhaloes enter the virial radius of the host at an
earlier time and their orbits are such that they remain within the
virial radius until the present day, being forever ``trapped'' in the
potential of the host. Backsplash subhaloes were introduced earlier
\cite[cf. Introduction, ][]{Gill.etal.05.3}. Both trapped and
backsplash subhaloes suffer mass loss, but it is instructive to ask
how this lost mass contributes to the tidal debris field within the
host halo. In particular, backsplash subhaloes lose a substantial
fraction of their mass as they pass through the denser environs of
their host (on average 40\%, \citet[][]{Gill.etal.05.3}), but how
important is their contribution to the tidal debris field? These are
interesting questions that will be addressed in this section.

\subsection{Trapped \& Backsplash Subhaloes} 
\label{sec:backsplash}
\begin{figure}
\begin{center}
        \epsfig{file=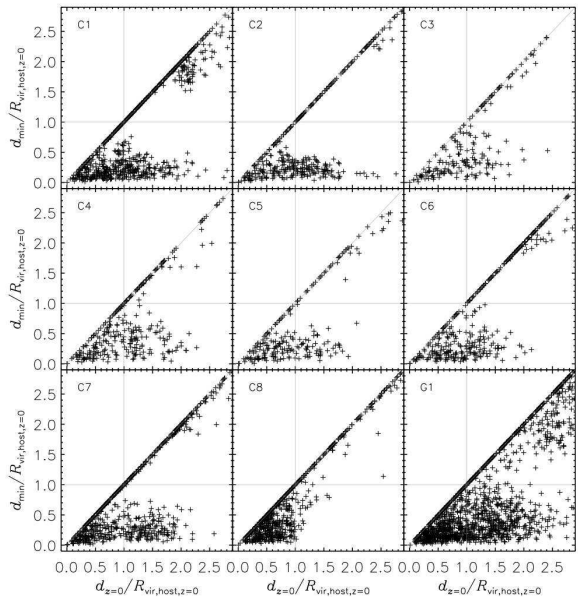, width=0.47\textwidth, angle=0}
\end{center}
\caption
{ Minimum distance of a subhalo versus present distance. Subhaloes in the lower 
right corner of the plot are 'backsplash galaxies': they have once been inside the
 host, but now are sitting outside the virial radius. Subhaloes in the upper right
 part have never been inside the host, but most of them are approaching the host 
halo ($d_{z = 0} = d_{\rm min}$, solid grey line). (compare with Figure 2 of Gill 
et~al. 2005))}
\label{f:satmindist}
\end{figure}

As noted by both \cite{Gill.etal.05.3} and \cite{Moore.Diemand.Stadel.04} 
a prominent population of backsplash galaxies is predicted to exist in 
galaxy clusters. We now demonstrate that a backsplash population is also 
predicted to exist in galaxies. In \Fig{f:satmindist} we show for each 
subhalo its minimum distance with respect to the centre of the host halo 
as a function of its distance with respect to the centre at the present 
day; both are normalised with respect to the host's virial radius at 
redshift $z=0$.

\Fig{f:satmindist} reveals that backsplash subhaloes are also present in 
our galaxy mass halo. The backsplash population occupies the bottom right 
rectangle, with $d_{z=0} \geq 1 R_{\rm vir,host,z=0}$ and $d_{\rm min} \leq 1 
R_{\rm vir,host,z=0}$. For the clusters, about 50\% of the subhaloes within a 
distance $d \in [R_{\rm vir, host},2\,R_{\rm vir, host}]$ once passed 
through the virial sphere of their respective host. For the galactic halo 
G1, this figure rises to 63\%. We note that the concept of the backsplash 
subhalo can partly be ascribed to the difficulty in choosing a good 
truncation radius for the host halo: the virial radius is more or less an 
arbitrary definition for the outer edge of a halo \citep[cf. 
][]{Diemand.Kuhlen.Madau.06} and it has been shown by others that bound 
objects (in isolation) may as well extend way beyond this formal 
point-of-reference \citep[][]{Prada.etal.06}.

\begin{table}
\begin{center}
\begin{tabular}{|l|c|c|c|}
\hline
host& $n\ (R_{\rm vir})$ & $n_{\rm bs}\ (2\,R_{\rm vir})$ & $n_{\rm bs}/n_{\rm all}\ (2\,R_{\rm vir})$\\\hline\hline
       C1 &  325	&	 149	  &	43\%\\
       C2 &  168	&	\s 98	  &	55\%\\
       C3 & \s 93	&	\s 54	  &	59\%\\
       C4 &  136	&	\s 84	  &	56\%\\
       C5 &  122	&	\s 50	  &	62\%\\
       C6 &  148	&	\s 75	  &	46\%\\
       C7 &  326	&	 163	  &	57\%\\
       C8 &  477	&	\s 22	  &	19\%\\[1ex]
       G1 &  606	&	 425	  &	63\%\\
\hline
\end{tabular}
\caption{Number of subhaloes inside the virial region of the host ($n$), backsplash subhaloes between 1 and 2 virial radii ($n_{\rm bs}$) and the fraction of backsplash subhaloes in that region (with respect to all subhaloes between 1 and 2 virial radii). Only subhaloes containing at least 20 particles are counted.}
\label{t:backsplash}
\end{center}
\end{table}

The numbers of backsplash subhaloes for each host are given in Table 
\ref{t:backsplash}, along with the number of trapped subhaloes. We note 
that the number of subhaloes is slightly larger than is given in 
\cite{Warnick.06}, reflecting differences in how subhaloes are identified. 
In the previous study we used \texttt{AHF}, which identifies a ``static'' 
distribution of subhaloes in a single snapshot, whereas in the present 
study we use the halo tracker \texttt{AHT} \cite[see][for more 
details]{Gill.etal.04.1}.

Interestingly, the fraction of backsplash subhaloes within one and two 
virial radii of the host varies between 43\% and 63\% for all hosts, 
except for the youngest system C8. There we find a rather low fraction of 
only 19\%. This host is not only the youngest, but also an ongoing triple 
merger providing the most 'unrelaxed' environment amongst our suite of 
hosts (cf. $Q$ value in Table~\ref{t:haloprop}). Thus maybe there are some 
potential backsplash subhaloes which simply did not yet have the time to 
leave their host again.

\subsection{Mass Loss from Trapped Subhaloes}

\begin{figure}
\begin{center}
        \epsfig{file=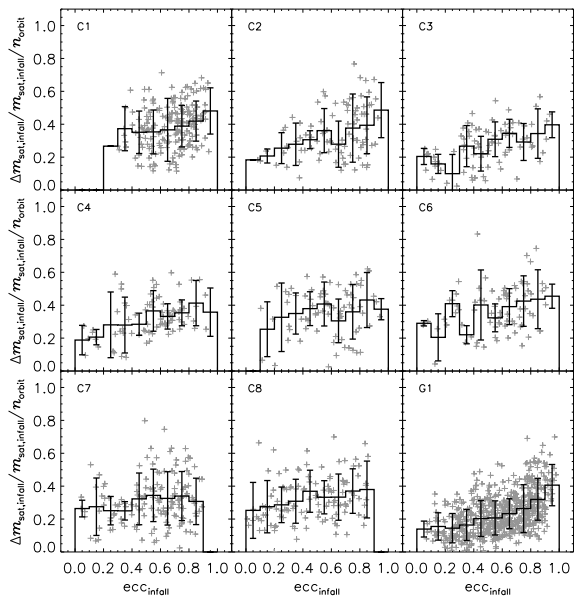, width=0.47\textwidth, angle=0}
\end{center}
\caption { Fractional mass loss per orbit since infall time as a
  function of infall eccentricity. Subhaloes were selected according
  to the criteria: $\ge 1$ orbit, $n_{\rm part} \ge 20$ particles
  today and they must reside inside the host today.  }
\label{f:satmassloss}
\end{figure}

Before investigating tidal debris fields it is necessary to quantify
the actual mass loss. Are our subhalo systems actually losing mass
and how does this mass loss relate to orbital parameters such as
infall eccentricity? To answer this question we show in
\Fig{f:satmassloss} that trapped subhaloes on more eccentric orbits
experience in general larger mass loss. We plot the total mass loss
since infall time normalised to the initial mass divided by the number
of orbits as a function of orbital eccentricity. The orbital
eccentricity is derived here from the first apo- and pericentre
distance of the subhalo's orbit, after the subhalo entered the
host region. This ``infall eccentricity'' might deviate strongly from
the present value, since the eccentricity usually is evolving within
time \cite[see e.g.][]{Hashimoto.Funato.Makino.03, Gill.etal.04.2,
  Sales.etal.07}.

While in general all of the trapped subhaloes lose about 30\% of their mass per 
orbit, some objects on radial orbits may lose up to 80\% per orbit or even 
more. We like to note that only surviving haloes ($\ge 20$ particles) enter 
these plots and completely disrupted ones were not considered, 
respectively. Furthermore, only objects with at least one orbit were taken 
into account for \Fig{f:satmassloss} as we needed one orbit to get a 
reliable measure for the (infall) eccentricity and the number of orbits.

\subsection{Mass Loss from Backsplash Subhaloes} 
\label{sec:mass_loss_backsplash}

\begin{figure}
\begin{center}
        \epsfig{file=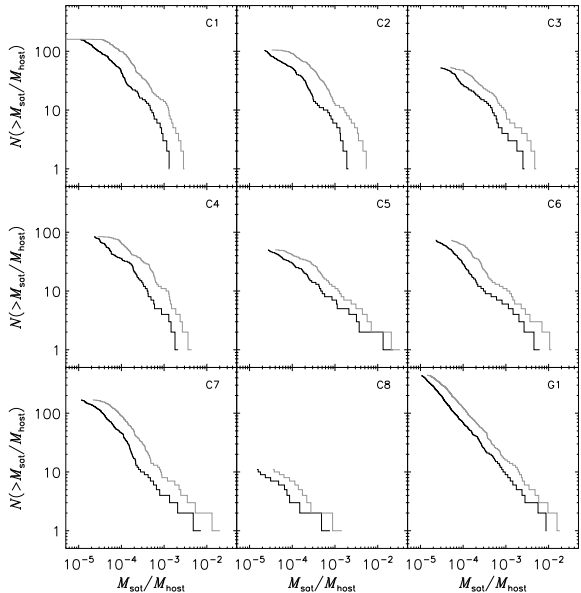, width=0.47\textwidth, angle=0}
\end{center}
\caption { The cumulative mass function of the backsplash galaxies at
  redshift $z=0$ (heavy lines) and when using the mass at the infall
  time of each individual backsplash galaxy (faint lines). }
\label{f:satbacksplash_massfct}
\end{figure}

\begin{figure}
\begin{center}
        \epsfig{file=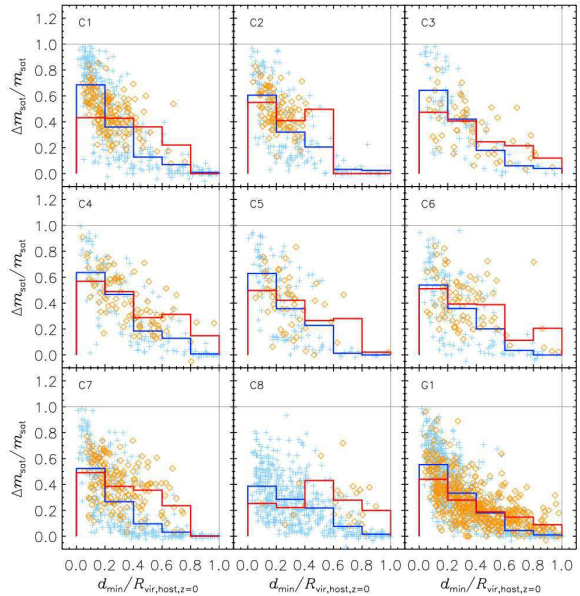, width=0.47\textwidth, angle=0}
\end{center}
\caption { Fractional mass loss from subhaloes since infall time
  versus the normalised minimum distance to the host. Only surviving
  subhaloes are plotted. Trapped subhaloes are presented by blue
  plus signs, backsplash galaxies are marked by orange diamonds. }
\label{f:satbacksplash_massloss_all}
\end{figure}

The results of the previous subsection show that trapped subhaloes suffer mass 
loss whose rate is enhanced for systems on more radial orbits. We now examine mass
loss from backsplash subhaloes.

In \Fig{f:satbacksplash_massfct} we show the cumulative mass function of all 
backsplash subhaloes at redshift $z=0$ (heavy lines) and compare with the 
mass function of these subhaloes at their time of infall onto their host 
(faint lines). The two sets of mass functions approximately preserve their 
shape and differ effectively only in horizontal offset -- the $z$=0 mass 
function can be obtained by a shift to the left of the mass function at 
infall. Further inspection of the figure reveals that \textit{every} 
backsplash galaxy loses on average half its mass as it travels within the 
virial radius of the host and out again, irrespective of its original 
(infall) mass, i.e. the mass functions presented for each host halo in 
\Fig{f:satbacksplash_massfct} are self-similar.

Although backsplash subhaloes may not count as ``true'' subhaloes of the 
host, they experience mass loss within the host and therefore contribute
to the tidal debris fields that we study in \Sec{sec:streams_sat} and 
\Sec{sec:trailing/leading}. We verify this in 
\Fig{f:satbacksplash_massloss_all}, similar in spirit to \Fig{f:satmassloss},
where we plot the total mass loss from a subhalo over its full orbit since 
infall time as a function of minimum distance in terms of current host radius. 
We separate trapped and backsplash subhaloes by plotting the former with blue plus signs 
 and the latter with orange diamonds. This figure demonstrates that 
subhaloes that have small pericentres with respect to the centre of the host
experience greater mass loss, regardless of whether the subhalo is 
trapped or backsplash. We show also the average mass loss at a given minimum 
halocentric distance by the histograms in \Fig{f:satbacksplash_massloss_all}. 
Interestingly, it is noticeable that backsplash subhaloes (orange) that grazed 
the outer regions of the host suffer greater mass loss than trapped subhaloes 
(blue) that reside in the same region. 

\begin{table*}
\begin{center}
\begin{tabular}{|l|c|c|c|c|c|c|}
\hline
host & $n_{\rm p}$ & $n_{\rm p,sat}/n_{\rm p}$ & $n_{\rm p,debris}/n_{\rm p}$ & $n_{\rm p,sat}/n_{\rm p,debris}$ &
$n_{\rm p,bspl}/n_{\rm p}$ & $n_{\rm p,bspl}/n_{\rm p,debris}$\\
\hline
\hline
C1  &   1810607   &  \s 7\%  &     20\% &  0.34  & 0.39\% &  \s 2\% \\
C2  &    884488   &  \s 8\%  &  \s 8\%  &  0.96  & 0.59\% &  \s 10\% \\
C3  &    672442   &  \s 5\%  &    21\%  &  0.23  & 0.36\% &  \s 2\% \\
C4  &    870732   &  \s 7\%  &  \s 10\% &  0.69  & 0.28\% &  \s 5\% \\
C5  &    744868   &  \s 4\%  &    22\%  &  0.18  & 0.58\% &  \s 5\% \\
C6  &    879644   &    11\%  &    25\%  &  0.44  & 0.52\% &  \s 3\% \\
C7  &   1798507   &    12\%  &  \s 2\%  &  4.89  & 0.16\% &    13\% \\
C8  &   1906744   &    36\%  &    20\%  &  1.81  & 0.02\% &  \s 0\% \\[1ex]
G1  &   2227123   &  \s 7\%  &    12\%  &  0.60  & 0.67\% &  \s 6\% \\
\hline
\end{tabular}
\caption{Number of particles inside the virial radius of the host (at
  $z=0$): $n_{\rm p} =$ total number, $n_{\rm p,sat} =$ the number of
  particles in still existing subhaloes ($\ge 20$ bound particles),
  $n_{\rm p,debris} =$ total number of particles in debris (including
  destroyed subhaloes and debris from backsplash galaxies), $n_{\rm
    p,bspl} =$ debris particles from backsplash subhaloes alone and
  the last column provides the ratio of the latter two numbers.}
\label{t:particlenumbers}
\end{center}
\end{table*}

This figure also underlines our previous assertion that backsplash
galaxies contribute to the overall tidal debris field. Table
\ref{t:particlenumbers} quantifies this: at most 13\% of the debris
particles found within the host originate from backsplash
subhaloes\footnote{We note a few instances where we find negative mass
  loss, i.e. subhaloes actually gained mass. This is due to marginal
  technical differences between the halo finder and the halo tracker:
  while the finder provides the tracker with the initial set of
  particles belonging to the (sub-)halo the tracker checks whether
  these particles are bound or unbound.  While the finder uses local
  potential minima as the centre of the halo the tracker re-calculates
  the centre as the centre-of-mass of the innermost 25 particles. If
  there is now a marginal shift this can lead to differences in the
  evaluation of the potential and hence the odd particle can be
  considered unbound by the tracker.  However, because it is in fact
  bound it will ``merge'' with the halo again mimicking mass
  growth. We carefully checked these instances and those objects
  appearing with negative mass growth are in fact subhaloes that did
  not suffer any mass loss at all.}.

Table~\ref{t:particlenumbers} gives the fraction of particles in
subhaloes, debris and backsplash galaxies: our hosts' masses in
general consist of about 5 to 10\% (in mass) from surviving subhaloes
while 10 or 20\% of the mass is contained in debris particles. A mere
0.5\% of the total host mass originates from debris of backsplash
galaxies. We further make the following observations about the subhalo
and debris fractions of individual hosts.

\begin{itemize}

\item We note that the subhalo fraction in C8 is high, reflecting its involvement 
	in a recent merger. Because a single massive subhalo persists (in fact, 
	one of the most massive progenitor of the host) that has not yet merged, 
	the subhalo mass fraction is close to 36\%. 

\item The debris fraction for host C6 amounts up to nearly 25\% because this host
      contains several massive subhaloes. We checked the initial subhalo mass 
      function, i.e. the mass function of all trapped subhaloes at the formation 
      time of the host, and indeed C6 contains initially more mass in subhaloes 
      than any of the other hosts. The subhalo destruction rate is similar for all
      hosts and so the high debris fraction reflects the number of massive 
      subhaloes that were present within the host at its formation time, rather 
      than reflecting enhanced disruption of subhaloes within the host. 

\item In contrast, C7 is a host with a very low debris content, although it 
      contains an average number of subhaloes. A detailed investigation of its
      temporal evolution revealed that neither the number of subhaloes nor
      the rate of subhalo disruption varies significantly with time. This is 
      consistent with the small measured debris fraction.

\end{itemize}

It is instructive to compare our findings against the results of, for example, 
\citet[][]{Diemand.Kuhlen.Madau.06} and \citet[][]{Gao.etal.04}. The former 
investigated mass loss per orbit as a function of pericentric distance (cf. 
their Fig.17) and arrived at the same conclusion: the closer a subhalo gets 
to the host's centre, the greater its mass loss. A similar result was obtained by
\citet[][]{Gao.etal.04} who found that subhaloes with large pericentric distance
retains a greater fraction of their initial (i.e. infall) mass (cf. Fig. 15).

\subsection{Summary -- Subhaloes}

Our study of the subhalo populations in the hosts reveals that backsplash subhaloes
are present in both galaxy and galaxy cluster haloes. Approximately half of the haloes
between $1$ and $2$ virial radii are actually backsplash subhaloes. Both trapped 
and backsplash subhaloes suffer mass loss as they orbit within the dense environs 
of the hosts -- between 30 to 80\% of their mass per orbit. The rate of mass loss 
is enhanced for those subhaloes that follow highly eccentric orbits and for those 
subhaloes whose orbits take them to small halocentric distances. This is in good 
agreement with the findings of previous studies \citep[e.g., ][]{Gao.etal.04, 
Diemand.Kuhlen.Madau.06}. 

This lost mass will contribute to the tidal debris field within the hosts. The 
fraction of a host's mass that comes from this lost mass ranges from a few 
percent (C7) to about 25\% (C6). This diversity reflects differences in the subhalo
populations that were present at the formation time of the host, and differences in
the accretion histories of the hosts. We note that the contribution of mass lost 
from backsplash subhaloes to the total debris inside the host halo region today 
amounts to at most 13\%. However, we conclude that \textit{when observing the 
debris of subhaloes and searching for possible corresponding subhalo galaxies, 
the region outside the virial radius of the host galaxy should also be considered.}

\section{Tidal Streams I:\\ Uncovering Subhalo Properties} 
\label{sec:streams_sat}

When a subhalo moves on its path around the centre of its host, tidal
forces (mainly from the host) tear it apart, usually generating
the formation of a leading arm ahead of the subhalo and a trailing arm
behind it. These arms are not always clearly distinguishable;
sometimes they are broadened to such a degree that only a rather
smooth sphere of unbound particles remains. How does the morphology of
the debris field depend on the subhalo and the host? Can we learn
anything about the host halo or the subhalo itself by studying \emph{only}
its debris field?

At this point it is instructive to consider the recent studies of 
\citet[][]{Penarrubia.etal.06} and \citet[][]{Sales.etal.07}. 
\citet[][]{Penarrubia.etal.06} investigated the formation and 
evolution of tidal streams in analytic time-evolving dark matter host halo 
potentials and showed that observable properties of streams at the present-day 
can only be used to recover the present-day mass distribution of the host -- 
they cannot tell us anything about the host at earlier times. However, if the full
 six dimensional phase space information of a tidal stream is available, it is 
also possible to determine the total mass loss suffered by the progenitor 
subhalo. \citet[][]{Sales.etal.07} showed that the survival of a subhalo in 
an external tidal field depends sensitively on its infall mass. One of our primary
interests is the relation between the debris field and the total mass loss of 
the progenitor subhalo as well as its orbital eccentricity.

In the following analysis we focus on subhaloes that are the
progenitors for tidal debris fields that contain at least 25
particles. We have taken care to ensure that low-mass subhaloes are
not artificially disrupted (because of finite time-stepping, force
resolution, particle number) by verifying that the fraction of a
subhalo's mass in the debris field does not correlate with the mass of
the subhalo at infall. If numerics were playing a role, then we would
expect lower mass systems to show preferentially higher mass fractions
in the debris field. We further remade all plots showing analysis of
the debris fields restricting the minimum number of particles in the
progenitor satellite to $N_{\rm min}=600$. We could not find any
substantial differences between results and hence adhere to the
original plots. We note that the plots obtained by using the more
restrictive limit on the satellite's mass appear like having randomly
sampled the original data; that is, they show the same
trends. However, we note that in some instances extreme outliers are
removed.

\subsection{Orbital and Debris Plane} \label{sec:fitting}

\subsubsection*{Definition}

\begin{figure}
\begin{center}
\begin{minipage}{0.495\textwidth}
        \epsfig{file=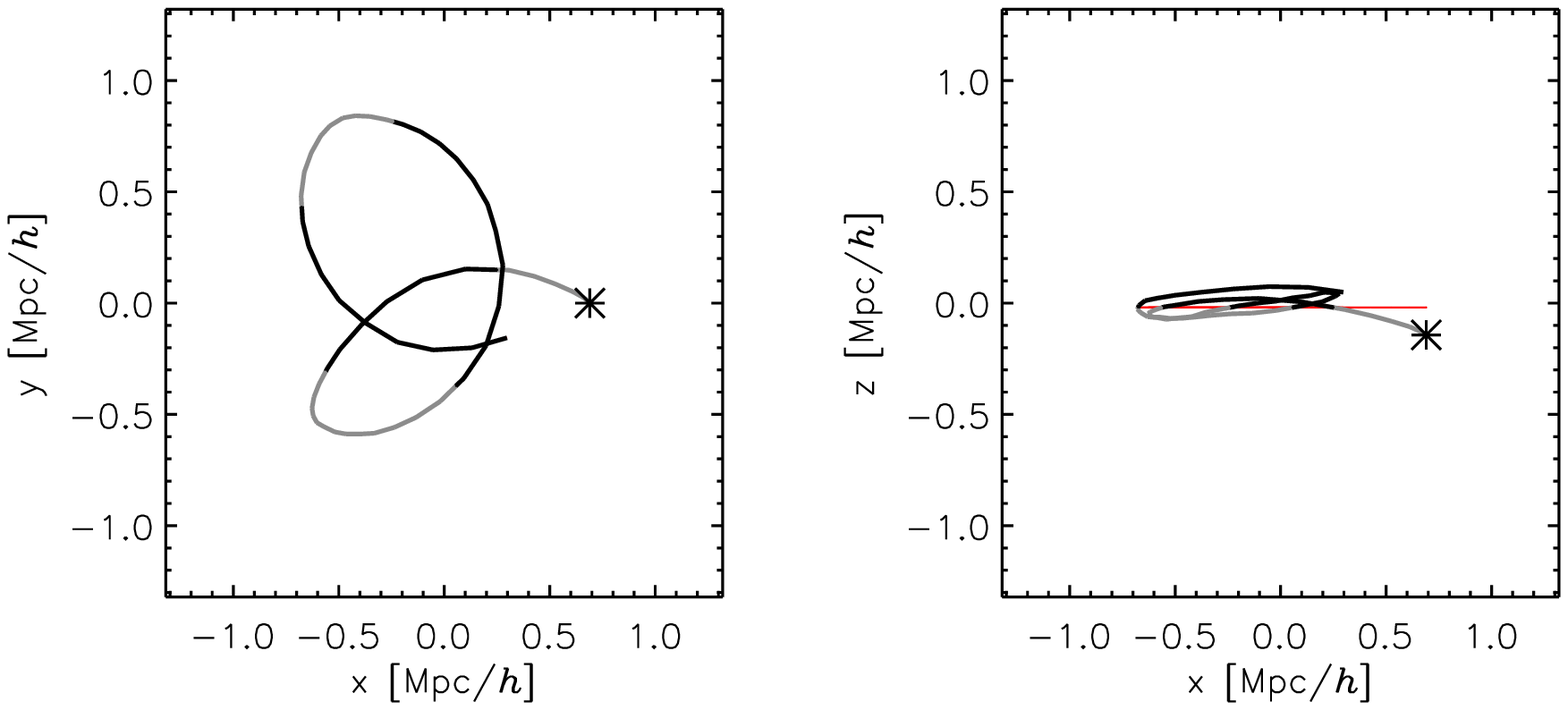, width=1\textwidth, angle=0}
        \epsfig{file=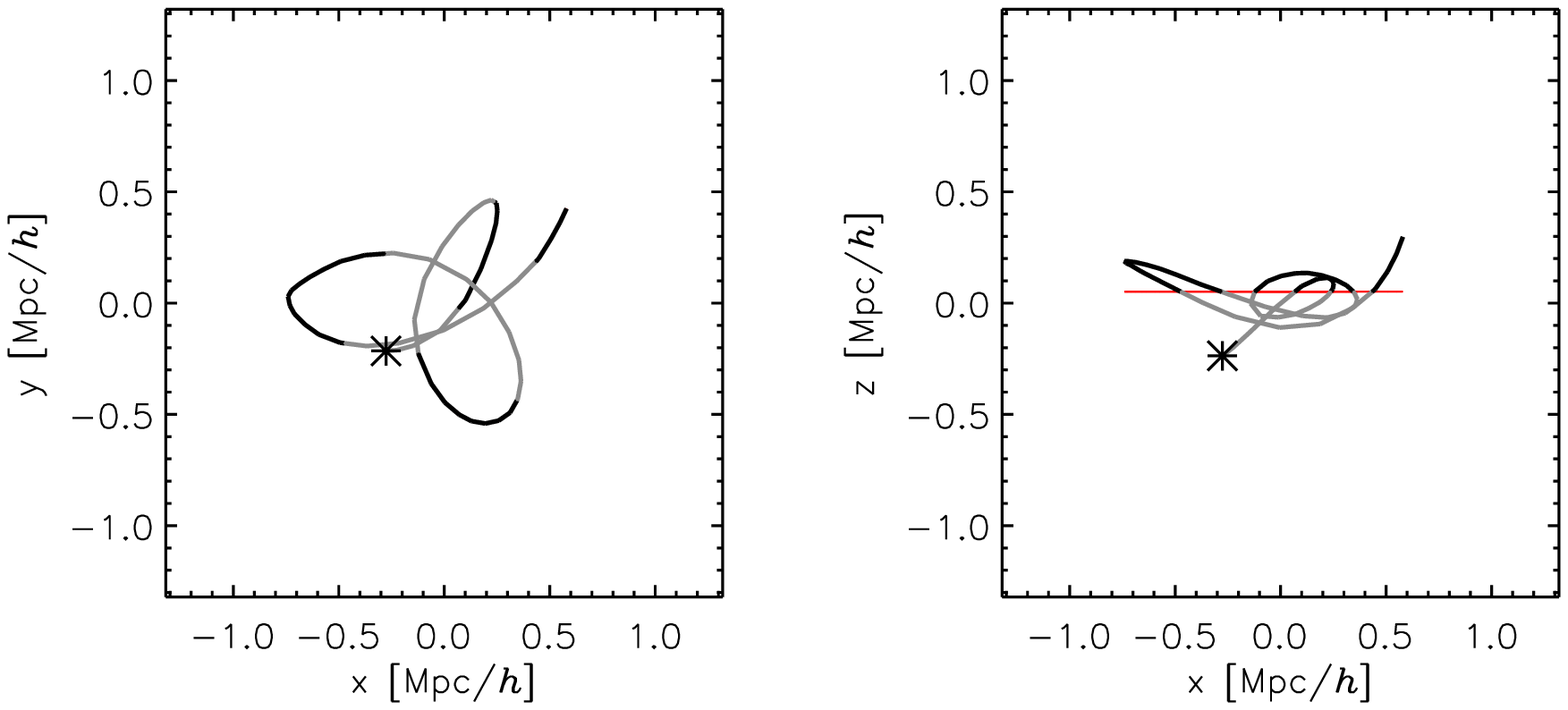, width=1\textwidth, angle=0}
        \epsfig{file=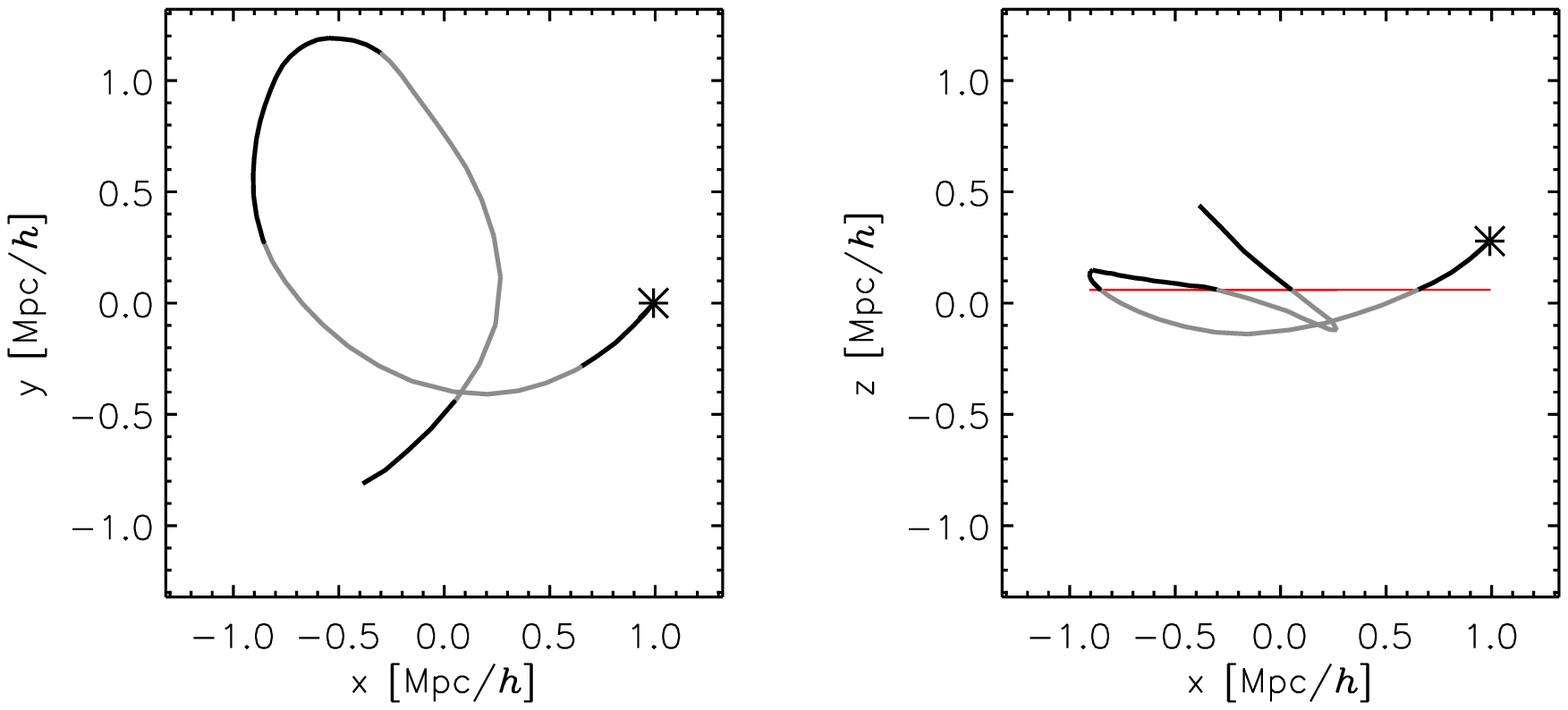, width=1\textwidth, angle=0}
\end{minipage}
\end{center}
\caption { Fitting a plane to the orbit of three sample subhaloes in
  host halo C1. The left panel shows the orbit projected to its best
  fit plane, the right panel shows a perpendicular view, directly
  looking at the edge of the plane (straight red line).  An asterisk
  marks the position of the subhalo today ($z=0$). The host halo's
  centre is located at (0,0,0).}
\label{f:planefit1}
\end{figure}

For an ideal spherical host potential the orbit of a subhalo
(and therefore its debris field) is constrained to a plane (i.e. the
orbital plane) because angular momentum is conserved. In
non-spherical potentials, however, the orbit precesses out of the
plane, and this leads to a thickening of the debris field in this
direction \cite[e.g.][]{Ibata.etal.01, Helmi.04,
  Johnston.Law.Majewski.05, Penarrubia.etal.06}. In addition to this
effect, we expect the debris fields of more massive subhaloes and of subhaloes on highly eccentric orbits to show a greater spatial spread because they are
more likely to lose more of their mass over a larger volume.

In order to quantify the deviation of both the subhaloes' orbit and 
debris field from the orbital plane, we first determine a ``mean orbital 
plane'', i.e. the best-fit plane to the orbit of the subhalo.
Utilising all available orbital points for each individual subhalo 
(since infall time), we applied the Levenberg-Marquardt algorithm 
\cite[see e.g.][Chapter 15.5]{NumRecC} to minimise the distances of the 
points (without loss of generality) in $z$-direction.
When using the general equation of a plane $E: ax+by+cz+d = 0$ (with the 
normal vector $\vec{n} = (a,b,c)$), we obtain for a given $x$ and $y$ the 
expected $z$-value:

\bqa
z &=& \alpha x + \beta y + \delta\\
\mathrm{with}&& \alpha = -\frac{a}{c}\\
	     && \beta  = -\frac{b}{c}\\
	     && \delta = -\frac{d}{c}\\\nonumber
	     && c \ne 0
\eqa

\noindent
Employing the Levenberg-Marquardt algorithm then delivers the coefficients 
$\alpha$, $\beta$ and $\delta$ describing our best fit plane.

Examples of our fitting results are given in \Fig{f:planefit1}. They 
illustrate how we can fit a plane to even a precessing orbit (cf. bottom 
panel). We have inspected visually many of our fits to orbital planes and 
we find that the ``mean orbital planes''  provide an adequate approximation
\footnote{The interested reader can look at \Fig{f:allmpsattidalarms} 
in \Sec{sec:trailing/leading} which shows examples of not only the subhalo orbit
but also the debris field in projections of the best-fit orbital planes.}.

\subsubsection*{Scatter about the orbit plane}
\begin{figure*}
\begin{center}

\begin{minipage}{0.495\textwidth}
        \epsfig{file=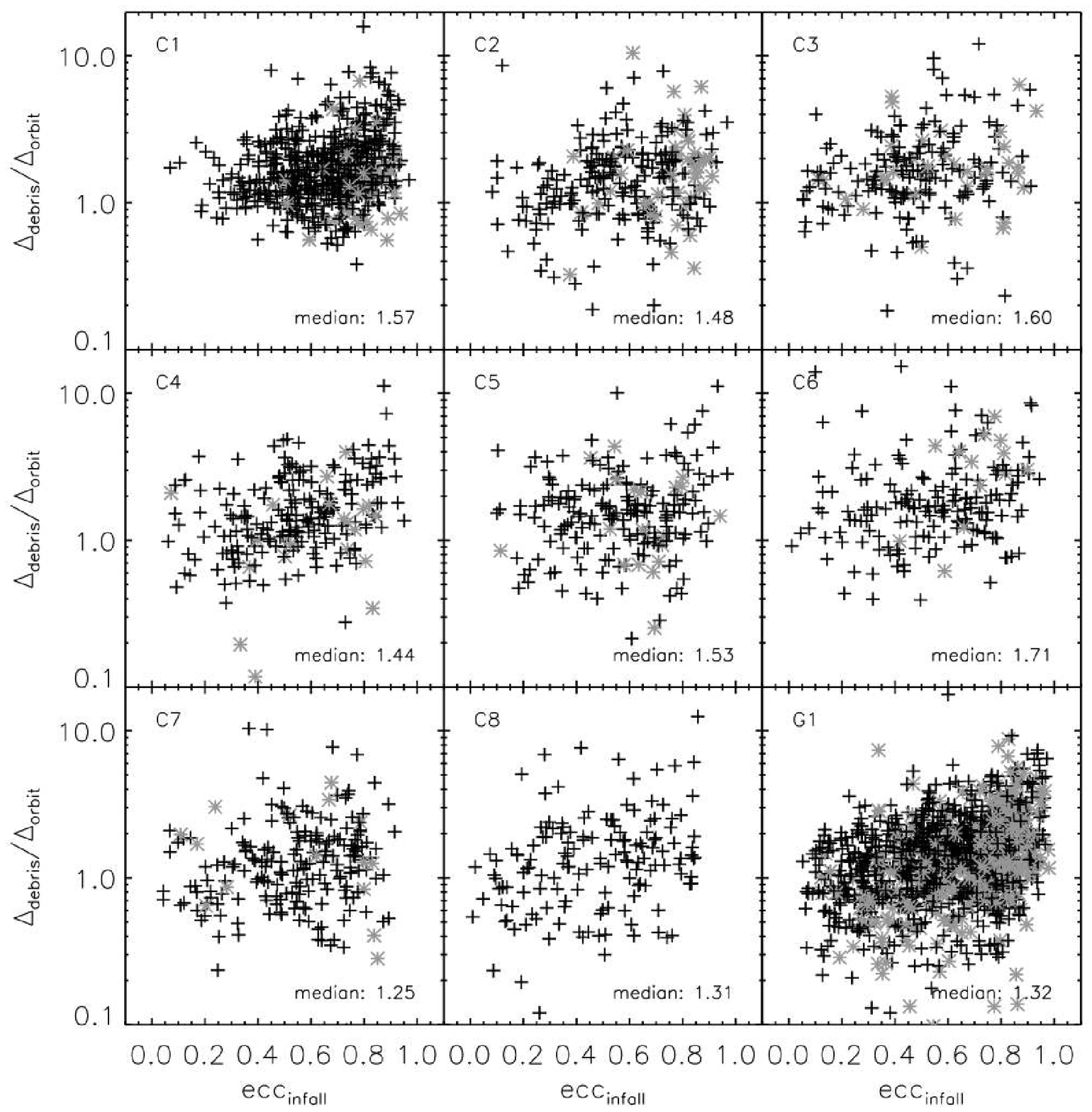, width=1\textwidth, angle=0}
\end{minipage}
\hfill
\begin{minipage}{0.495\textwidth}
        \epsfig{file=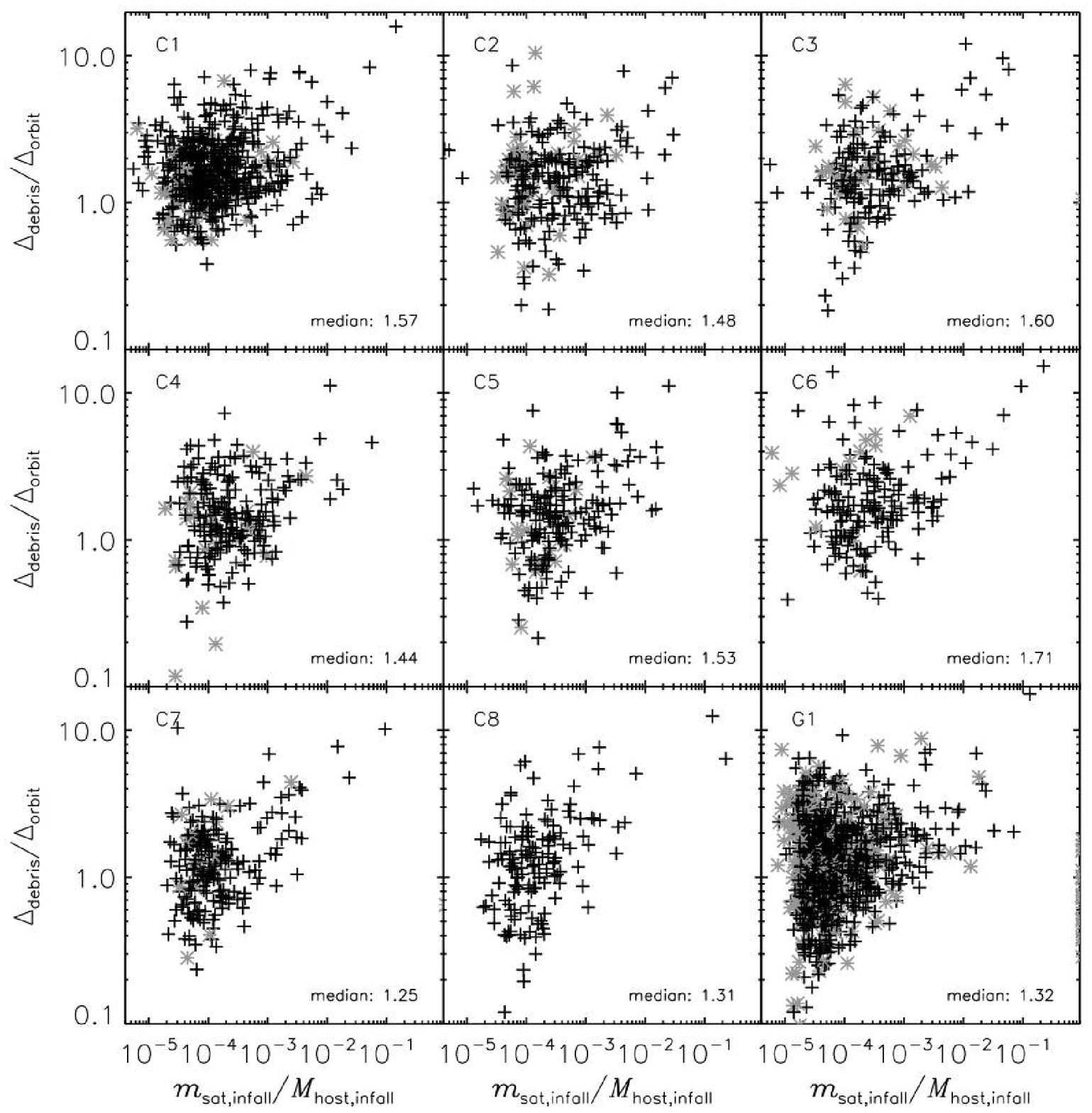, width=1\textwidth, angle=0}
\end{minipage}

\end{center}
\caption { The deviation of debris particles from the best fitting
  \emph{orbital plane} (with respect to the deviation of the orbit
  itself) as a function of infall eccentricity (left panel) and infall
  mass (right panel) for subhaloes with at least one
  orbit. Backsplash subhaloes are marked with grey asterisks. The
  median values of the distribution (trapped and backsplash galaxies
  together) are given in the lower right corner of each panel.}
\label{f:orbitplanedev}
\end{figure*}

Given the best-fit orbital plane, we now have a means to quantify the 
deviation of the orbit itself from this `mean' plane as well as the 
scatter of the debris particles about this plane. Using the orbital plane 
given by the equation $E: ax+by+cz+d = 0$, we determine the perpendicular 
distance $\Delta_i$ of each orbit point/debris particle to the plane via 
the (general) equation
\bq
\delta_i = \left| \frac{a\,x_i+b\,y_i+c\,z_i+d}{a^2+b^2+c^2} \right|
\eq
The mean deviation of the orbit $\Delta_{\rm orbit}$ is determined using
\bq
\Delta^2_{\rm orbit} = \frac{1}{n-3} \sum_{i=0}^{n} \delta_i^2	\quad.
\eq
The factor $(n-3)$ corresponds to the number of degrees of freedom, i.e. 
`number of data points ($n$)' $-$ `number of free parameters (3)'.

For the scatter of the debris, we use an analogous formula for the 
mean deviation:
\bq
   \Delta^2_{\rm debris} = \frac{1}{n_{\rm debris}-1} \sum_{i=0}^{n_{\rm debris}} \delta_i^2	\quad.
\eq

\Fig{f:orbitplanedev} shows the resulting debris scatter normalised to
the deviation of the orbit itself for all chosen subhaloes in each
simulation as a function of the infall eccentricity and the infall
mass. Only subhaloes that have undergone at least one complete orbit
inside the host halo are shown, since only then a reliable (infall)
eccentricity determination based upon (first) apo- and pericentre of
the orbit is possible. Disrupted subhaloes, or rather their debris,
are also included in these plots. Subhaloes that once resided inside
the host but are today found outside the host's virial radius,
i.e. the backsplash galaxies (see \Sec{sec:backsplash}), are marked by
grey asterisks. Interestingly, these are often the points deviating
most from our expectation: we expect to observe a correlation of
deviation with (infall) eccentricity, i.e. the larger the orbital
eccentricity the greater the deviation of the tidal debris from the
orbit. 

Only a marginal trend is apparent that is most pronounced for
the oldest systems (i.e. C1 and G1): C1, on the one hand, is one of
those with a prolate shape, deviating more strongly from sphericity
than most others. Thus, the large deviation of debris from the orbit
plane could be a result of the flattened host halo as suggested by
controlled experiments of disrupting subhaloes in flattened
analytical host potentials \citep[e.g.][]{Ibata.etal.03, Helmi.04,
  Johnston.Law.Majewski.05, Penarrubia.etal.06}. G1, on the other
hand, contains the most subhaloes, thus encounters of debris streams
with other sub-clumps are very likely, causing a heating of the stream
which spreads the debris even more \cite[][]{Moore.etal.99,
  Ibata.etal.02, Johnston.Spergel.Haydn.02, Penarrubia.etal.06}.

The right panel of \Fig{f:orbitplanedev} suggests that the
trend is significantly more pronounced when plotting the
deviation from the orbit plane against (infall) mass of the
subhalo. Massive subhaloes always lead to large deviations of the
debris about the orbital plane, whereas the debris of low mass
subhaloes shows all kinds of deviations. We cannot construct a 
unique mapping of ``debris deviation'' onto ``progenitor mass''
by this method alone, but we note that it agrees with the findings of
\citet[][]{Penarrubia.etal.06}, who claim that it is possible to infer
the amount of mass loss from the stream's progenitor using the full
phase-space distribution of a tidal stream. We will return to this
point later when studying the energies of stream particles.

\subsubsection*{Scatter about the debris plane}
\begin{figure*}
\begin{center}
\begin{minipage}{0.495\textwidth}
        \epsfig{file=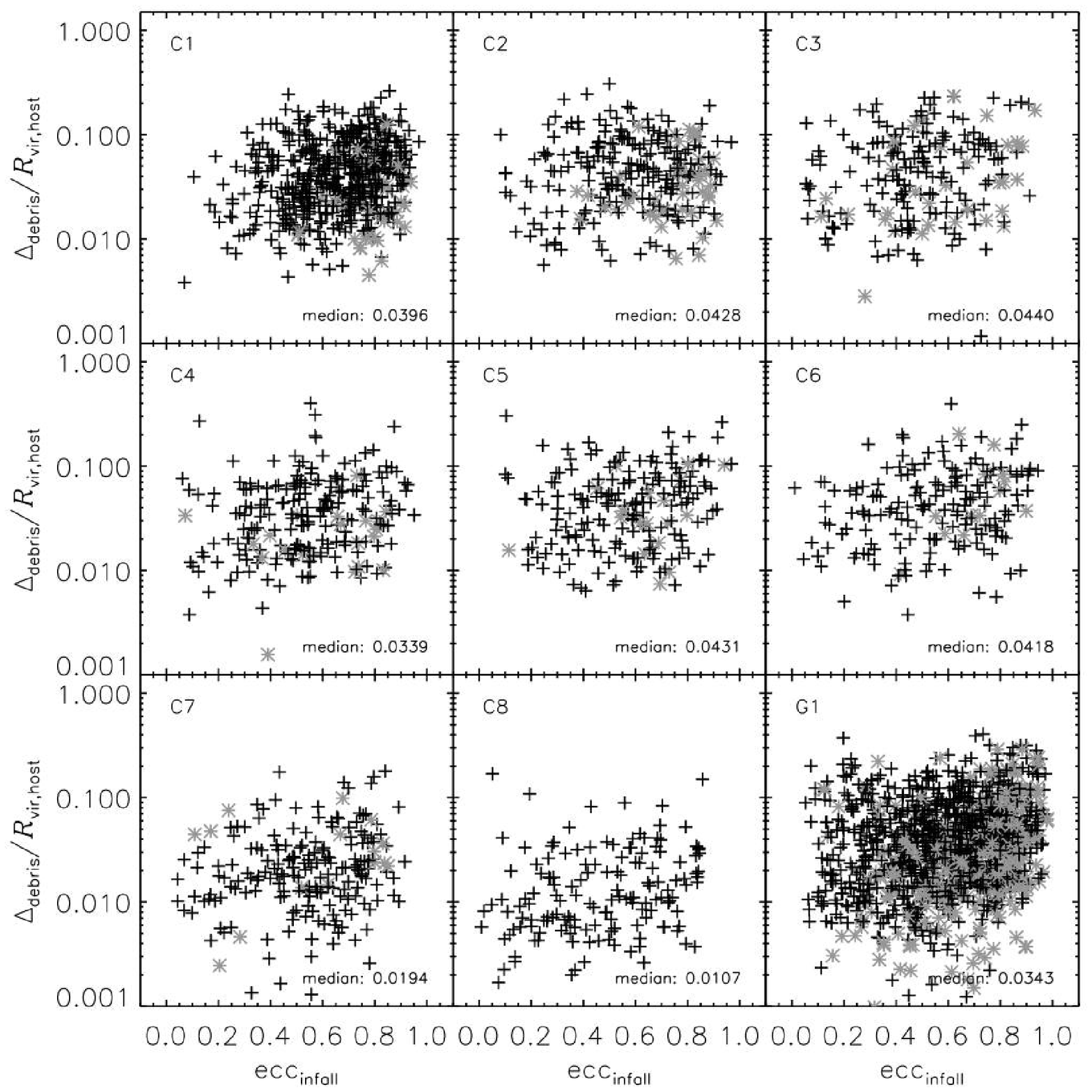, width=1\textwidth, angle=0}
\end{minipage}
\hfill
\begin{minipage}{0.495\textwidth}
        \epsfig{file=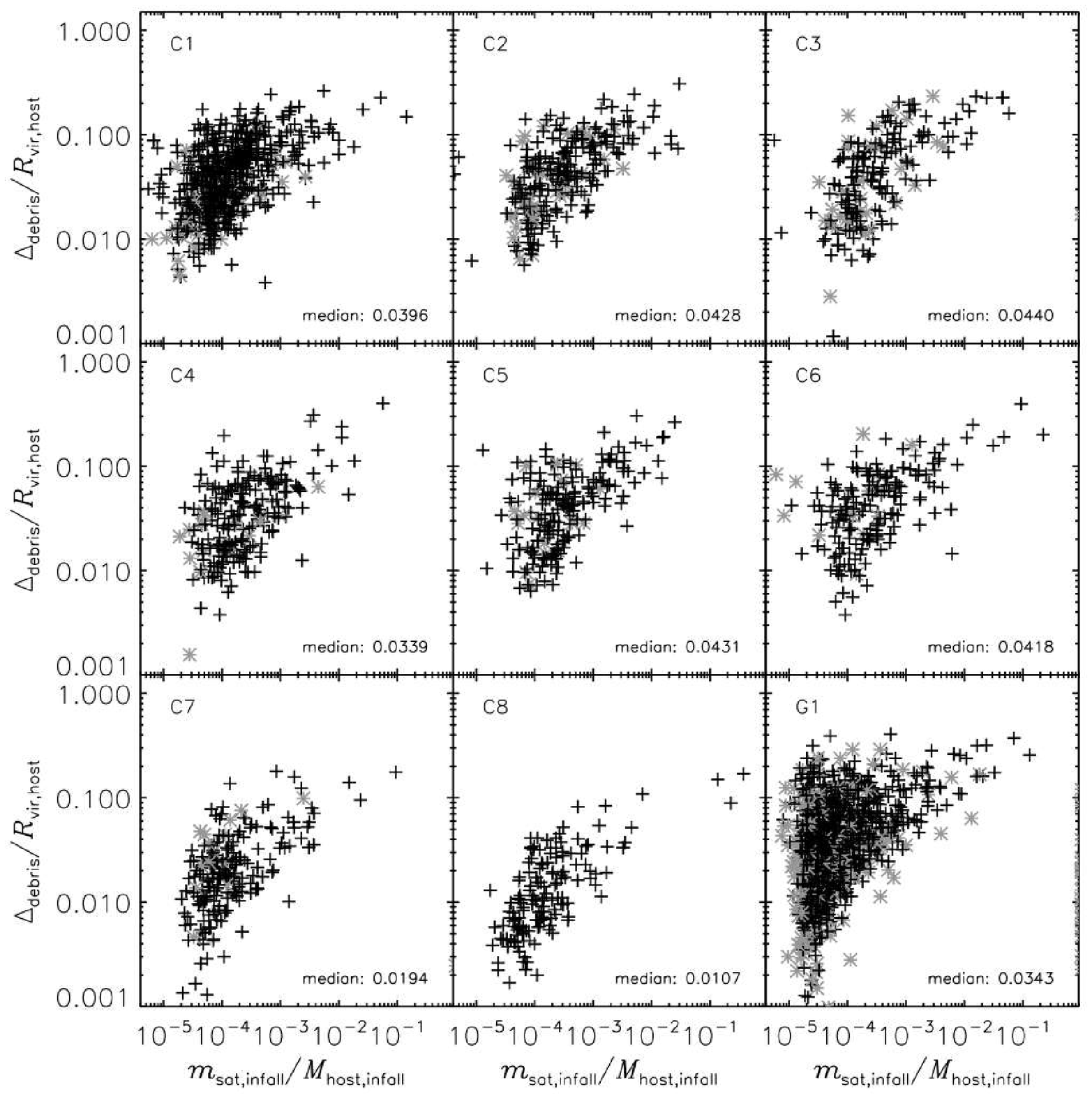, width=1\textwidth, angle=0}
\end{minipage}

\end{center}
\caption
{ The deviation of debris particles from the best fitting \emph{debris plane} (normalised to the virial radius of the respective host) as a function of infall eccentricity (left panel) and infall mass (right panel) for subhaloes with at least one orbit. Backsplash subhaloes are marked again with grey asterisks and the median values of the distributions are shown in the lower right corners.}
\label{f:orbitdebrisplanedev_Rvir}
\end{figure*}

An observer will not be able to recover the orbital path of a
subhalo galaxy (without, for instance, semi-analytical modeling of
its trajectory) and thus no information about the orbital plane will
be directly available to him (or her). However, he (or she) could either define
a plane based on the present subhalo's position and velocity
(i.e. the plane defined by the angular momentum vector) or fit a plane
to the detected debris field and then determine the deviation of
the debris field from this best-fit plane. We have adopted the latter approach
and the result can be viewed in \Fig{f:orbitdebrisplanedev_Rvir} where we show 
the scatter of the debris particles about the best-fit debris plane versus infall
eccentricity and mass. We note that the deviations of
debris from a plane defined in this way are smaller than the
deviations from the orbital plane. Again, only a marginal trend of
increasing deviation for eccentric orbits is apparent. However, we now
observe a stronger correlation between infall mass and debris
deviation. Thus, by measuring the deviation of debris from the debris
plane itself, it is possible for an observer to predict the most probable 
mass of the subhalo; in contrast, recovering the subhalo's eccentricity 
will be more ambiguous.

The question arises as to why the correlation between (infall) mass and
the debris variation is stronger in the case of the debris plane
when compared to the orbital plane. Furthermore, we do not observe an
``improvement'' in the correlation with orbital (infall)
eccentricity. We attribute this to the fact that the stream particles
follow naturally the debris plane and hence show greater deviation
from the orbital plane (that does not necessarily coincide with the
debris plane); the debris plane is defined such that it minimises the
scatter of the stream particles about that plane. Therefore, the
correlation already apparent in \Fig{f:orbitplanedev} strengthens when
reducing the scatter in the deviation as is done when considering the
debris plane rather than the orbital plane.

One may further argue that the destruction of subhaloes depends on
their (infall) concentration and that more massive systems have lower
concentrations \citep[e.g., ][]{Bullock.etal.01}. We confirm the
expected scaling between mass and concentration, although we find
no correlation between (total) mass loss and (infall) concentration. 
Therefore, the trends seen in \Fig{f:orbitplanedev} and 
\ref{f:orbitdebrisplanedev_Rvir} do not represent simple manifestations of
the mass--concentration relation.

\subsection{Tube Radius}
\begin{figure*}
\begin{center}
   \begin{minipage}{0.495\textwidth}
        \epsfig{file=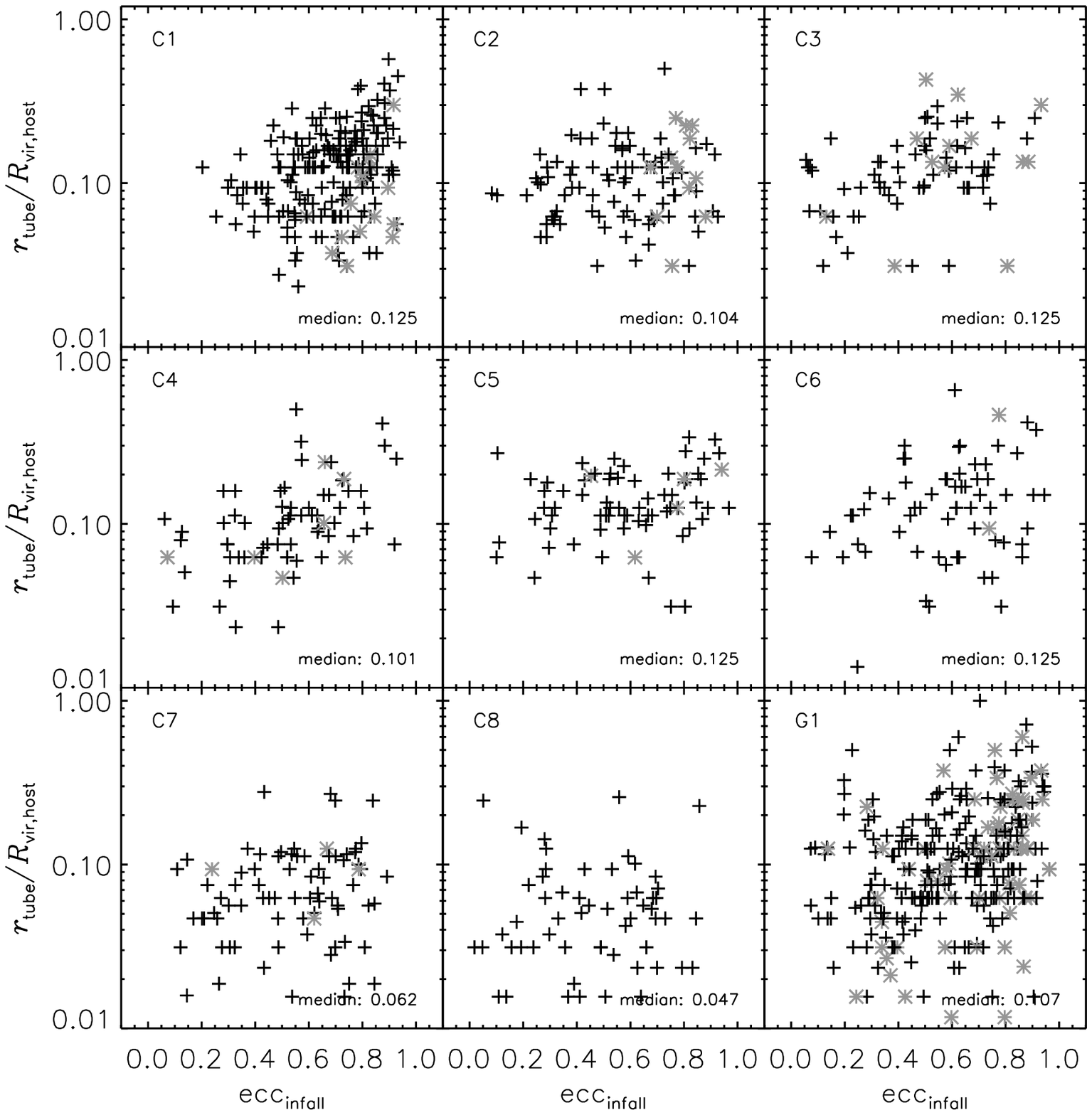, width=1\textwidth, angle=0}
   \end{minipage}
   \hfill
   \begin{minipage}{0.495\textwidth}
        \epsfig{file=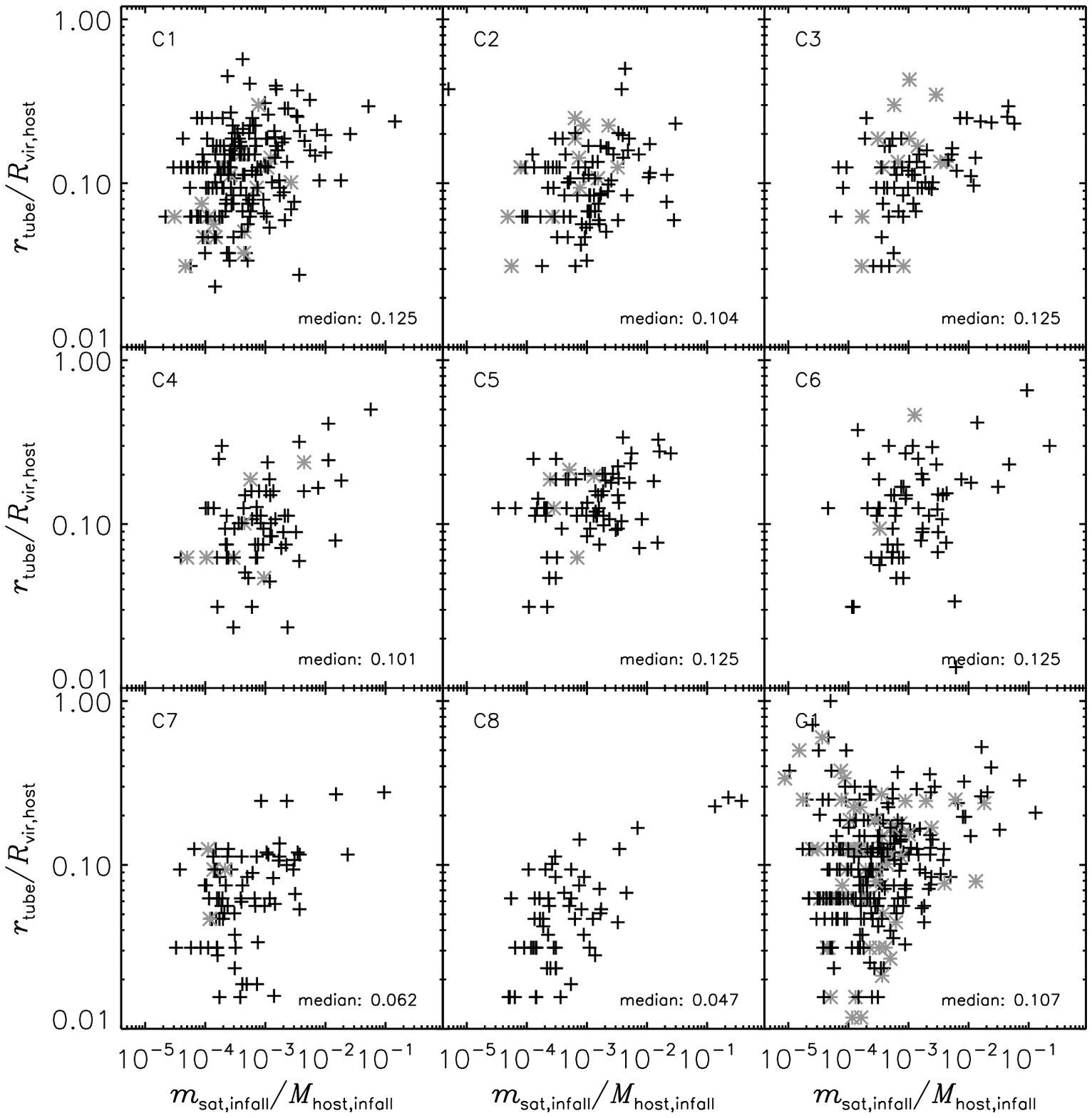, width=1\textwidth, angle=0}
   \end{minipage}
\end{center}
\caption { The tube radius as a function of infall eccentricity (left
  panel) and infall mass (right panel), normalised to the virial
  radius of the host today. Only subhaloes with at least one orbit
  and with a well-defined tube radius (\# particles inside tube =
  $68.3\% \pm 2\%$) are plotted. Grey asterisks correspond to
  backsplash subhaloes again. Numbers in the lower right corner of
  each panel show the median values for both, trapped subhaloes and
  backsplash subhaloes together.}
\label{f:orbittuberadius}
\end{figure*}

Even though the deviation from the orbital and debris plane may
provide a means to infer the (infall) mass of a disrupting subhalo
galaxy, we do not know whether the debris field is confined to the
orbit or becomes completely detached from it. The work by
\cite{Montuori.etal.06} on tidal streams of globular clusters suggests
that, in general, streams are good tracers of the orbit of their
progenitor, at least on large scales. However,
\cite{Choi.Weinberg.Katz.07} argue that the mass of the progenitor has
a great influence: a massive subhalo gravitationally attracts the
debris field, pulling it back and therefore causing a stronger bending
of the arms. Thus, the streams should deviate strongly from the
orbital path, the larger the mass of the progenitor subhalo is.

Furthermore, a large difference between the orbit and tidal streams may also 
indicate that the debris is just spreading over a larger area. This could 
then again be related to the infall eccentricity: subhaloes on spherical 
orbits will probably keep their streams quite close to the orbit, whereas 
subhaloes on radial orbits could have their debris spread to a much wider 
degree.

To quantify how debris relates to the orbit we introduce a ``tube
analysis'' -- we generate a tube about the orbital path of the
subhalo, adjusting its radius $r_{\rm tube}$ such that $1 \sigma$ 
(68.3\%) of all debris particles are enclosed by the tube. Larger tube 
radii would then indicate that debris particles are deviating more 
strongly from the orbit of the subhalo. A detailed explanation of the 
``tube analysis'' method and how to overcome its complications is presented in
Appendix~\ref{app:tubemethod}.

The resulting tube radii (normalised to the host's virial radius) as a 
function of eccentricity and mass are plotted in \Fig{f:orbittuberadius}. 
Subhaloes are ignored, if the tube radius did not converge satisfactorily
(more than 2\% deviation from the desired $1\sigma$ fraction of particles, 
mostly due to a small number of unbound particles). Again, only subhaloes 
with at least one full orbit are taken into account.

The expected trend of increasing tube radii with mass and eccentricity
is a weak one at best and we conclude that such a tube analysis is
less helpful in revealing subhalo properties than investigating the
spread perpendicular to the debris plane. However, it nevertheless
shows that the deviation of debris from the orbital path can be quite
large. We therefore conclude that debris is not a well defined tracer
of the orbital path of the progenitor subhalo as suggested by
\citet[][]{Choi.Weinberg.Katz.07}. 

We note also that in addition to studying how tube radii relate to subhalo
properties, we examined how median tube radii related to the shape of
the host. However, we were unable to identify any correlation.

\subsection[Radial velocity]{Radial Velocity}

\begin{figure}
\begin{center}
        \epsfig{file=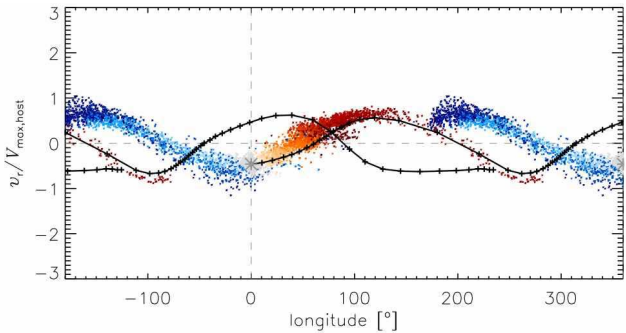, width=0.47\textwidth, angle=0}
\end{center}
\caption { The radial velocity of subhalo \#32 in host C1
  (normalised to the maximum of the host's rotation curve) versus
  longitude for an observer sitting in the centre of the host
  halo. Grey dots are bound particles, the black line shows the
  orbital path of the subhalo. Leading arm particles are coloured
  blue, trailing arm particles are shown in red with the brightness
  corresponding to the time since the particle became unbound (see
  \Sec{sec:trailing/leading} for more details). Note that we
  periodically wrapped the longitude extending the presented range
  from -180 to 360.}
\label{f:radvel.sat32}
\end{figure}

So far we have utilised the 3D position (and velocity) information
available to us, but is interesting to ask how a subhalo stream would
appear to an observer. An observer can, in principle, measure radial 
velocities of material in the tidal stream of a subhalo as well as 
its position on the sky. Let us imagine an observer positioned
at the centre of a host halo, measuring the
radial velocity of the (debris) particles of a certain subhalo in
our simulations. We fix our coordinate system such that the plane best
fitting the debris lies at zero latitude (i.e. zero height). A
longitude of 0 shall correspond to the (remnant) subhalo's position
today. This enables us to plot an `observed' radial velocity
distribution versus longitude for a subhalo, as shown in
\Fig{f:radvel.sat32}. Note that only debris particles lost
\emph{after} the infall of the subhalo into the host are considered;
see \Sec{sec:trailing/leading} for more details. From this figure it is
apparent that the arms are following a sine-like curve, just like the
orbit (black line). The trailing arm\footnote{Details on how to
  separate trailing and leading arm will follow in
  \Sec{sec:trailing/leading}.} (red) sits neatly on top of the orbit
path in this example and the leading arm (blue) seems like a perfect
continuation of the orbit to future times. This
particular subhalo corresponds to one of the cases in the tube
analysis that had a small tube radius. Its particulars are
$N_{\rm orbit}=2.2$, ${\rm ecc}_{\rm infall}=0.458$, $M_{\rm
  infall}=8$, $M_{z=0}=1.6$, $M_{\rm trailing\ arm}=3.7$, $M_{\rm
  leading\ arm}=2.7$ (where masses are measured in $10^{11}$\hMsun).

Such plots contain a remarkable wealth of information about the
subhalo and its orbit. The maximum/minimum radial velocity are a
direct measure of the \emph{eccentricity} of the orbit: for circular
orbits we would expect a constant line at $v_{\rm r} = 0$; the more
eccentric an orbit is, the larger are its minimum/maximum radial
velocity (in absolute values). Furthermore they hint at the
\emph{number of orbits} a subhalo has already undertaken: each
period which the sine-like curve of the debris describes corresponds
to one half (debris) orbit. The phase-shift of successive sine-curves
in these plots could also be used to infer the \emph{precession} of
tidal debris in the debris plane, which may be correlated with the
triaxiality of the host halo. Also, the spread of each arm in radial
velocity could also tell us something about the \emph{mass} of the
subhalo's progenitor: more massive subhaloes will have a broader
distribution of debris, in spatial terms as well as radial velocity
\cite[e.g.][]{Choi.Weinberg.Katz.07}.

\begin{figure*}
\begin{center}
   \begin{minipage}{0.495\textwidth}
        \epsfig{file=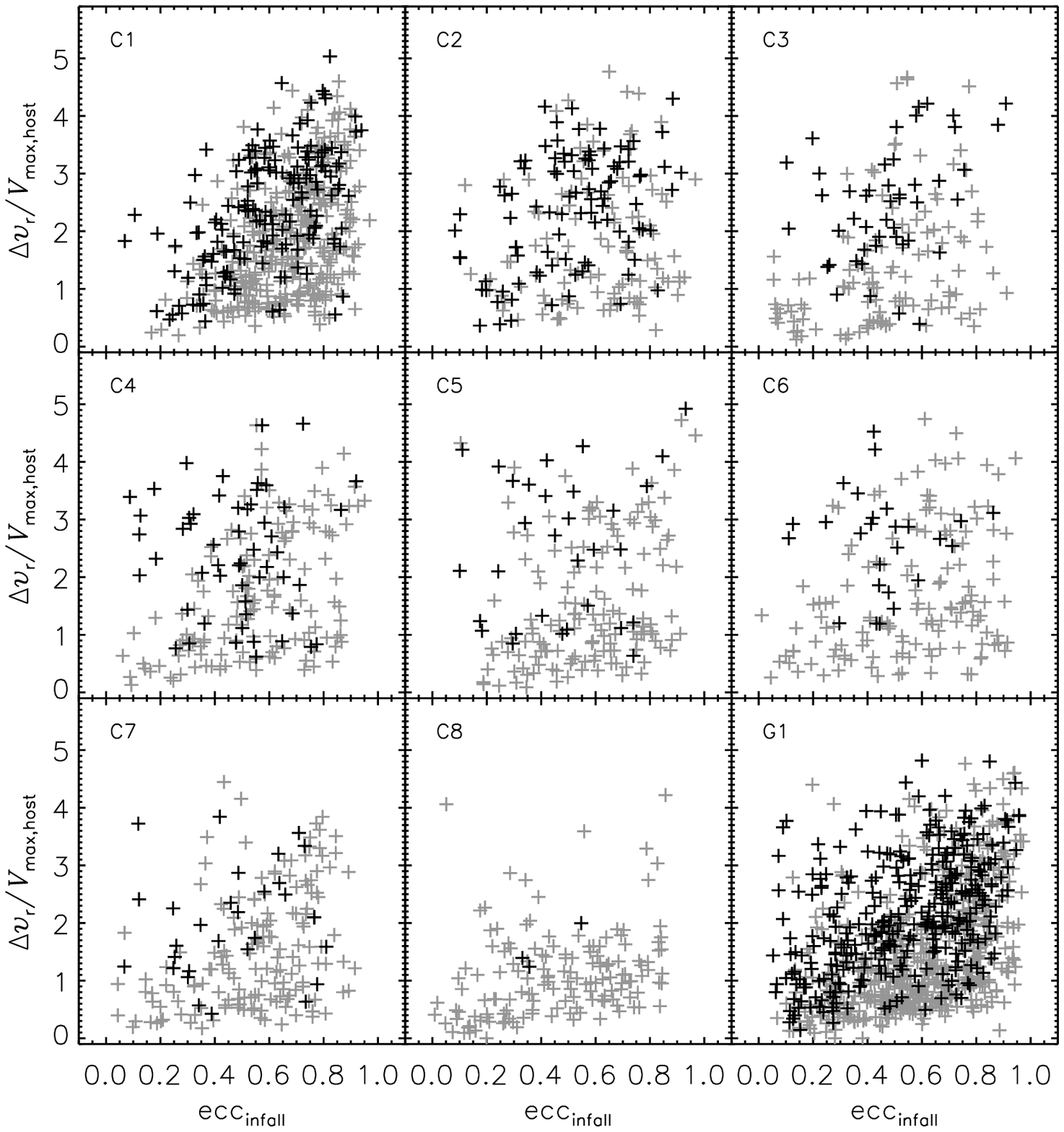, width=1\textwidth, angle=0}
   \end{minipage}
   \hfill
   \begin{minipage}{0.495\textwidth}
        \epsfig{file=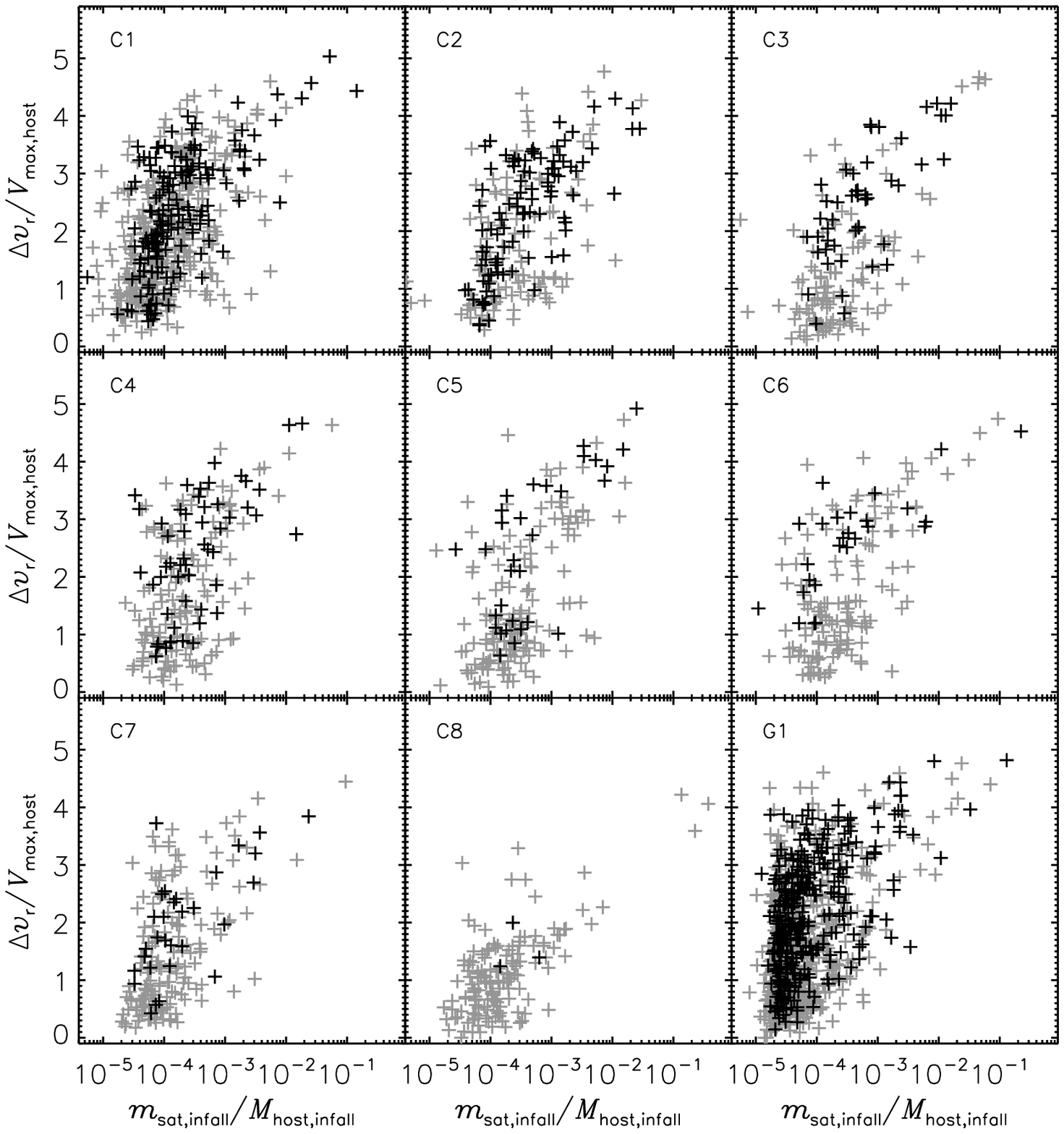, width=1\textwidth, angle=0}
      \end{minipage} \end{center} \caption { The spread in radial
      velocity of unbound particles (normalised to the maximum
      velocity of the rotation curve of the corresponding host) as a
      function of infall eccentricity (left panel) and normalised
      initial subhalo mass (right panel). The grey crosses correspond
      to subhaloes with $N_{\rm orbit} < 3$ while the black crosses
      represent subhaloes with $N_{\rm orbit} \ge 3$.}
\label{f:radvel_spread}
\end{figure*}

However, for the moment we concentrate on the total spread of the
radial velocity

\begin{equation}
 \Delta v_r = v_r^{\rm max}-v_r^{\rm min} \ ,
\end{equation}

\noindent and its relation to infall eccentricity and progenitor mass
again. Figure \ref{f:radvel_spread} shows the difference between
smallest and largest radial velocity measured for the corresponding
subhalo's debris as a function of infall eccentricity (left panel) and
initial subhalo mass (right panel). Indeed we observe the expected
trend, although the scatter is very large. We can also see that
subhaloes with more than one orbit (darker plus signs in
Figures~\ref{f:radvel_spread}) show a stronger signal: the more orbits
undergone, the greater the probability that the difference in maximum
and minimum radial velocity is determined by the orbital eccentricity
rather than random scattering of the debris. But we also need to bear
in mind that the debris field originating from subhaloes that have
undergone a greater number of orbits will tend to be more massive, and
therefore we would expect to make a more precise determination of
$\Delta v_r$.

Using a more elaborate method for estimating the spread in radial 
velocity, such as using the mean values from the ten largest and smallest 
velocities or binning the distribution and taking the 
largest and smallest bin with a certain minimum number of particles, does 
not improve our results.\\

We now turn to the correlation between the spread in radial velocity
and the progenitor's infall mass (right panel of
\Fig{f:radvel_spread}); we do not expect to observe a strong
correlation, because only the width of the streams themselves should
depend on the mass, whereas the total spread correlates more strongly
with orbital parameters. However, \Fig{f:radvel_spread} reveals quite
a pronounced mass dependence: more massive haloes show a larger variety
in radial velocities. Yet this correlation vanishes when looking at
the radial velocity spread of the orbit itself (not shown here
though). Thus, the signal found in \Fig{f:radvel_spread} suggests that 
the debris does not merely follow the orbit but spreads out to some 
angular extent that depends on its
mass (compare also with the example radial velocity plot in
\Fig{f:radvel.sat32}.). Again this shows that the stream itself is more
sensitive to the mass of its progenitor subhalo than the orbital
properties, as already suggested by the results of the previous subsections.

\subsection{Summary -- Uncovering Subhalo Properties}

We set out to recover properties of the progenitor subhalo from its
tidal debris field alone. Therefore we focused upon correlations between the tidal
debris field and the progenitor subhalo's infall eccentricity and mass. 

Although the expected correlation with infall eccentricity is weak, the 
statistics that we have presented may nevertheless help to recover the 
orbit of an observed subhalo. Both methods (scatter about the debris plane and 
tube analysis) rely on measuring the debris field, thus allowing for a 
determination of the subhalo's eccentricity. We therefore envision that a 
combination of both the debris scatter about the best fit debris plane and 
the tube analysis applied to existing stream data may help to refine models 
for the reconstruction of orbital parameters.

Relating the tidal streams with the original mass of the subhalo, we
observed a noticeable correlation when using the deviation
of the debris from the mean debris plane. This opens the opportunity
to infer the original mass of a (currently) disrupting subhalo such
as, for instance, the Sagittarius dwarf spheroidal
\cite{Ibata.etal.94}. Using the relation between mass loss and
eccentricity as presented in \Fig{f:satmassloss}, it may be possible
to reconstruct the eccentricity of the orbit indirectly as soon
as we know the original and present mass of the corresponding
subhalo.

In addition, we presented an analysis of radial velocity plots
as they might appear for an observer in the centre of a host. The
spread in the radial velocity of the orbit itself is obviously
strongly correlated with the (infall) eccentricity (not shown here),
but it is less pronounced for the spread in debris
(cf. \Fig{f:radvel_spread}). It nevertheless shows a correlation with
the mass of the progenitor, which is attributed to the fact that the
debris of massive subhaloes actually spreads to a greater degree than
for low-mass objects.

\section{Tidal Streams II:\\ Separating Leading \& Trailing Arms}
\label{sec:trailing/leading}

So far the debris particles have been characterised by ``bulk''
properties, such as deviation from the orbital plane. No 
distinction has been made between leading and trailing arm or the 
time since the debris particles became unbound. In this section we
investigate differences in the properties of both the leading and
trailing arms, which makes it necessary to tackle the rather
challenging task of assigning particles to the trailing or 
leading debris fields.

\subsection{Age} 
We begin by defining the age of stream particles, in order to
understand the build-up of the stream. We classify a particle as
belonging to the stream at the moment at which it becomes unbound from
its progenitor subhalo, which occurs when their velocity exceeds 1.5
times the escape velocity of the subhalo (see section \ref{sec:HaloTracker}).
This age information is used to colour-code some of the following
plots.

\subsection{Membership to leading/trailing arm} \label{ssec:armseparation}
\begin{figure*}
\begin{center}
\begin{minipage}{0.495\textwidth}
        \epsfig{file=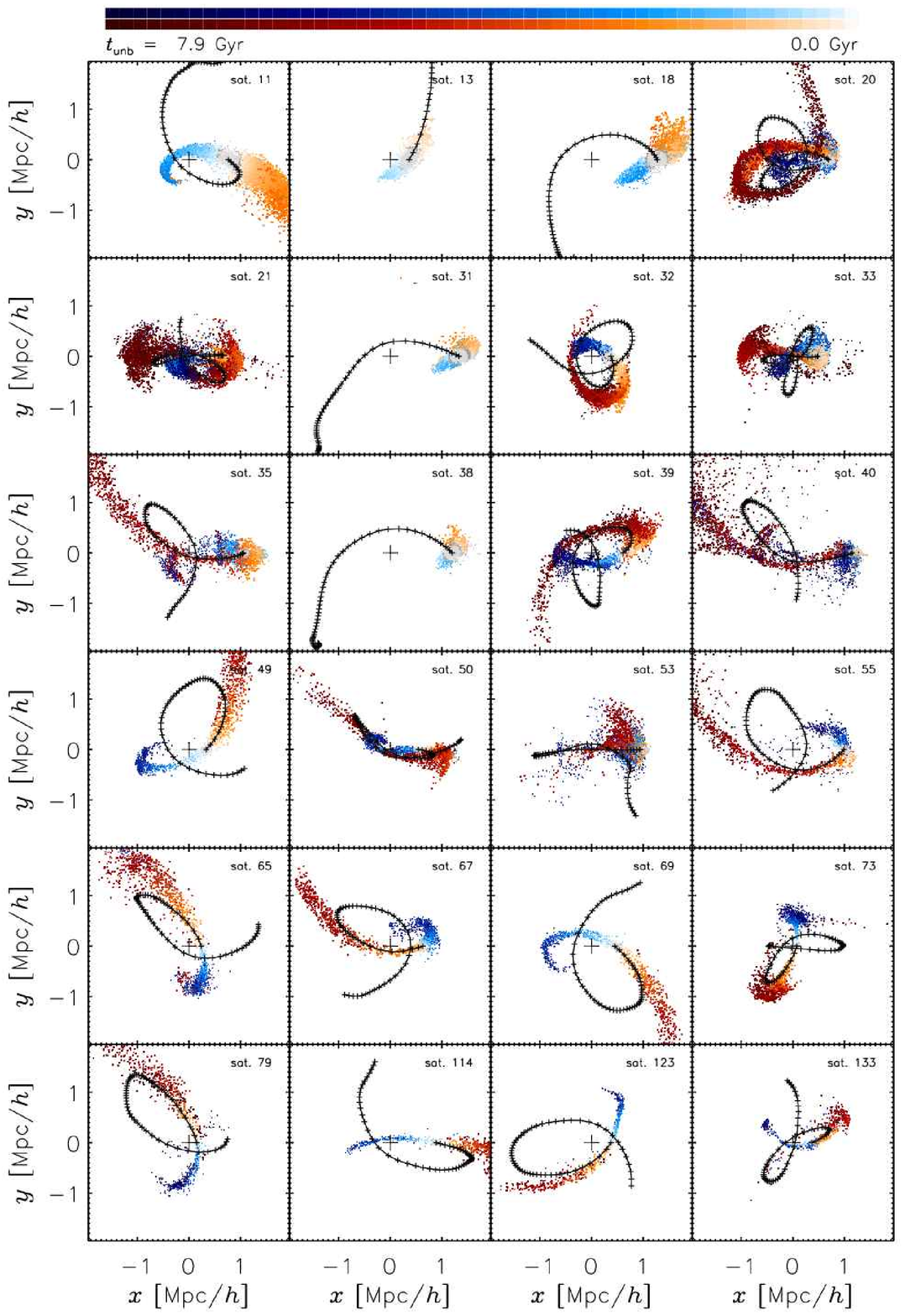, width=1\textwidth, angle=0}
\end{minipage}
\hfill
\begin{minipage}{0.495\textwidth}
  \epsfig{file=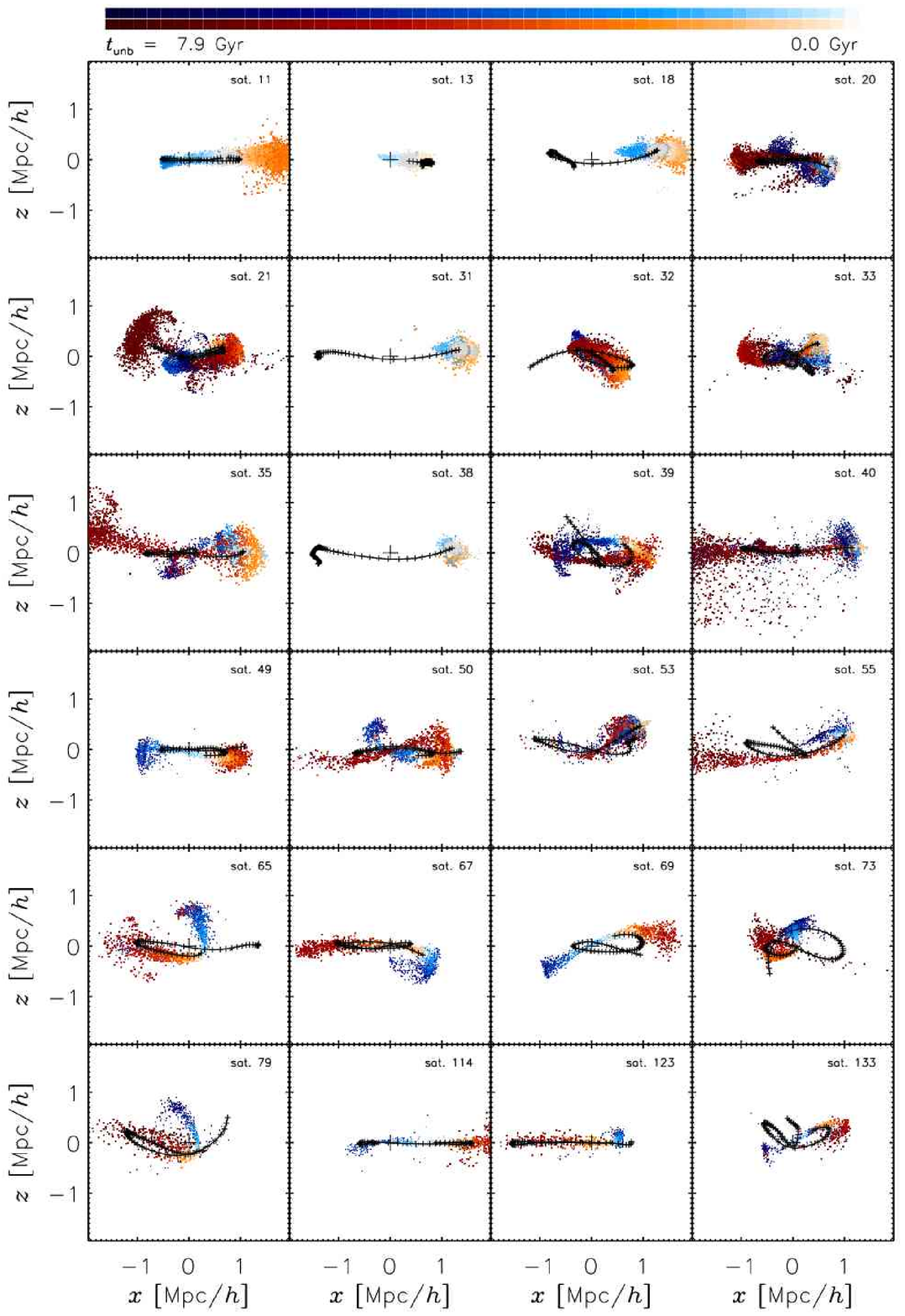, width=1\textwidth, angle=0}
\end{minipage} \caption { Classifying stream particles according to
  age and leading/trailing arm. Shown are the particles of a
  representative sample of subhalo galaxies orbiting in host C1
  projected to their orbital planes (left panel) and as an edge-on
  view (right panel). The black line represents the path of the orbit.
  Grey particles are still bound to the subhalo, red particles are
  marking the trailing, blue ones the leading arm. The colours vary
  from dark to bright, indicating the age of the respective stream.}
\label{f:allmpsattidalarms}
\end{center}
\end{figure*}

We require not only the age of stream particles but also a means to
determine which of the arms, i.e. leading or trailing, the particle
belongs to. For subhaloes with
clearly distinguishable arms, one can separate leading and trailing
stream easily by eye; however, an automated method that can
\emph{objectively} classify particles is both desirable and necessary,
especially when analysing literally hundreds of subhaloes. We have
developed such a method, which we now present.
 
For each snapshot, we mark particles leaving the subhalo in the 
``forward'' direction as leading and those leaving in
``backward'' direction as trailing arm particles, where we define
``forward'' and ``backward'' as follows.
If a particle leaves the subhalo, then -- in a simplified picture -- 
the subhalo has lost its influence on this particle; the host now
plays the most important role and hence all subhaloes shall be
neglected. In addition, the velocity of a recently unbound particle 
will still be close to the velocity of its progenitor subhalo. In the
absence of any forces the particle would follow trajectory defined
by the direction of the velocity vector, but it feels the gravitational 
pull of the host and therefore its direction is defined relative to the
vector

\begin{equation} 
  \vec{d} = \frac{v^2}{r}  \frac{\vec{v}}{v} - \frac{GM_{\rm host}(r)}{r^2} \frac{\vec{r}}{r}.
\end{equation}

\noindent
If a particle leaves the subhalo in this direction, it is tagged
leading, otherwise trailing. For a subhalo moving at high velocity,
a stream will form mostly along the orbit ($v^2$-term prevails),
whereas the stream of a slow subhalo is more attracted by the host
centre \cite[see e.g.][]{Montuori.etal.06}. 

We note that using either the direction of the subhalo's velocity
($\vec{v}$) or the direction to the host ($\vec{r}$) alone when
distinguishing between leading and trailing arm particles does \textit{not}
give satisfactory results. Streams cannot be separated in this way
because particles are not leaving simply in the direction of $\vec{r}$
or $\vec{v}$.

Our leading/trailing classification method works well, as
indicated by the example subhaloes presented in
\Fig{f:allmpsattidalarms} (see also the colors in the radial velocity
plot, \Fig{f:radvel.sat32}, where the same classification procedure
was used). It fails, however, for those subhaloes which have just
recently fallen into the host. Here the subhaloes may have already
lost mass before entering the host, due to interactions with adjacent
subhaloes or other overdensities. Particles lost in such a way cannot
be described by our (nevertheless simple) picture since the influence
of other subhaloes is too large compared with the host halo's
influence. We therefore restrict the remaining analysis of debris to
particles lost after their (progenitor) subhalo entered the host.

We note that \Fig{f:allmpsattidalarms} provides further evidence that 
our procedure for fitting orbits to a plane as outlined in 
\Sec{sec:fitting} works well for a variety of orbits.

\subsection{Masses in streams}
\begin{figure*}
 
 \begin{minipage}{0.495\textwidth}
        \epsfig{file=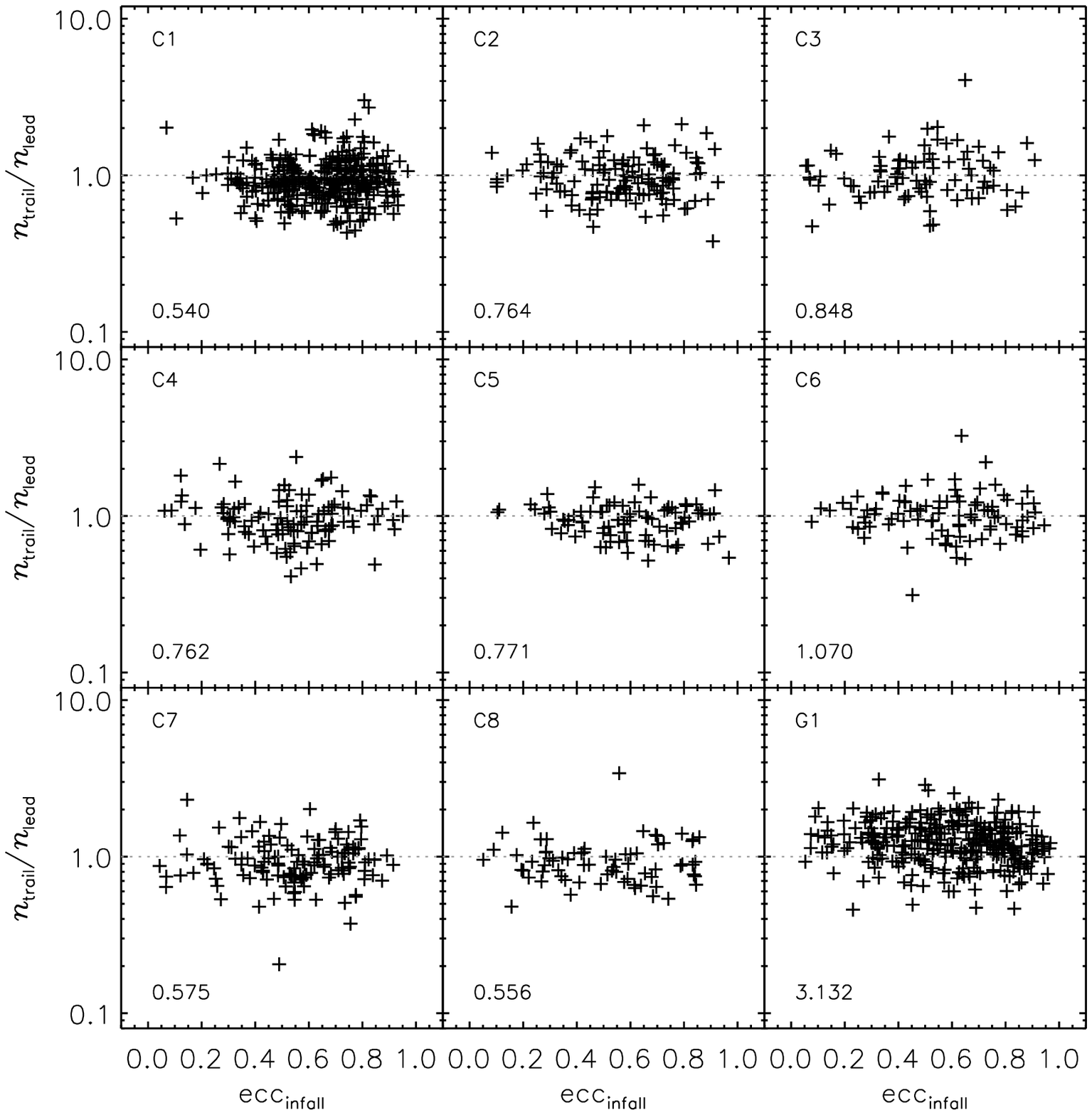, width=1\textwidth, angle=0}

 \end{minipage}
 \hfill
 \begin{minipage}{0.495\textwidth}
        \epsfig{file=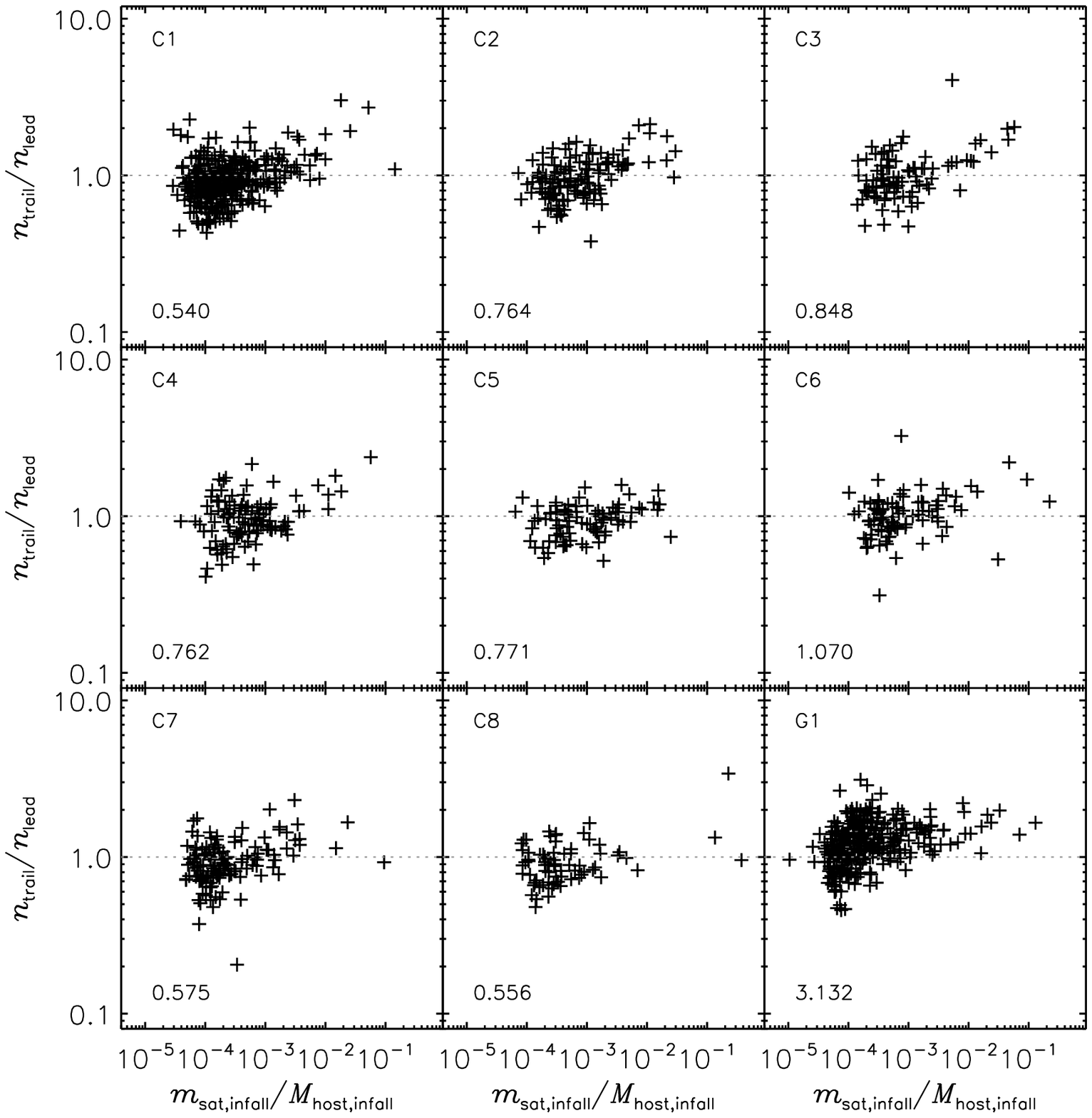, width=1\textwidth, angle=0}
      \end{minipage} \caption { Ratio of the number of leading and
        trailing arm particles for subhaloes inside the host with at
        least one orbit, plotted versus the infall eccentricity (left
        panel) and infall mass (right panel). The number in the lower
        left corner of each panel gives the ratio of the number of
        subhaloes with a value above and below 1.}
\label{f:sattidalarms_mass}

\end{figure*}

We are now in the position to explore differences in trailing and
leading arm in detail and start with explicitly calculating the ratio
of the mass in both arms. \Fig{f:sattidalarms_mass} shows the ratio of
the number of particles in the leading and trailing arm versus the
infall eccentricity and mass of each subhalo. Again only subhaloes
inside the host today and with at least one complete orbit (inside the
host) are taken into account. Further, only subhaloes with at least
25 particles per arm are considered.\footnote{Interestingly, when
  including subhaloes with less massive arms, the number of points
  below 1 increases (more subhaloes with massive leading arm)!}

There appear to be slightly more particles in the leading than in the
trailing arm in each of our cluster sized haloes while the reverse is
true for the galactic halo G1. This is interesting because of the
recent claim of \cite{Kesden.Kamionkowski.06}, who argue that if dark matter is
coupled to dark energy, there must exist another force acting on dark
matter particles besides gravitation. This additional force could manifest itself 
in non-equilibrium systems such as the tidal tails of subhaloes, leading to an
imbalance in the mass of the leading and trailing arms. According to a
simplified picture of tail formation in cosmologies without such
coupling, both arms should contain the same amount of mass. Even
though our simulations do \textit{not} couple dark matter to dark
energy, we observe a trend for an imbalance in the masses of leading
and trailing arm. We therefore conclude that the proposition put
forward by \cite{Kesden.Kamionkowski.06} cannot hold in general.

In addition, we do not observe any tendency for the ratio of masses in leading 
and trailing arms to correlate with infall eccentricity, but we do note a trend 
for more massive subhaloes to lose more of their mass to the trailing arm. As
\cite{Choi.Weinberg.Katz.07} have pointed out, the gravitational attraction of 
the subhalo on the tidal arms may cause an asymmetry in leading and trailing arm.
The strength of this effect will increase with the mass of the subhalo, which
can explain the mass dependence we observe. However, it is not clear
why this ratio reverses for the galactic halo, although we note that
the reversal is evident only for the least massive subhaloes.

\subsection{Velocity dispersion in tidal streams}
\begin{figure}
\begin{center}
  \epsfig{file=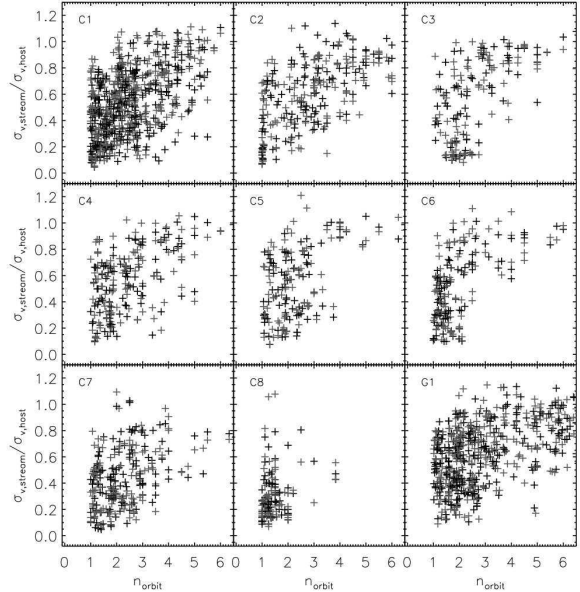,
    width=0.47\textwidth} \caption{The velocity dispersion of trailing
    (black) and leading (dark grey) arm, normalised to the velocity
    dispersion of the host, as a function of the number of orbits.
    Since 'partial' orbits at the beginning and end of the considered
    time interval are also taken into account, the number of orbits is
    not limited to integers.}
\label{f:sattidalarms_vdisp_orbits}
\end{center}
\end{figure}

\begin{figure}
\begin{center}
  \epsfig{file=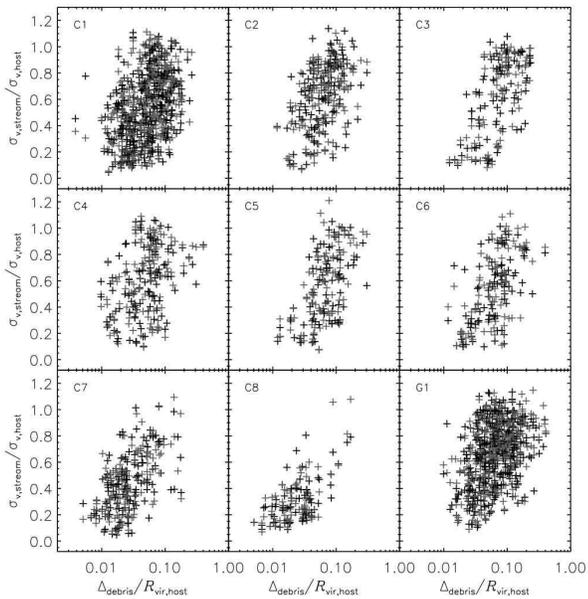,
    width=0.47\textwidth} \caption{The
    velocity dispersion of trailing (black) and leading (dark grey)
    arm as a function of the deviation of debris from the best fit
    debris plane (compare with \Fig{f:orbitdebrisplanedev_Rvir}).}
  \label{f:sattidalarms_vdisp_planedev} \end{center} \end{figure}

\begin{figure*}
 
 \begin{minipage}{0.495\textwidth}
        \epsfig{file=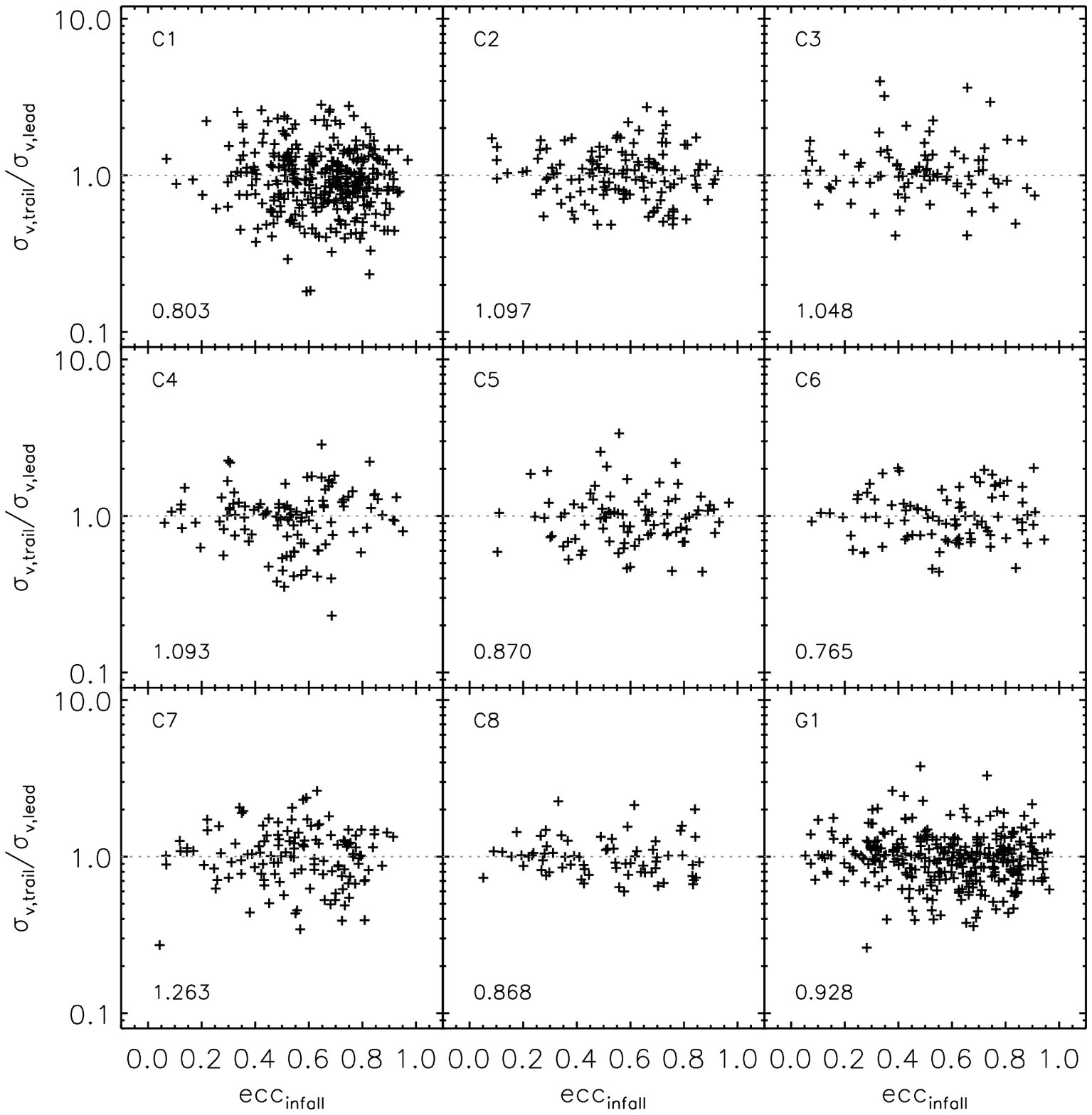, width=1\textwidth, angle=0}
 \end{minipage}
 \hfill
 \begin{minipage}{0.495\textwidth}
        \epsfig{file=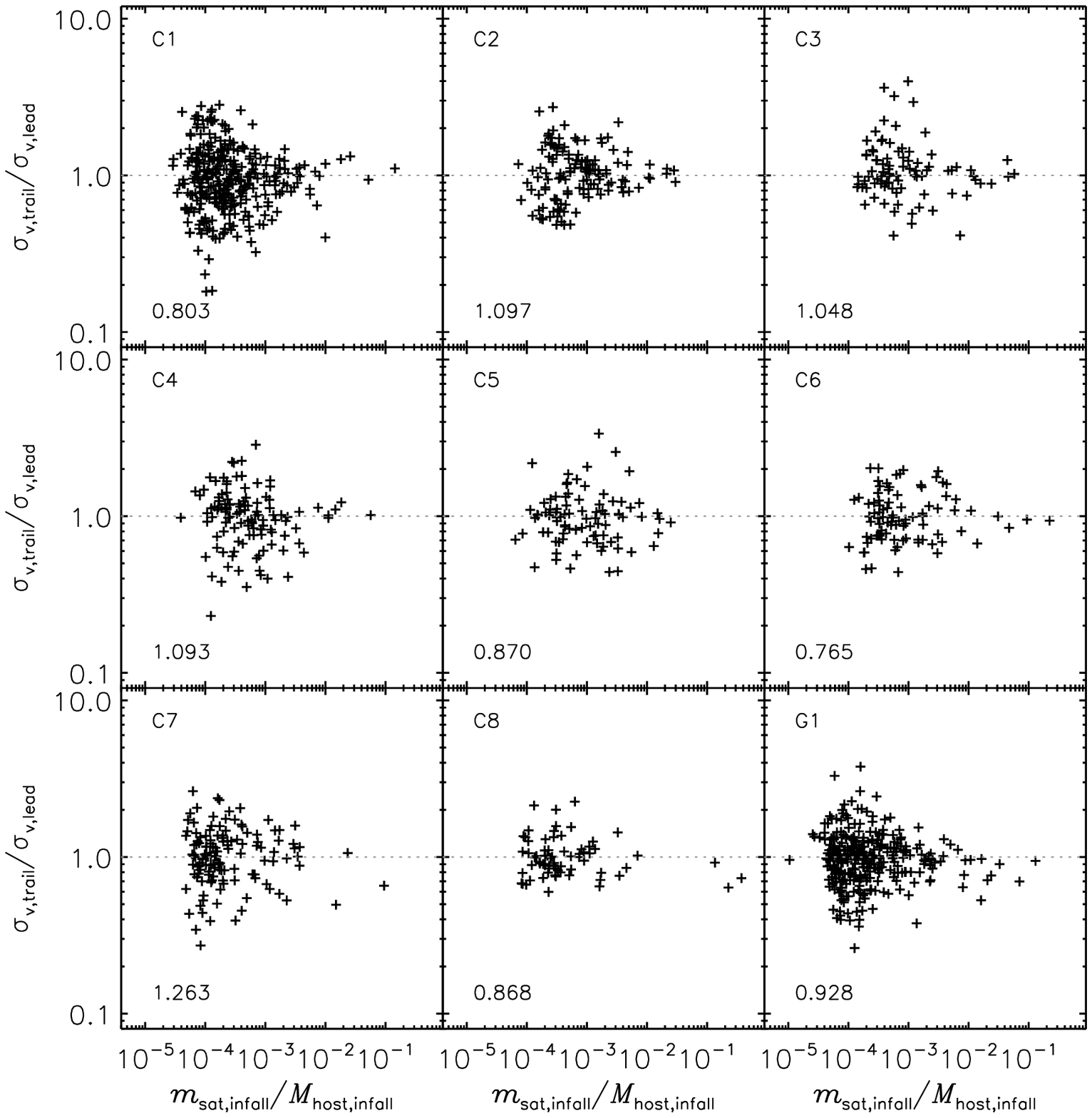, width=1\textwidth, angle=0}
      \end{minipage} \caption { Ratio of the velocity dispersion in
        leading and trailing arm for subhaloes in the host with at
        least one orbit as a function of infall eccentricity (left
        panel) and infall mass (right panel).}
\label{f:sattidalarms_vdisp}
\end{figure*}

The velocity dispersion of streams is another property that can reveal 
important insights to the history of the subhalo and characteristics of 
the host. For example, \cite{Johnston.Hernquist.Bolte.96} note that a high 
velocity dispersion in the tail is suggestive of a massive progenitor
as well as a high internal velocity dispersion in the original
subhalo. However, the dispersion can increase further if the stream
orbits in a ``lumpy'' host halo: encounters between the stream and
sub-clumps will lead to dynamical heating of the stream. Furthermore, if the 
shape of the host deviates strongly from spherical symmetry, the subhalo 
and thus also its debris particles will start to precess, leading to
yet further spread of the debris, increasing its spatial dispersion.

We start by exploring the relation between the number of orbits and
the velocity dispersion in both the leading and trailing debris
arm. The result can be viewed in \Fig{f:sattidalarms_vdisp_orbits}
where we find a general increase in the velocity dispersion with the
number of orbits -- the greater the number of orbits, the higher the velocity 
dispersion of the debris field. This figure is accompanied by
\Fig{f:sattidalarms_vdisp_planedev} where we show the relation between
the velocity dispersion of debris and its scatter about the best fit
debris plane (cf. \Sec{sec:fitting}). It underlines the trend that a
larger deviation of the debris from the best fit plane correlates with a
higher velocity dispersion and vice versa. Viewed in conjunction with
\Fig{f:orbitplanedev} (right panel) we recover the findings of
\cite{Johnston.Hernquist.Bolte.96} that a larger progenitor
automatically leads to a higher velocity dispersion.\\

Are there differences in the velocity dispersions of the leading and
trailing arms, induced by the host halo? In
\Fig{f:sattidalarms_vdisp} we show the ratio of the velocity
dispersions in both tidal arms as a function of the subhalo's infall
eccentricity and mass again. We observe that the mean ratio is
always of order unity with a roughly constant scatter when plotting
against infall eccentricity, but the scatter decreases for initially
more massive subhaloes (right panel of \Fig{f:sattidalarms_vdisp}).
There appears to be only marginal differences between the leading and
trailing arm with respects to the internal velocity dispersion. As
they are spatially separated entities we therefore suspect interactions
with sub-clumps in the host halo to have little influence unless both
arms accidently encounter the same number of subhalo. It is more
likely for the host to play the dominant role in the determination of
the velocity dispersion of these two detached debris field.

\subsection{Energy of tidal debris}

We have learned that the masses of leading and trailing arm can differ,
but their velocity dispersions are similar. While this may make it
difficult to separate leading and trailing arms using raw velocity
dispersions, we wish to understand whether we can use kinematic
information to separate particles (or stars) into leading and trailing
arms, as an observer might wish to do.

To this end, we present a method that allows classification into 
leading and trailing arms based upon the energy and 
angular momentum of tidal debris. Inspired by the work of 
\cite{Johnston.98} who found a bi-modal energy distribution representative 
of leading (decrease in energy) and trailing arm (increase in energy) we 
calculate for each debris particle its change in total energy \cite[also 
see][]{McGlynn.90, Johnston.etal.99, Johnston.Sackett.Bullock.01, 
Penarrubia.etal.06}. When combining the resulting distribution with our 
knowledge about the association of particles with leading/trailing arms
we recover the findings of \cite{Johnston.98} that there is a clear
separation between the two arms in all our 'live' host haloes -- at
least for the majority of subhaloes. In what follows, we focus on the
properties of streams at $z$=0.

\subsubsection*{Energy of debris particles}

The energy of a particle is simply the sum of its potential and kinetic
energy. While it is straightforward to calculate the kinetic energy,
the potential energy is always somewhat more complicated. We are left
with the option to either use the \texttt{AHF} analysis of the
simulation which leaves us with (more or less) spherically cut dark
matter haloes or go back to the original simulation and derive the full
potential used to integrate the equations of motion. As the former
relies on assumptions that are not necessarily true for all our
systems (i.e. we know quite well that our cosmological dark matter
haloes are not spherical and better modeled by a triaxial density
distribution) we chose to follow to the latter approach. This provides 
us with the most accurate measure for the potential
energy we can get. We also mention that we estimate both
the potential and kinetic energy in the rest frame of the underlying
host halo. We understand that this approach works against any
hypothetical observer, but we wish to maximise the possibility that
there is \textit{any chance at all} to separate particles into 
leading and trailing arm.

\subsubsection*{Energy scale} 

According to \cite{Johnston.98} the changes in the total energy are of the 
order of the orbital energy of the subhalo itself. In order to scale the 
variation in the particles' energies in a similar fashion we define this 
scale to be

\begin{equation} \label{eq:escale}
 \epsilon = \epsilon_{\rm pot} + \epsilon_{\rm kin}\ , \
\end{equation}

\noindent
where $\epsilon_{\rm pot}$ and $\epsilon_{\rm kin}$ are the specific
potential and kinetic energies of a test particle at the position and
with the velocity of the (remnant) subhalo at redshift $z=0$. Note
that for $\epsilon_{\rm pot}$ we interpolate the potential
as returned by the simulation code \texttt{AMIGA} to the test
particles' position.

\subsubsection*{Energy distribution}
\begin{figure*}
\begin{center}
        \epsfig{file=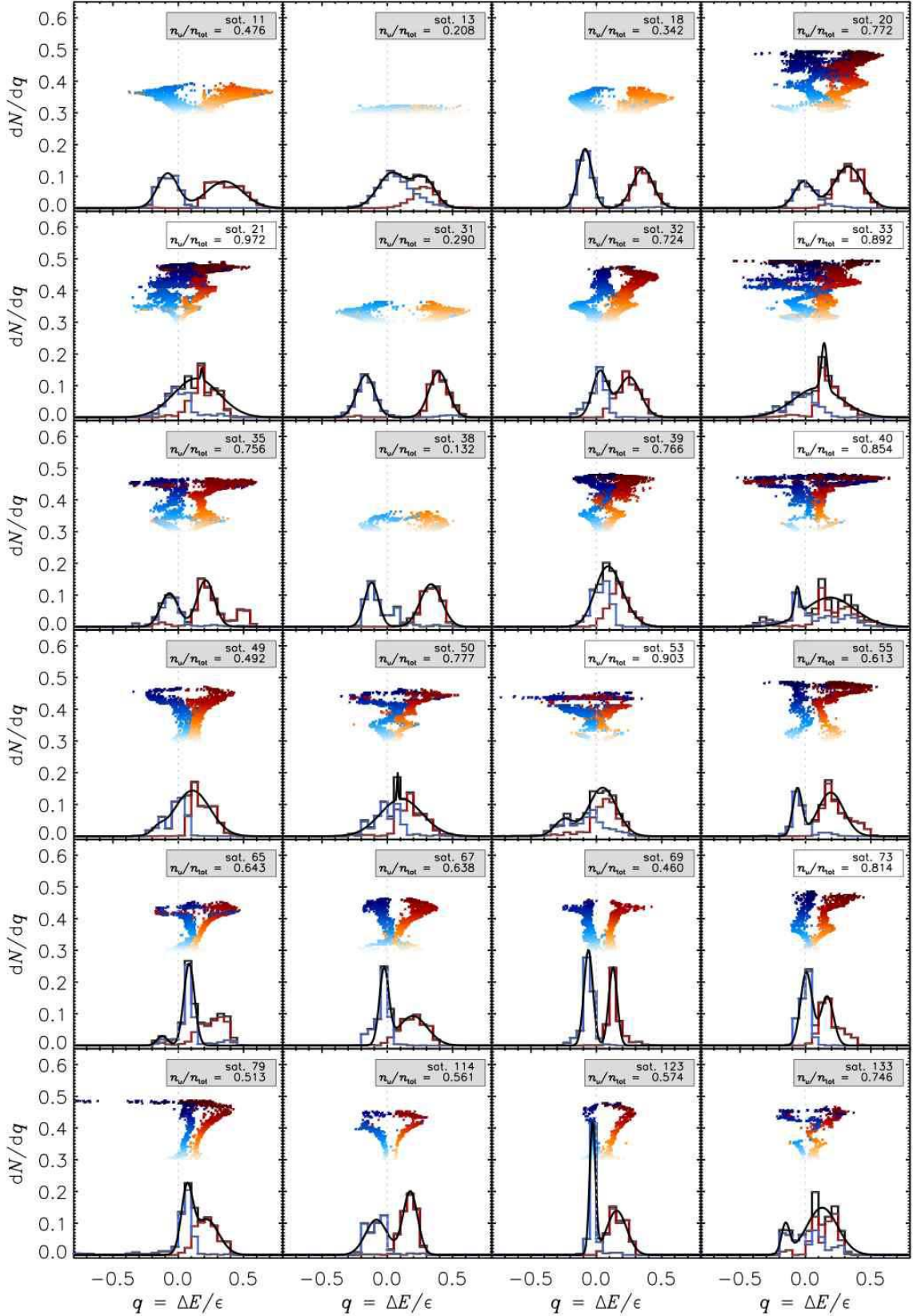, width=0.9\textwidth, angle=0}
\end{center}
\caption
{ The distribution of energy change of debris particles for a random sample of subhaloes in the host C1. Please refer to the main text for details.}
\label{f:streamenergy_z0}
\end{figure*}

In order to be able to make predictions for an observer with access to the 
full 6D phase-space information for debris particles/stars and a realistic 
model for the host halo (in order to derive potential energies), we 
restrict the following analysis to quantities measured at $z=0$. We 
define the relative energy change of a debris particle to be

\begin{equation} \label{eq:q}
 q = \frac{\Delta E}{\epsilon} = \frac{E-\epsilon}{\epsilon}\ , \
\end{equation}

\noindent
where $E$ represents the particles total energy and $\epsilon$ the
energy scale as defined above.

The resulting distribution for the same sample of subhaloes as already 
presented in \Fig{f:allmpsattidalarms} can be viewed in 
\Fig{f:streamenergy_z0}. This figure actually contains a substantial 
amount of information beside the energy distribution and hence requires 
careful explanation.

First of all, \Fig{f:streamenergy_z0}, shows the distribution of $q$ as 
defined by \Eq{eq:q}. According to \cite{Johnston.98} we should find a 
bi-modal distribution, at least for those subhaloes that are \textit{not} 
yet completely disrupted. We therefore added the fraction of mass in the 
stream of the total mass to the legend (i.e. $n_{\rm u}/n_{\rm tot}$, 
where $n_{\rm u}$ is the number of unbound particles and $n_{\rm tot}$ the 
total number of particles left in the subhalo and $n_{\rm u}$ together). 
To further highlight non-disrupted subhaloes the legend box in the upper 
right corner is shaded in grey for objects where less than 80\% of all 
particles are forming the debris field ($n_{\rm u}/n_{\rm tot}<0.8$). As 
we are in the unique position to separate leading and trailing arm debris 
using the method described in \Sec{ssec:armseparation} we not only 
present the combined energy distribution (black histograms) but also the 
individual distributions for trailing (red histograms) and leading arm 
(blue histograms). The colour-coded dots above the histograms are the 
relative energies $q$ of each individual debris particle where the 
saturation of the colour corresponds to stream age as defined in 
\Sec{ssec:armseparation}, too.

While many of the energy distributions are clearly bi-modal (even for 
disrupted subhaloes), we notice that the dip in between the peaks is not 
necessarily located at $q=0$, which just means that our choice of the 
energy scale is not good enough. Nevertheless, we claim that (i) this 
distribution can be obtained observationally once the full 6D phase-space 
information (for at least a substantial amount of debris particles/stars) 
along with a sufficient model for the underlying host potential is 
available and (ii) subsequently be used to separate leading from trailing 
arm even if the debris is wrapped around the host and the progenitor 
subhalo had a number of orbits, respectively. One may claim that our 
model for the host potential is out of reach for an observer as we are 
using the \textit{exact} potential values as returned by the simulation 
code, but we also applied the less sophisticated approach of using a 
simple spherically symmetric potential and were unable to detect 
significant differences in the respective energy distributions. 

Another noticeable feature in \Fig{f:streamenergy_z0} are the ``swings'' in 
the particle distribution: these most probably correspond to the orbital 
motion of the subhalo. If we were scaling the particle's energies with 
the corresponding energy scale when the particle became unbound or with 
the corresponding pericentre energy (as suggested by \cite{Johnston.98}), 
these ``swings'' might vanish and we could possibly achieve a ``cleaner''
distribution. However, this energy scale would not be directly accessible 
to an observer.

\subsubsection*{Separating leading and trailing arm via energy}
\begin{figure}
\begin{center}
  \epsfig{file=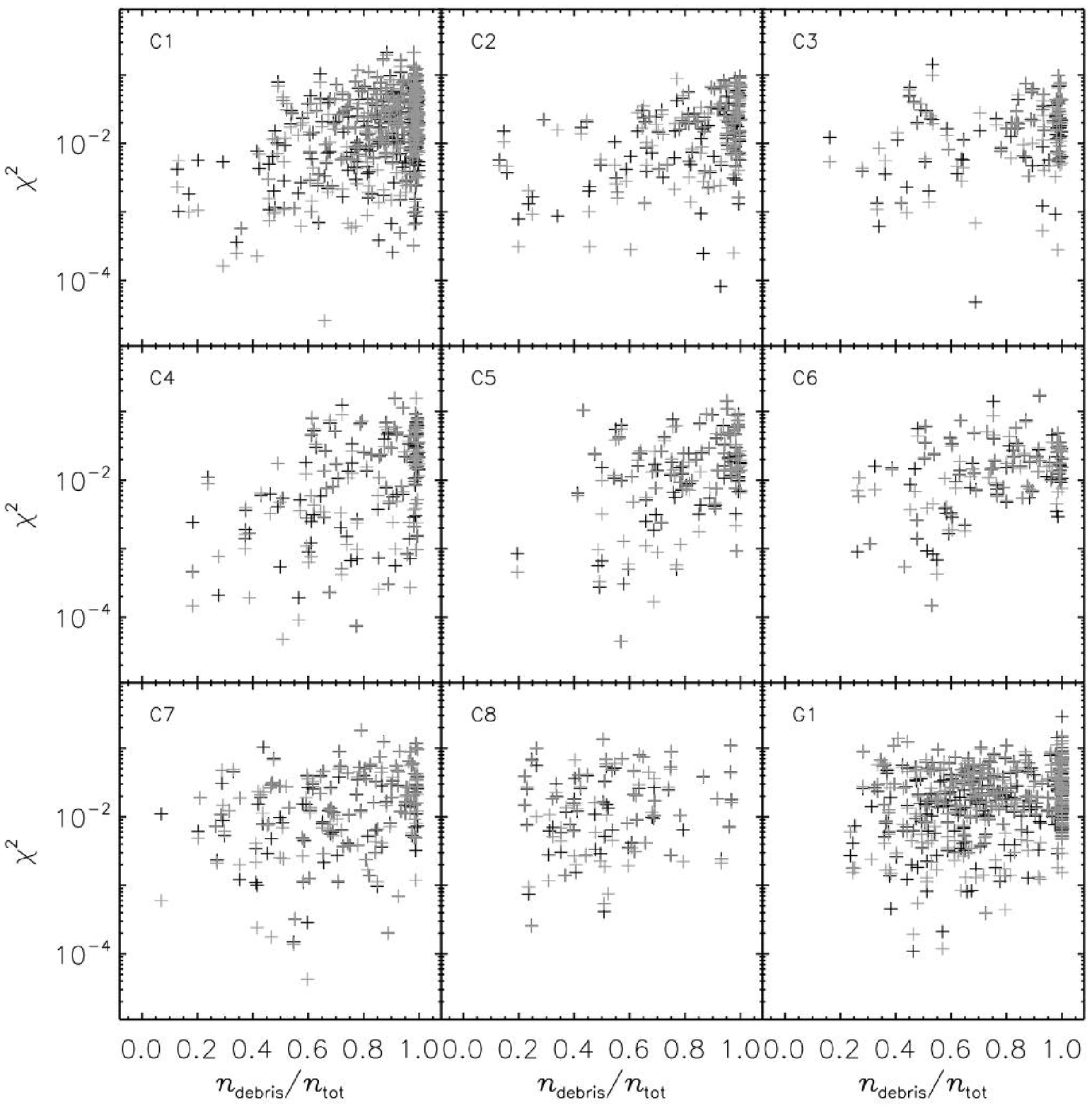,
    width=0.47\textwidth, angle=0} \end{center} \caption {
  Reproducibility of leading and trailing arm energy distribution by
  decomposing a double-Gauss-fit: The goodness $\chi^2$ for trailing
  (black) and leading (grey) arm is plotted versus the fraction of
  debris particles. }
\label{f:streamenergy_stats_ndebris}
\end{figure}

\begin{figure}
   \epsfig{file=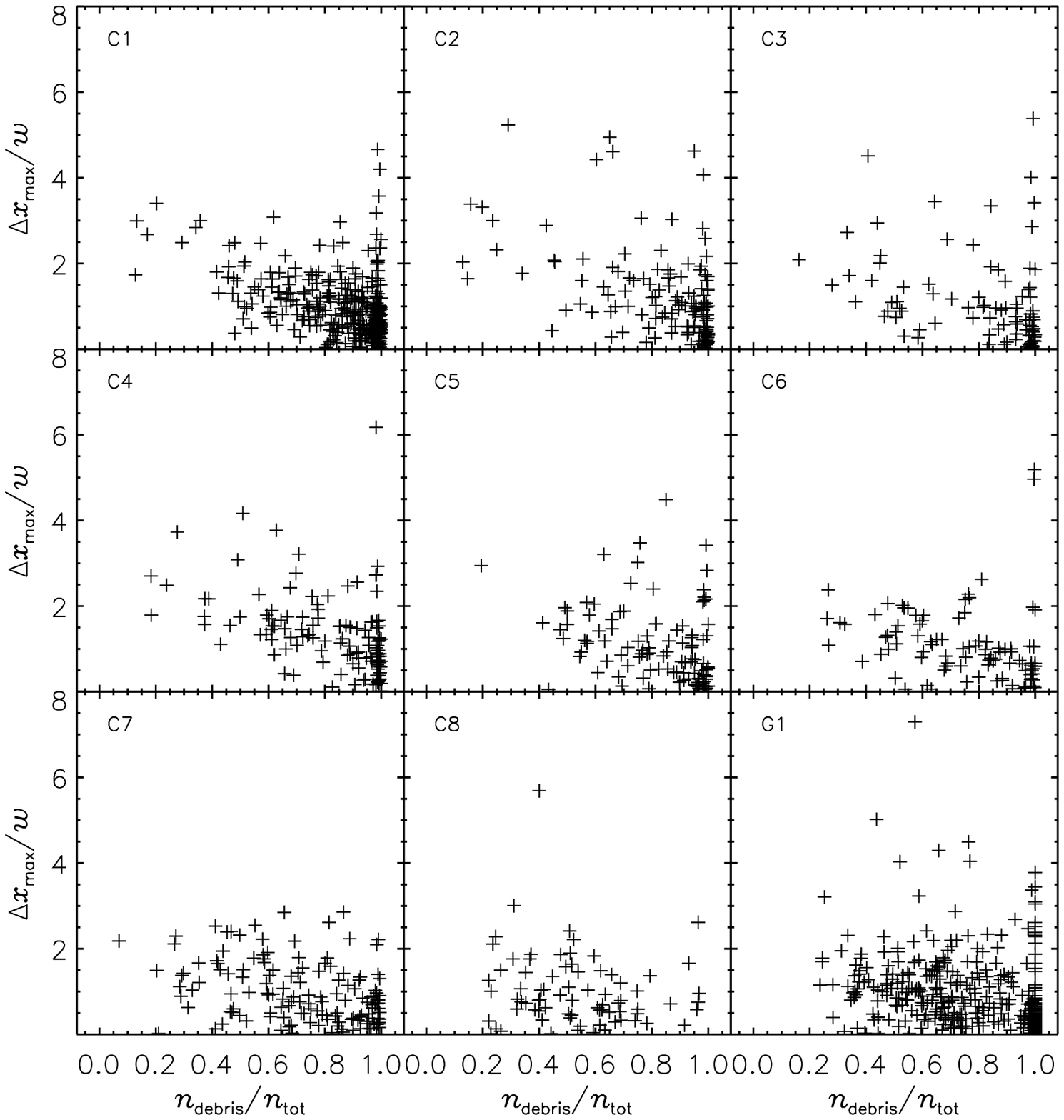, width=0.48\textwidth}
   \caption{Separation of maxima and the fraction of debris
     particles. Grey: 80\% or more particles in debris, black: 20\% or
     more particles are still bound.}
\label{f:streamenergy_stats_peaksep_ndebris}
\end{figure}

In order to check the credibility of our proposed method to classify a 
particle/star by its (relative) energy and assign it to either the leading 
or trailing arm we need to check the statistical significance of this 
technique. To this end we fitted a 'double Gaussian' to the energy 
distribution of the subhaloes in all our nine host haloes

\begin{equation} \label{eq:dgaussian}
\frac{dN}{dq} = A_1 e^{{-\frac{1}{2} \left( \frac{x-x_1}{\sigma_1} \right)^2}}
              + A_2 e^{{-\frac{1}{2} \left( \frac{x-x_2}{\sigma_2} \right)^2}} \ , \
\end{equation}

\noindent
where $A_1, A_2, x_1, x_2, \sigma_1,$ and $\sigma_2$ are free
parameters. The resulting best fit curves to the combined distribution
$dN/dq$ for a number of sample subhaloes are given by the solid black
lines in \Fig{f:streamenergy_z0}.

\begin{figure*}
\begin{center}
 \begin{minipage}{0.495\textwidth}
        \epsfig{file=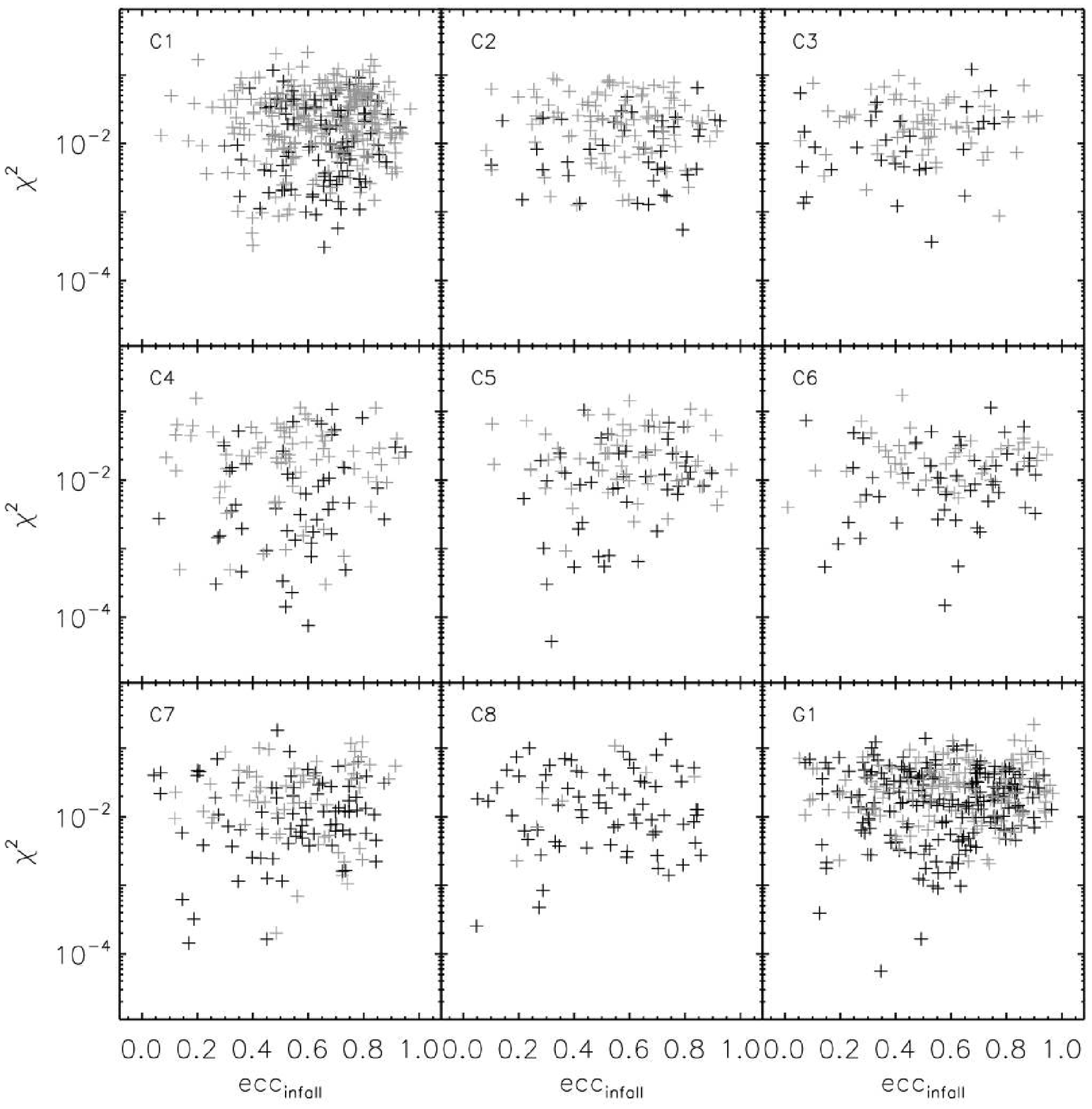, width=1\textwidth, angle=0}
 \end{minipage}
 \hfill
 \begin{minipage}{0.495\textwidth}
        \epsfig{file=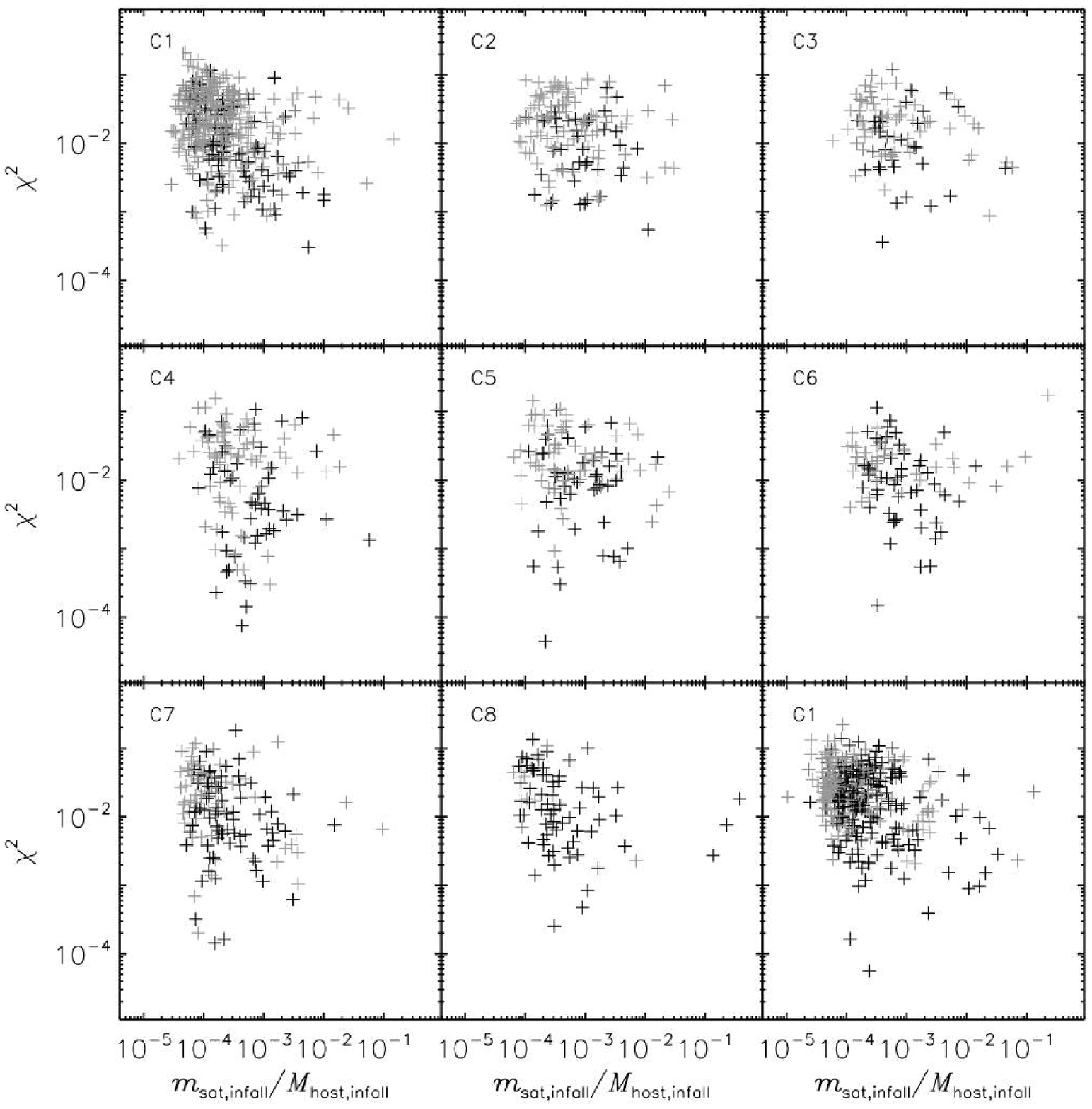, width=1\textwidth, angle=0}
      \end{minipage} \end{center} \caption { Correlation of the infall
      eccentricity (left panel) and infall mass (right panel) with the
      quality of the decomposition of the fit of the energy
      distribution into leading and trailing arm distribution. Grey
      (black) symbols mark disrupted (alive) subhaloes where the
      criterion for disruption corresponds to $n_{\rm
        debris}/n_{\rm tot} > 0.8$ again. }
\label{f:streamenergy_stats}
\end{figure*}

To gauge the integrity of our claim that it is possible to
separate leading and trailing arm using the energy distribution, we
make use of the fact that we are in the unique situation to 
separate leading and trailing arm particles again. We therefore calculate for each arm the energy distribution individually. We
further decompose the fit to the combined energy distribution based
upon all particles in the debris field (cf. \Eq{eq:dgaussian}) into
two single Gaussian (i.e. one should describe the leading and the
other the trailing arm). We check the validity of the
decomposed Gaussians by comparing them to the corresponding ``true''
energy distributions derived via our ``exact'' leading/trailing arm
classification method. Hence, for
each subhalo we calculate two $\chi^2$ values representative of
the squared deviations between the decomposed fit and the true
(binned) distributions for leading and trailing arm. The results can
be viewed in Figures~\ref{f:streamenergy_stats_ndebris},
\ref{f:streamenergy_stats_peaksep_ndebris},
and~\ref{f:streamenergy_stats} to be explained in more detail below.

\cite{Johnston.98} stated that the bimodality is in general more
pronounced for subhaloes that are \textit{not} disrupted, and for
disrupted subhaloes a single-peaked distribution is expected.  We
therefore start by inspecting the quality of our double-Gaussian
fits by plotting the respective $\chi^2$ values against the fraction
of mass in the debris field, i.e. number of particles $n_{\rm debris}$
in the debris field over number of particles $n_{\rm tot}=n_{\rm
  debris} + n_{\rm sat}$ where $n_{\rm sat}$ are the remnant particles
in the subhalo; a ratio of $n_{\rm debris}/n_{\rm tot}=1$ refers to
a completely disrupted subhalo. The result is presented in
\Fig{f:streamenergy_stats_ndebris} where we observe a (marginal) trend
for more disrupted subhaloes (i.e. larger $n_{\rm debris}/n_{\rm
  tot}$ ratios) to have higher $\chi^2$ values in agreement with our
expectation and \cite{Johnston.98}, respectively. However, we find no
difference for leading (grey symbols) and trailing (black symbols)
arm.

In \Fig{f:streamenergy_stats_peaksep_ndebris} we plot the
separation of the maxima of the two Gaussians (derived via fits to the
distributions stemming from the separation method and not the
decomposition of \Eq{eq:dgaussian} versus the debris fraction $n_{\rm
  debris}/n_{\rm tot}$ again. The peak separation is further
normalised to the sum of the half width at half maximum of each
curve. We can confirm a small trend towards smaller separations for
more disrupted subhaloes, i.e. the maxima are approaching each other
and leading to less distinguishable curves for more disrupted
subhaloes, as expected.

Finally we turn to our two subhalo properties again,
namely the infall eccentricity and infall mass, examining if or how 
they relate to the (bi-modal) energy distribution. We plot
in \Fig{f:streamenergy_stats} the correlation of the infall
eccentricity (left panel) and infall mass (right panel) with the
quality of the decomposition of the fit of the energy distribution to
a double Gaussian $\chi^2$. We (again) observe little relation with
eccentricity while there is a trend for more massive subhaloes to
lead to better fits and hence a clearer separation between leading and
trailing arm. Please note that we separated the subhaloes in
\Fig{f:streamenergy_stats} according to disrupted (grey symbols) and
alive (black symbols), where our criterion for disruption was fixed at
$n_{\rm debris}/n_{\rm tot} \geq 0.8$.

\subsection{Summary -- Leading vs. Trailing Arms}
In this section we investigated trailing and leading arms individually
working out differences and similarities between the debris
fields. Somewhat surprisingly, most of the properties we considered 
show little (if any) variation between the arms (e.g. velocity
dispersion), but there does appear to be a mass difference between
them. Our main conclusion is that even for complicated geometries of
both arms it is in principle possible to separate individual
particles/stars amongst leading or trailing arms by means of their
energy distributions. This in turn allows to (observationally) weigh
both arms and verify our claim that their masses are different.

In short, our results of the analysis of leading and trailing arm
properties can be summarised as follows:

\begin{itemize}

\item We presented a method to separate leading and trailing arm
  particles in cosmological simulations.

\item We confirmed that the method to separate leading and trailing
  arm utilising the energy distribution works well in cosmological
  simulations.

\item Trailing and leading arm can be separated observationally, when
  investigating the energy of debris particles: the energy
  distributions can be fitted by Gaussians, where the peak for the
  leading arm lies at smaller energies than the peak for the trailing
  arm.

\item The leading arm contains slightly more mass than the trailing
  arm for more than half of the subhaloes in our cluster haloes,
  whereas the opposite relation holds for the galactic halo. For more
  massive subhaloes, the fraction of mass in the trailing arm is
  increasingly higher.

\item We could not find differences in the velocity dispersions of
  leading and trailing arm.

\end{itemize}

\section{Summary and Conclusions}	
\label{sec:summary}


In this paper, which is the first in a series, we have investigated
the tidal disruption of subhaloes (or \emph{satellite galaxies}) 
orbiting in cosmological dark matter host haloes. We used nine high 
resolution cosmological $N$-body simulations of individual galaxy- and 
cluster-mass haloes, each of which contain in excess of $\sim 1$ million 
particles within the virial radius, and within which a few hundred subhaloes 
can be reliably resolved. The time sampling of our simulations is such
that we can track subhalo orbits in detail, which allows us to both
determine orbital planes for subhaloes and to follow the mass loss
they suffer as a function of time.

In the first part of the paper (i.e. \Sec{sec:hosts}) we concluded
that, while all of our host haloes are sufficiently relaxed, they 
exhibit a range of \emph{integral} properties.  For example, they 
show a spread in formation times, mass assembly histories and shapes.
These are precisely the kind of host properties that we expect to affect the
properties of tidal debris field stripped from orbiting satellites.

The second part of the paper 
(i.e. Sections~\ref{sec:subhaloes}--\ref{sec:trailing/leading})
focused on the subhaloes and their tidal disruption. Our main
conclusions may be summarised as follows;

\begin{itemize}

\item We predict that ``backsplash'' satellite galaxies should be found not
      only in the outskirts of galaxy clusters but also around galaxies.

\item Backsplash subhaloes contribute as much as 13\% of the mass of the total 
      tidal debris field within the host. This suggests that the region outside 
      the virial radius of the host galaxy should be considered also when
      observing the debris of satellites and searching for possible
      corresponding satellite galaxies.
  
\item Tidal stream properties correlate well with progenitor satellite properties.
      For example, we find a relationship between the scatter about the 
      (best-fit) debris plane and the infall mass of the satellite, as well 
      as the spread in radial velocity and the infall eccentricity.

\item We devised a method to separate leading and trailing debris arm in 
      cosmological simulations.

\item The masses of leading and trailing arms differ.

\item We do not find differences in the velocity dispersions of
      leading and trailing arms.

\item We present a method that uses energy distributions to separate leading 
      and trailing arms and that can, in principle, be employed by observers.

\item The formation of the tidal debris field is extremely complex in ``live'' 
      host haloes such that there appears to be little correlation between debris
      properties and host halo properties (age, mass assembly history, dynamical 
      state, shape).

\end{itemize}

We set out to address two central questions key to understanding tidal streams;
\begin{itemize}
\item \emph{Can observations of tidal streams be used to infer the properties
  of their parent satellite?}
\item \emph{Can tidal streams reveal properties of the underlying
  dark matter distribution of the host halo?} 
\end{itemize}

These have proven to be very weighty questions to address, and so we
focused on the first question in this paper. As the bullet points
above indicate, the answer would appear to be yes. For example, there
is a clear correlation between mass loss and orbital infall
eccentricity, and between deviation of the tidal debris field from its
orbital plane and the infall mass of the progenitor
satellite. Addressing the second question is less straightforward. We
have examined our data for correlations between stream properties and
host properties as summarised Table~\ref{t:haloprop}, but evidence for
correlations with host mass accretion history, dynamical state and
shape are extremely weak at best.  Indeed, the only suggestion of a
correlation that we find is between the mass ratio of leading to
trailing arm and the virial mass of the host; while all our cluster
mass haloes have more mass in the leading debris arm, the relation is
reversed in the galaxy mass halo. 

Further, it is interesting to note that the galaxy halo G1 appears to
be distinct from the eight galaxy cluster haloes (with the possible
exception of C1) in terms of how its bulk properties relate to its
stream properties. However, it is not clear whether this is a
systematic effect, given that G1 is the only realisation of a galactic
type halo. For example, the differences are not so striking if one
compares G1 with C1. We are in the process of
building a sample of galaxy haloes to supplement G1.\\

We now conclude with some general remarks. Previous studies of individual 
satellites disrupting in analytic host potentials using controlled $N$-body 
experiments have hinted at simple and direct links between, for example, 
the flattening of the host potential and the dispersion of the debris field 
\citep[e.g., ][]{Ibata.etal.03, Helmi.04, Penarrubia.etal.06}. However, our
experience suggests that the situation is vastly more complicated in ``live'' 
host haloes, where the host undergoes a complex mass accretion history, contains
a wealth of substructure and cannot be easily modeled as a simple ellipsoid.
\citet[][]{Penarrubia.etal.06} pointed out that all present-day
observable stream properties can only constrain the present mass
distribution of the host halo, independent of its past evolution. We
would argue strongly that even present-day properties of the host halo
will be difficult to infer from stream properties, given the
complexity of interactions responsible for the emergence of debris
fields. There are multiple processes driving these effects and so
while we note tentative correlations, we require further simulations
to help us understand what is happening. This will represent the focus
of the following paper in this series. We point out that this
situation is akin to the findings of \citet[][]{Gill.etal.04.2}; they
presented a detailed study of the dynamics of satellite galaxies in
similar host haloes and were also unable to establish correlations to
the host properties.

We close with a cautionary note: this paper is \textit{not} to be 
understood as a definitive quantitative analysis of tidal streams in
dark matter haloes. Rather it shows that certain (unexpected)
correlations between the debris field and satellite properties
exist. Our study therefore represents the first step towards our
understanding of whether or not tidal streams from disrupting satellite 
galaxies can be used to deduce robustly properties of their host dark 
matter haloes and of the orbits of the satellites.

\section*{Acknowledgments}

We thank Anatoly Klypin for providing the galaxy-mass simulation data
and for several stimulating discussions. We enjoyed invaluable debates
with Stuart Gill, Stefan Gottl\"{o}ber and Volker M\"{u}ller. KW and
AK acknowledge funding through the Emmy Noether Programme by the DFG
(KN 755/1). CP acknowledges the support of the Australian Research
Council funded ``Commonwealth Cosmology Initiative'', DP Grant
No. 0665574. The simulations of the eight cluster-mass haloes were
carried out on the Swinburne Supercomputer at the Centre for
Astrophysics~\& Supercomputing, Swinburne University. The simulation
of the galaxy-mass halo and all of the analysis presented in this
paper was carried out on the Sanssouci cluster at the AIP, where also
all the post-processing and analysis has been undertaken.

\bibliographystyle{mn2e} 

\appendix

\section{Measuring Halo Shapes: A New Approach} 
\label{app:shapemethod}

\subsection{Motivation for our Approach}
\label{app:shapemethods_motivation}
There is a rich literature on methods for the determination of halo
shapes. Many of these studies utilise methods that identify subsets of
particles and deduce axis ratios and orientations of best fitting
ellipsoids to the haloes from the eigenvalues and eigenvectors of the
moment of inertia tensor of the particles \cite[e.g.,
][]{Frenk.etal.88,Jing.Suto.02} \footnote{At this point we need to
  mention the novel method of \citet{Springel.White.Hernquist.04}
  \citep[also used in~][]{Hayashi.Navarro.Springel.07} which does not
  use the moment of inertia of the particle distribution to determine
  halo shape and orientation. Rather, it explicitly fits an ellipsoid
  to the potential estimated on a fine mesh.}. However, the predicted
shape depends on what is measured. Typically particles are selected
according to radius \cite[e.g.][]{Frenk.etal.88} but they can also be
selected according to local density \cite[e.g.][]{Jing.Suto.02} or
potential \cite[e.g.][]{Chris.PhD.03}, with differing results. The
presence of substructure within haloes also affects the measured shape
\cite[e.g.][]{Bailin.Steinmetz.05}. We wish to develop a method that
produces an unbiased estimate of a dark matter halo's shape and
orientation and that is unaffected by substructure.

\subsection{Isodensity and Equipotential Surfaces}
\label{app:isoequi}
It can be argued that the shape of the potential rather than the mass
density provides a more meaningful measure of the shape of the halo for
our purposes, because it is the potential that determines the orbits of
satellites and tidal debris \citep[e.g.][]{Hayashi.Navarro.Springel.07}. 
We use a novel technique to define both isodensity and equipotential
surfaces while simultaneously allowing identifying a corresponding
(spherical) radius \Riso\ with the given surface. 

Our method for locating particles on isodensity/equipotential surfaces
is as follows;

\begin{itemize}

\item We start by calculating the local value of the density and
  potential at each particle position. To this extent we invoke again
  the original adaptive grid hierarchy used during the actual
  simulation, obtain the density and potential on these grids and
  interpolate to the particle positions.

\item We select a fiducial radius \Riso. We then successively increase
  the thickness $\Delta R$ of a spherical shell centred about the
  fiducial radius \Riso, i.e. the shell extends from $R_{\rm iso}^{\rm
    min}=R_{\rm iso}-\Delta R$ to $R_{\rm iso}^{\rm min}=R_{\rm
    iso}+\Delta R$. The value of $\Delta R$ is determined in a way
  that we require to include a certain fraction of the total number of
  halo particles in the shell; in our case this amounts to 10\%.

\item For all the particles in the shell we calculate the mean value
  of the density (potential). This now defines the threshold
  isodensity (equipotential) value at our fiducial radius \Riso.

\item From all simulation particles in the vicinity of the host
  (i.e. out to 2.5 \Rvir) we now pick those ones whose local density
  (potential) value lies within a range of 10\% about the mean value.

\end{itemize}

\subsection{Shape Determination}
\label{app:shapes_determination}
Having identified the subset of particles with the desired local
density (potential) at \Riso\ (cf. \Sec{app:isoequi}), we compute
their moment of inertia tensor $I_{jk}$.  The eigenvalues and
eigenvectors of $I_{jk}$ are related to the axis ratios and orientation,
respectively, of the best-fitting ellipsoid to the particles.  For a
distribution of discrete point masses, $I_{jk}$ is expressed as

\bq
\label{eq:Inertia} 
I_{jk} = \sum_{i=1}^N m_i (r^2_i \delta_{jk} - x_{ij}x_{ik}) \quad\\ 
	{\rm with}\ j,k = \{1;2;3\} \quad, 
\eq 

\noindent 
where $m_i$ is the mass of particle $i$, $N$ the number of particles
and $r_i = \sqrt{x_{i1}^2+x_{i2}^2+x_{i3}^2}$ is the distance of
particle $i$ from the centre of mass of the particles\footnote{We note
  that the same ellipsoid fit can be obtained from the tensor 
\bq
  \label{eq:commoninertiatensor}
	{\cal M}_{jk} = \sum_{i=1}^N m_i x_{ij}x_{ik}\ ,
\eq
\noindent which has been used widely in previous studies. This differs
from $I_{jk}$, which has an additional $r^2_i \delta_{jk}$. Both forms
provide axis ratios and orientations that are identical (though the
individual eigenvalues are different).}.

The novelty in our method of determining the shape is that we weigh
particles by their local density when computing $I_{jk}$ as follows;

\bq
\label{eq:inertiaII}
I_{jk} = A \cdot \sum_{i=1}^N \frac{1}{\rho_i} (r^2_i \delta_{jk} - x_{ij}x_{ik}) \quad\ \  {\rm with}\ j,k = \{1;2;3\}.
\eq

\noindent 
Here $\rho_i$ is the local density of particle $i$ and the
normalisation factor $A$ is given by, 

\bq
\label{eq:normdens}
A = \frac{1}{\sum_i {1/\rho_i}} \cdot N\ .\\
\eq

\noindent
This is necessary to ensure that especially particles in equipotential
shells, which can have a substantial spread in their local densities,
have approximately equal weighting in the summation. If this weighting
is neglected, particles in higher density regions will dominate the
summation and will bias our estimate of the axis ratios of the
best-fitting ellipsoid. The formula for obtaining the correct axis
lengths $a>b>c$ from the eigenvalues of this inertia tensor is given
in Appendix~\ref{app:axislengths} below. 

As our host haloes contain a considerable amount of substructure which
may distort the determination of the host's shape, we performed the
shape analysis with as well as without removing all subhalos (i.e. all
bound spherical halos found by the \AHF\ analysis) from the host. We
found that these subsystems do in fact have a great influence on the
shape (especially on the shape of the potential) and thus we removed
them for a credible and unbiased shape determination.

Our shape determination of all nine host haloes is presented in
Appendix~\ref{app:halo_shapes} where we show the best fit ellipsoid to both
the equipotential (upper panels) and isodensity (lower panels)
contours at $R_{\rm iso} = R_{\rm vir}$.

We further calculate the
usual measure for triaxiality $T=(a^2-b^2)/(a^2-c^2)$ and list the
values for both $T$ and $c/a$ (sphericity measure) at $R_{\rm iso} =
R_{\rm vir}$ in Table~\ref{t:haloprop}.

\subsubsection{Comparison to Previous Methods}
The simplest method determines the shape and orientation of a halo using 
all particles within a spherical volume or shell at a given radius 
\cite[e.g.][]{Frenk.etal.88,Kasun.Evrard.05,Hopkins.Bahcall.Bode.05, 
Bailin.Steinmetz.05}. While this method robustly recovers the orientation 
of the halo, the resulting axis ratios tend to be biased towards larger 
values (i.e. haloes are predicted to be rounder).

An alternative iterative approach to the problem again determines the 
shape and orientation of a halo using all particles within a spherical 
volume or shell, but this initial surface or shell is now deformed in the 
direction along the principal axes of the best fitting ellipsoid; this 
process is repeated until convergence is achieved \cite[e.g.][]{Katz.91, 
Dubinski.Carlberg.91,Warren.etal.92,Bullock.02,Allgood.etal.06,Maccio.etal.06}. 
Both \cite{Jing.Suto.02} and \cite{Bailin.Steinmetz.05} have noted that 
iterative methods have difficulty in achieving convergence in simulations 
in which haloes are very well resolved and contain a population of 
satellites. Satellites tend to lead to a distortion of the shape, and this 
is most pronounced in the outermost parts of host haloes where recently 
accreted satellites are most likely to be found.

The impact of substructure can be reduced by working with a ``reduced 
inertia tensor'';
\bq
\label{eq:reducedinertiatensor}
\hat{{\cal M}}_{jk} = \sum_{i=1}^N \frac{m_i x_{ij}x_{ik}}{r_i^2}\ .
\eq
\noindent Here each particle is weighted by the inverse square of its
distance to the centre of the halo. While this recovers accurately the
orientation of the ellipsoid, the axis ratios are systematically
overestimated and thus haloes are predicted to be more spherical than
they actually are \cite[e.g.][]{Bailin.Steinmetz.04}.

Both of these approaches have in general been applied in studies in which 
the best resolved haloes contain several tends of thousands of particles 
at most, in which it is not possible to defined thin shells containing 
many particles with similar local densities or potentials. A couple of 
studies \citep{Jing.Suto.02,Chris.PhD.03} have looked at the shapes of 
well resolved haloes, each containing several hundred thousand particles 
within the virial radius. They identified shells of particles selected by 
local density and potential and determined the best fitting ellipsoids. 
While this approach is similar to the one we have adopted, we note that 
these authors selected particles according to a local density or potential 
threshold.

\subsubsection{Advantages of Our New Approach}
The method presented in Appendix~ref{app:isoequi} enables us to select 
particles explicitly according to both their radius with respect to the 
centre of the halo and their local density (potential). This is desirable 
for a number of reasons.

\begin{itemize}

\item It overcomes the problems of convergence faced by iterative
  methods and it avoids the artificial bias towards more spherical
  shapes introduced by the reduced inertia tensor
  (equation~\ref{eq:reducedinertiatensor}). 

\item By focusing on thin shells, we can obtain uncorrelated results
  for the shapes in the inner and outer parts of haloes.
  
\item By fixing the radius at which we select particles rather than
  fixing the local density or potential threshold of particles 
  \citep[as in][]{Jing.Suto.02,Chris.PhD.03}, we can compare directly 
  the shape of the local density and potential at the same radius. 
  
\item We can compare the shapes of different haloes at the same
  fraction of the virial radius, which depends solely on virial mass,  
  rather than at a similar local density or potential, which will
  depend on concentration, recent merging history, larger scale
  environment, etc....
  
\end{itemize}

\section{Measuring Halo Shapes: From Eigenvalues to Ellipsoid Axes}
\label{app:axislengths}

When we diagonalise the moment of inertia tensor (as defined by 
equation~\ref{eq:Inertia}) for a thin shell of particles, we obtain the 
``principal moments'',
\bqa
\lambda_1 &=& I_{x'x'} = M \sum_{i=1}^N (y'^2_i + z'^2_i)\\ 
\lambda_2 &=& I_{y'y'} = M \sum_{i=1}^N (z'^2_i + x'^2_i)\\
\lambda_3 &=& I_{z'z'} = M \sum_{i=1}^N (x'^2_i + y'^2_i).
\eqa
\noindent Here $x'$, $y'$, $z'$ are the coordinates in the principal axes 
frame, $M$ is the total mass of the shell.

For a thin ellipsoidal shell with major, intermediate and minor axes $a > 
b > c$, the corresponding principal moments are,
\bqa
\label{eq:thinellvalue}
\lambda_{\rm ell,1} &=& \frac{1}{3} M (b^2 + c^2)\\
\lambda_{\rm ell,2} &=& \frac{1}{3} M (c^2 + a^2)\\
\lambda_{\rm ell,3} &=& \frac{1}{3} M (a^2 + b^2)
\eqa
\noindent If the thin shell of particles can be described as an 
ellipsoidal shell, then the coordinate axes $a$, $b$ and $c$ can be chosen 
such that $\lambda_{\rm ell,i} = \lambda_i$. It follows that
\bqa \label{eq:ellipsoid_axes}
a &=& \sqrt{\frac{3(-\lambda_1 + \lambda_2 + \lambda_3)}{2M}}\\
b &=& \sqrt{\frac{3(\lambda_1 - \lambda_2 + \lambda_3)}{2M}}\\
c &=& \sqrt{\frac{3(\lambda_1 + \lambda_2 - \lambda_3)}{2M}},
\eqa
\noindent which allow us to relate our thin shells of particles to their 
best fitting ellipsoids.

\bsp

\section{Host Halo Shapes\\(C1-C8 and G1)}
\label{app:halo_shapes}

Here we present our shape determination of all nine host haloes by
showing the best fit ellipsoid to both the equipotential (upper
panels) and isodensity (lower panels) contours at $R_{\rm iso} =
R_{\rm vir}$.

Even though most ellipsoids fit rather well, we also note that for
some host haloes there is an offset with respect to the particle
distribution (especially C6, C8, G1).  In these cases, the density
centre of the host does not correspond to the centre of isodensity
(equipotential) shells. In most these cases (i.e. C6 and G1) there is
a more or less pronounced 'bump' visible in the equipotential plots
(upper panels), indicating an adjacent substructure currently merging
with the host halo.  Though substructure was removed, small
overdensities in the environment of the subhalo still remain and
distort the equipotential surfaces.  \cite{Gao.White.06} recently
studied asymmetries in the inner regions of $\Lambda$CDM haloes,
claiming that recent accretion events and deviations from equilibrium
are responsible for such an offset.

\begin{figure}
  \begin{center}
    \epsfig{file=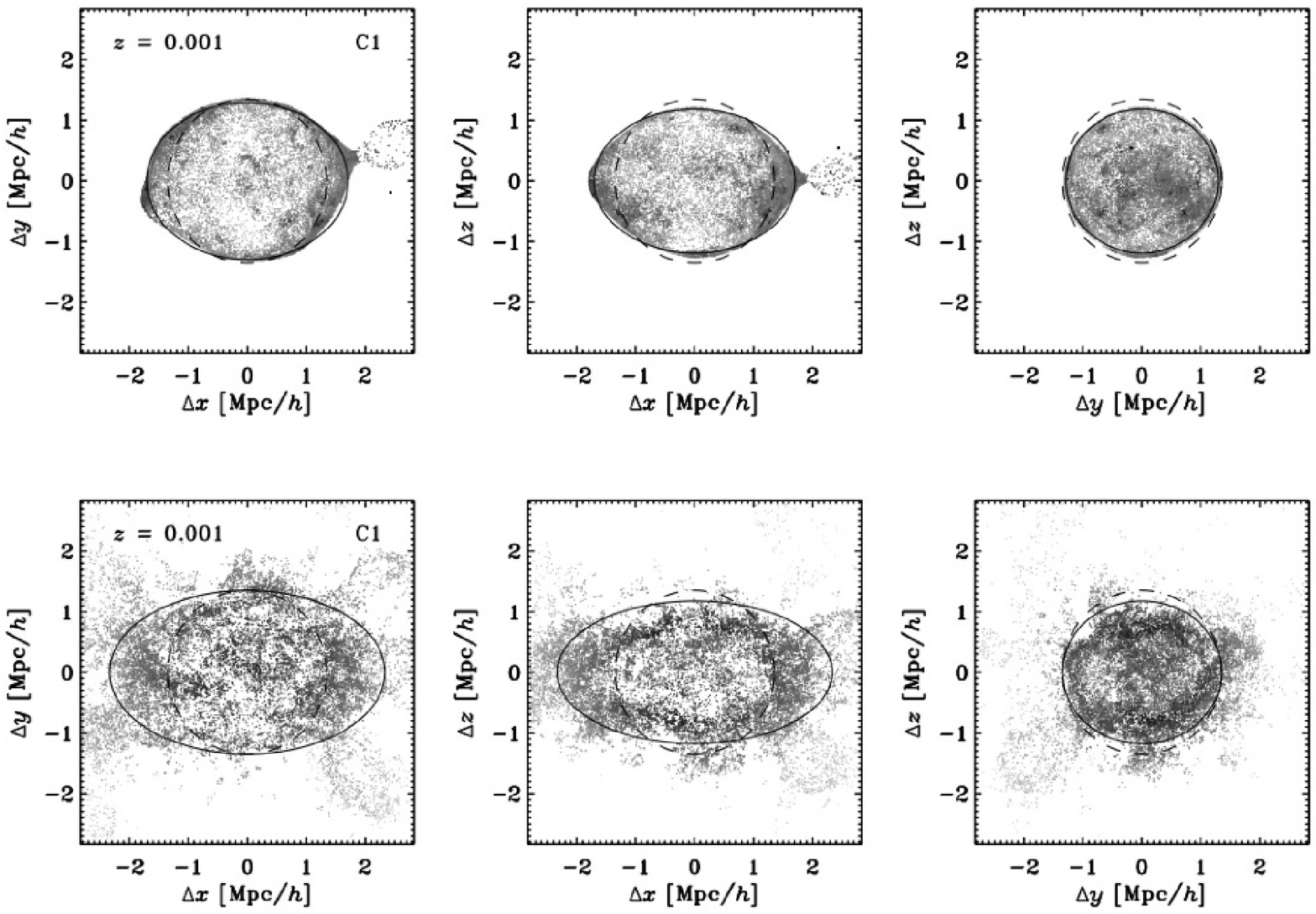, width=0.46\textwidth, angle=0}
  \end{center}
  \caption
  { Best fit ellipsoids (solid lines) for halo \simu{C1}{01-01} at
    \Riso=\Rvir. The top row shows particles in the equipotential
    shell, projected to the planes of the ellipsoid (eigenvector
    coordinate system). The bottom row shows the best fit ellipsoid
    for the corresponding isodensity range. The dashed circles mark
    the virial radius. Particles bound to satellites have been removed
    \emph{and} remaining particles weighted by their local density
    prior to fitting.}
      \label{f:hostshape.01-01.NoSub.adweight}
\end{figure}

\begin{figure}
\begin{center}
   \epsfig{file=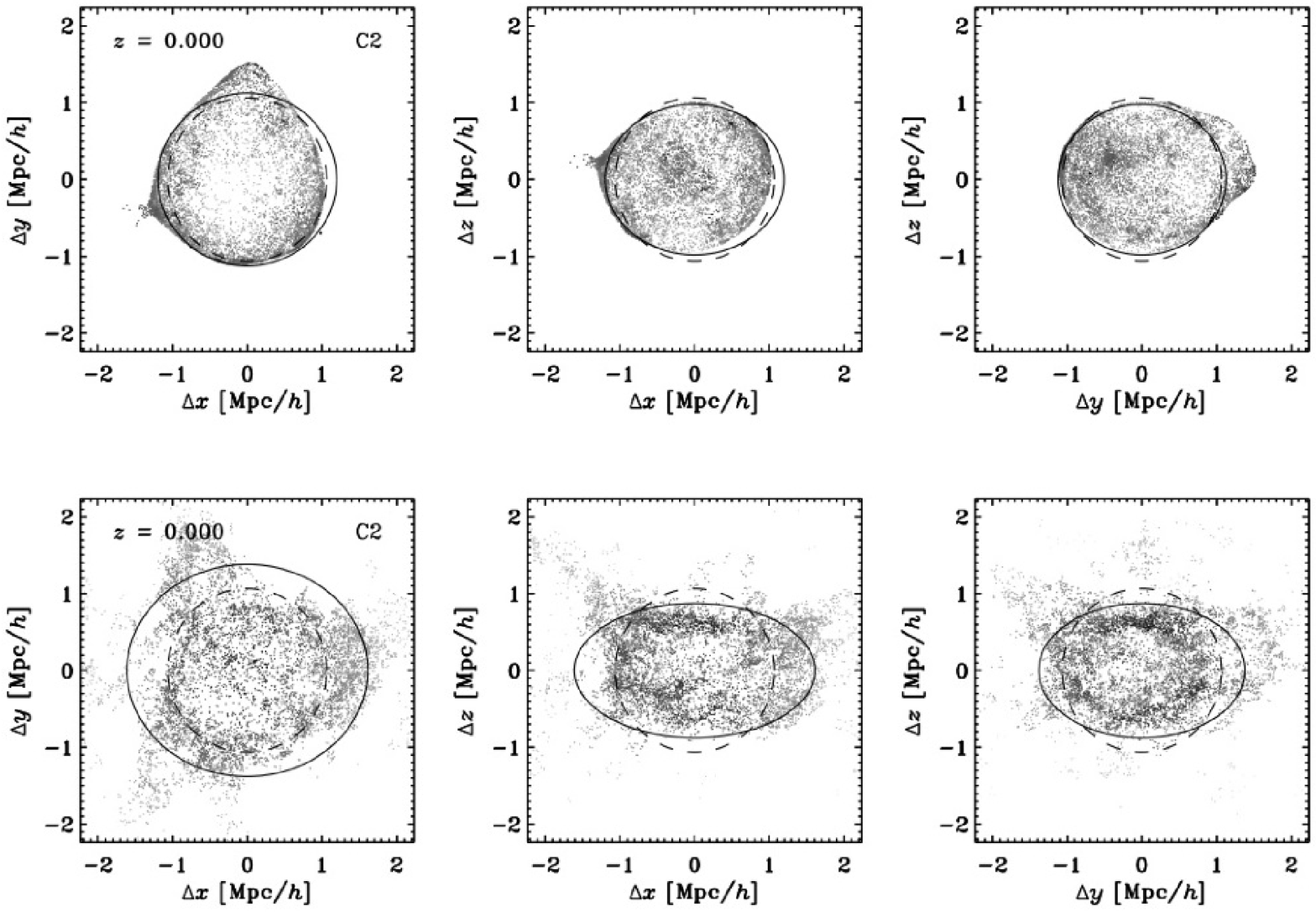, width=0.46\textwidth, angle=0}
\end{center}
\caption
{ Ellipsoid fit for the cluster \simu{C2}{04-03}.}
\label{f:hostshape.04-03}
\end{figure}

\begin{figure}
\begin{center} 
   \epsfig{file=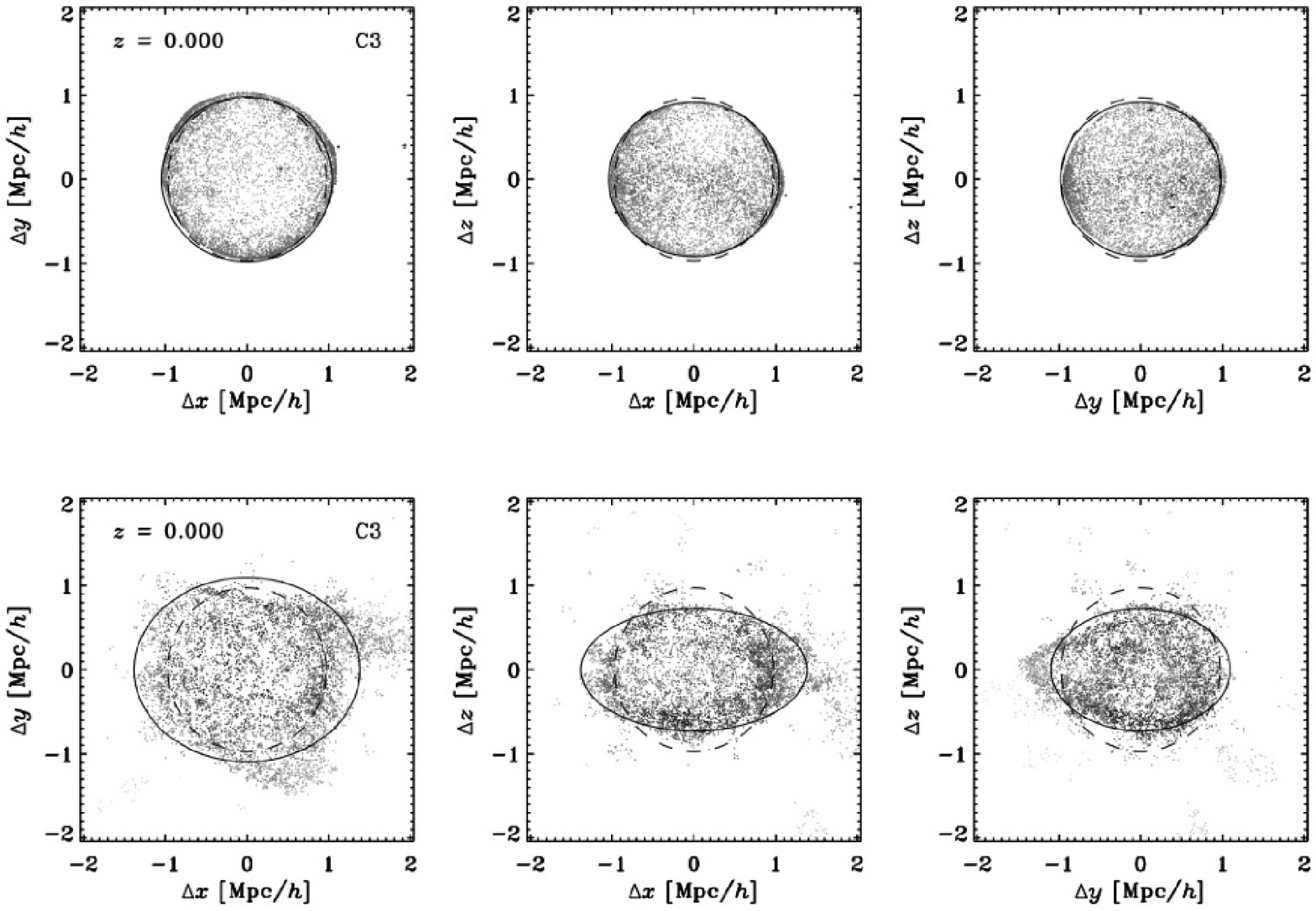, width=0.46\textwidth, angle=0}
\end{center}
\caption
{ Ellipsoid fit for the cluster \simu{C3}{01-07}.}
\label{f:hostshape.01-07}
\end{figure}

\begin{figure}
\begin{center}
   \epsfig{file=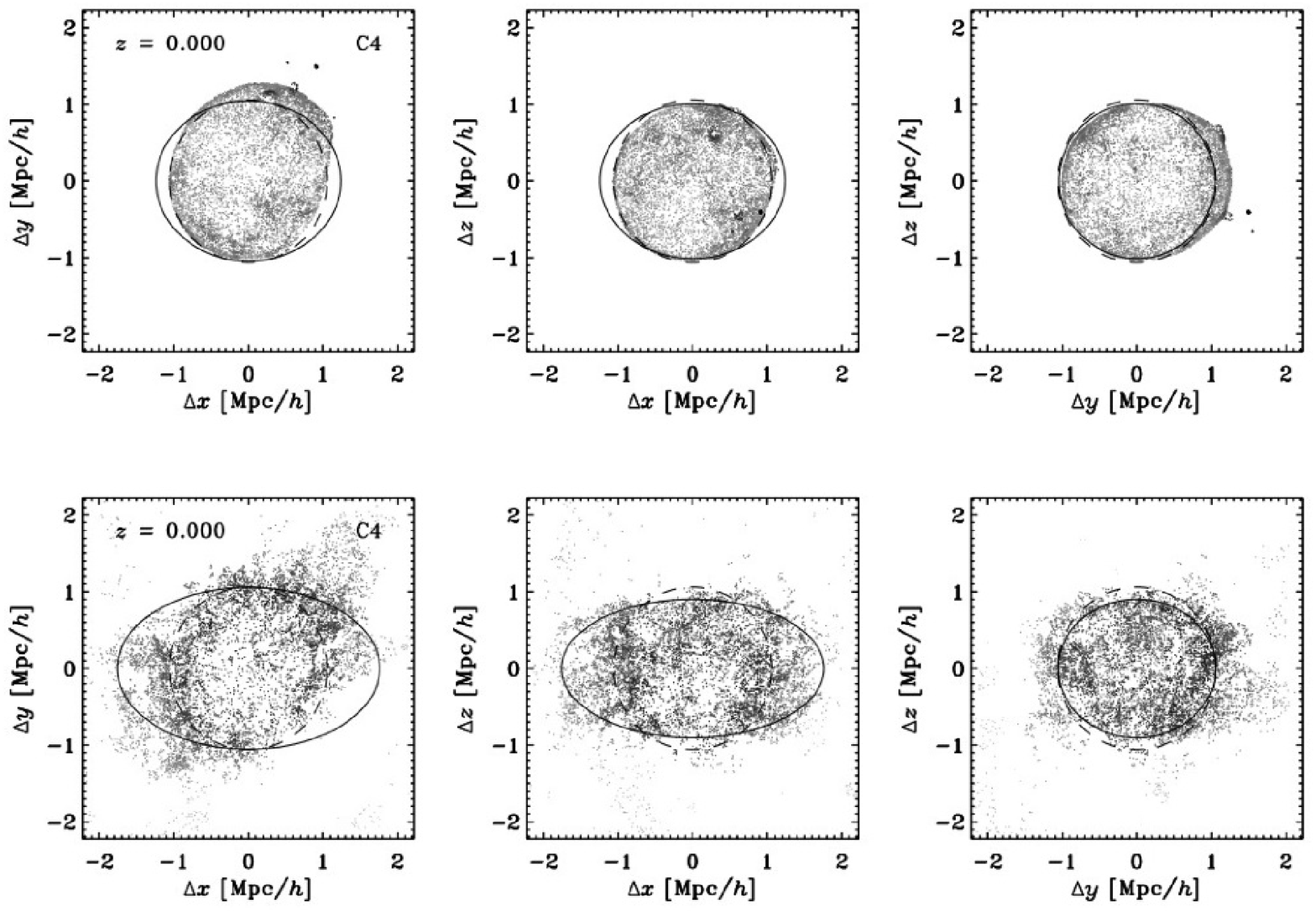, width=0.46\textwidth, angle=0}
\end{center}
\caption
{ Ellipsoid fit for the cluster \simu{C4}{02-07}.}
\label{f:hostshape.02-07}
\end{figure}

\begin{figure}
\begin{center}
   \epsfig{file=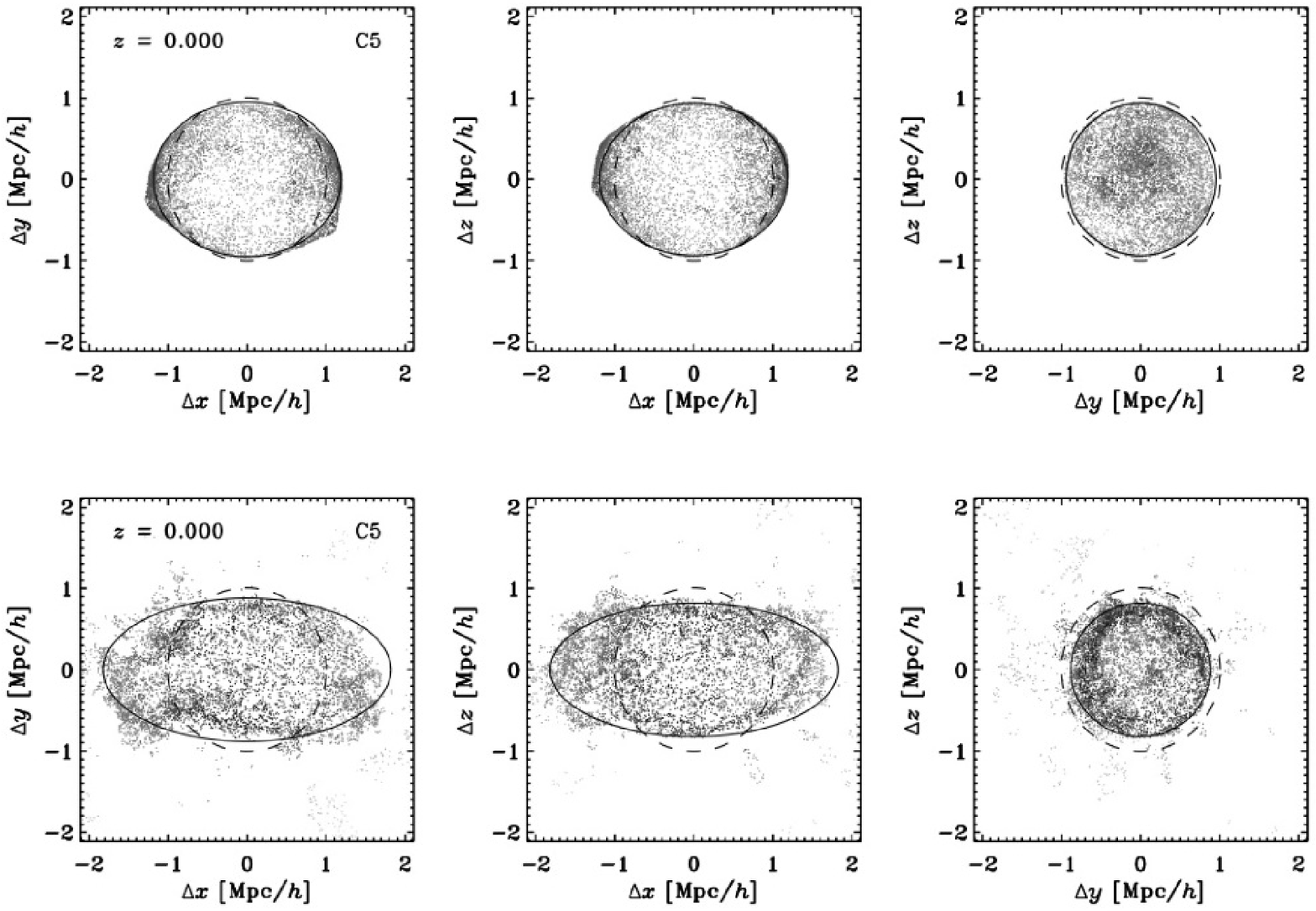, width=0.46\textwidth, angle=0}
\end{center}
\caption
{ Ellipsoid fit for the cluster \simu{C5}{03-05}.}
\label{f:hostshape.03-05}
\end{figure}

\begin{figure}
\begin{center}
   \epsfig{file=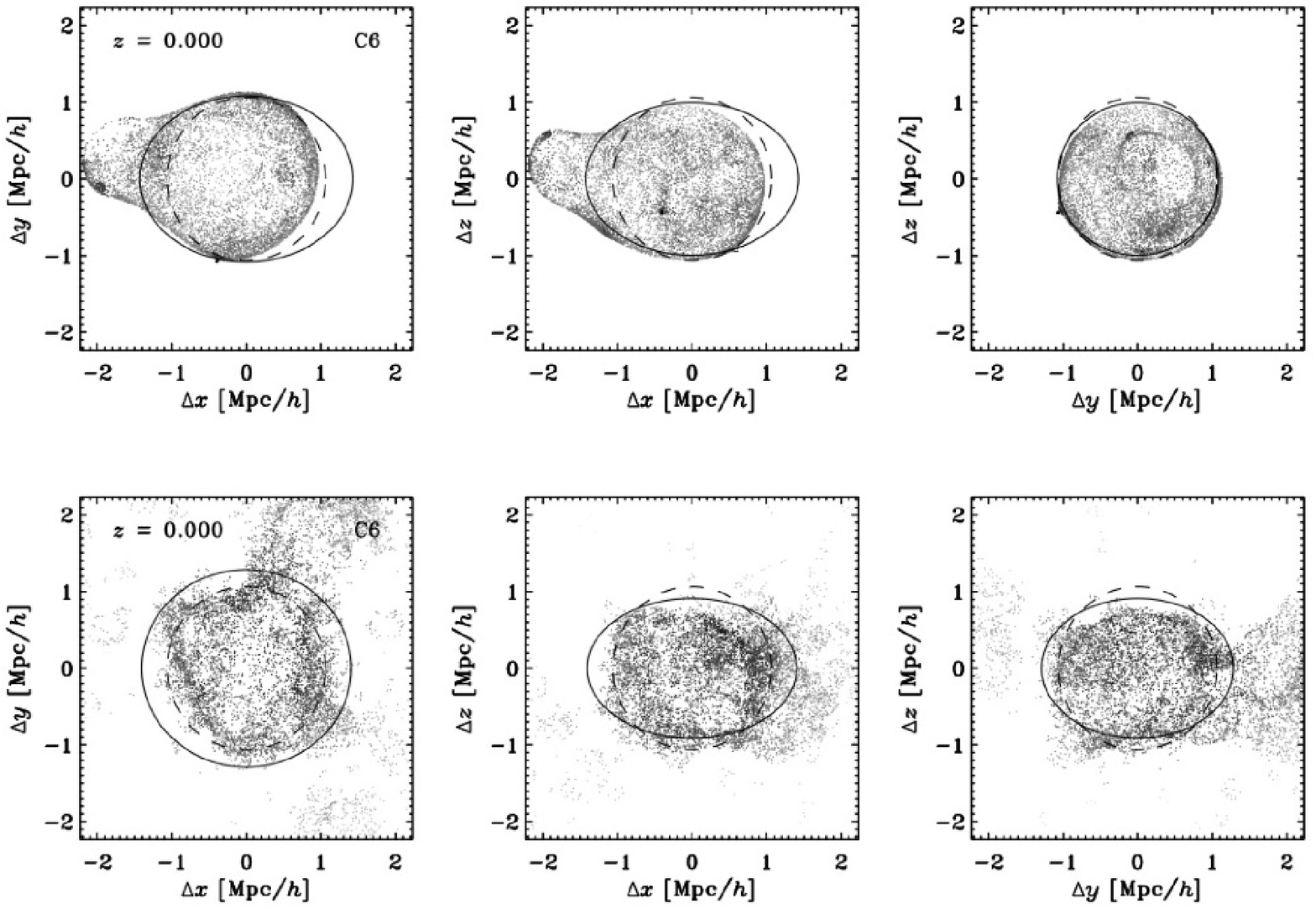, width=0.46\textwidth, angle=0}
\end{center}
\caption
{ Ellipsoid fit for the cluster \simu{C6}{04-04}.}
\label{f:hostshape.04-04}
\end{figure}

\begin{figure}
\begin{center}
   \epsfig{file=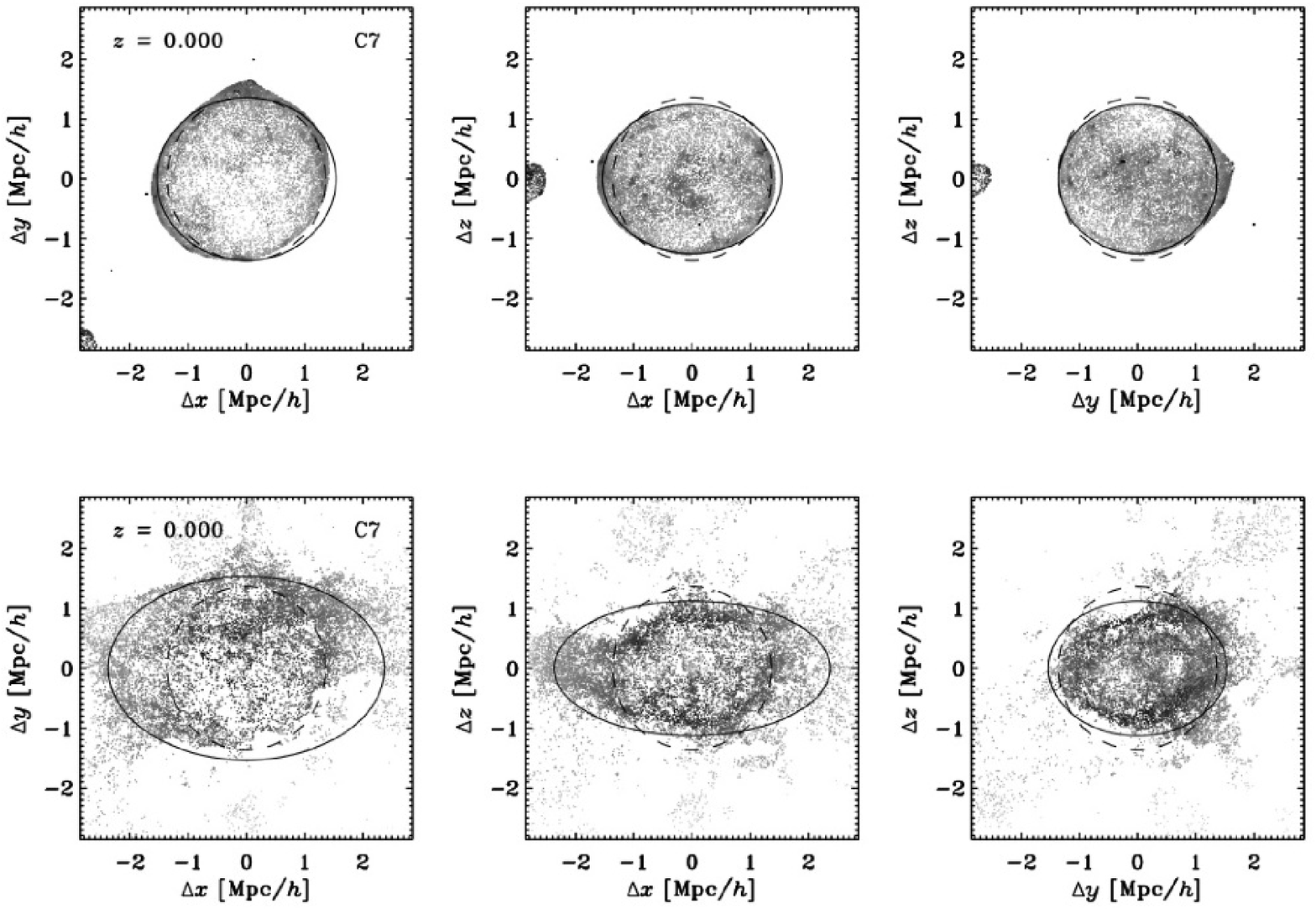, width=0.46\textwidth, angle=0}
\end{center}
\caption
{ Ellipsoid fit for the cluster \simu{C7}{01-02}.}
\label{f:hostshape.01-02}
\end{figure}

\begin{figure}
\begin{center}
   \epsfig{file=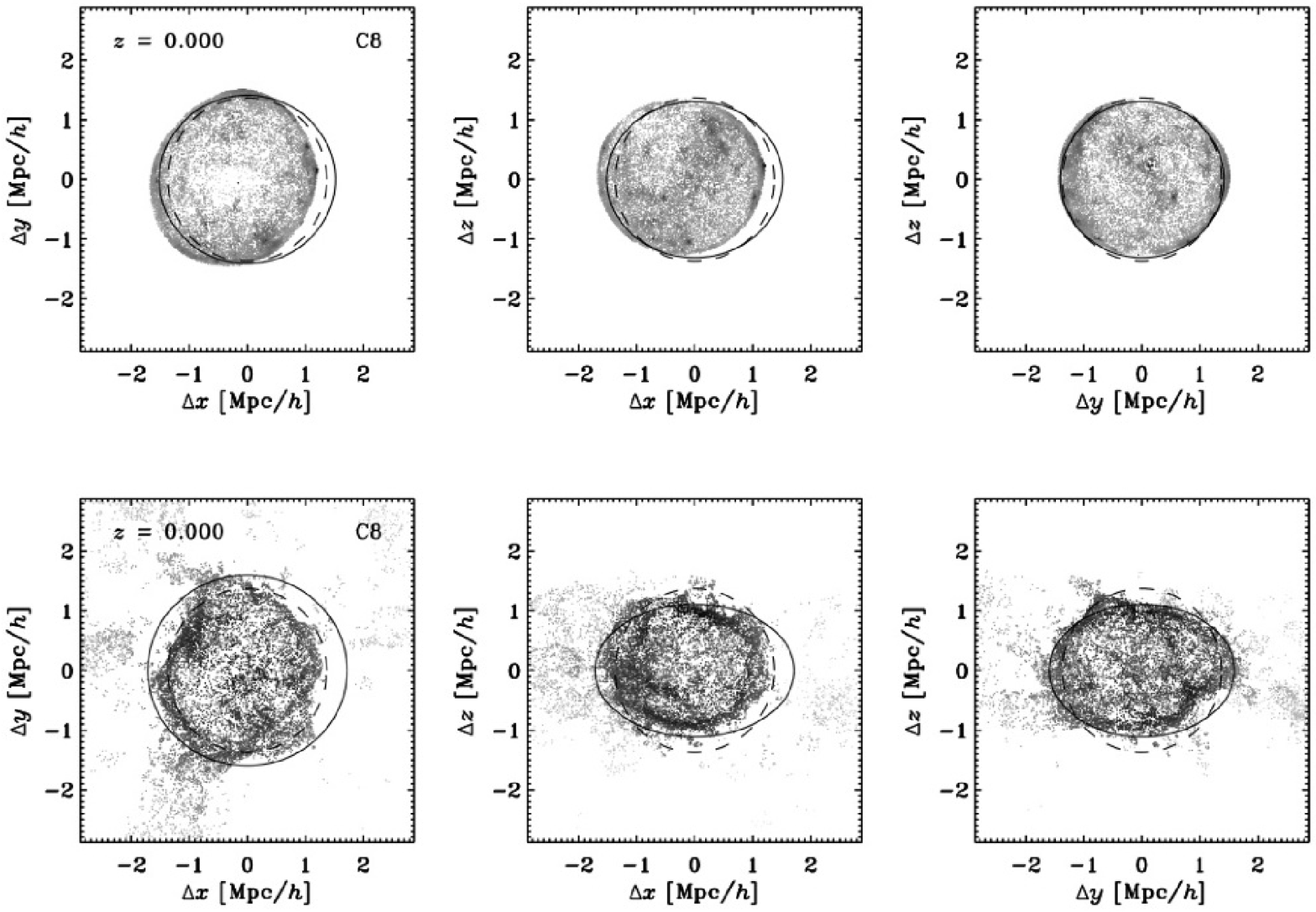, width=0.46\textwidth, angle=0}
\end{center}
\caption
{ Ellipsoid fit for the cluster \simu{C8}{01-10}.}
\label{f:hostshape.01-10}
\end{figure}

\begin{figure}
\begin{center}
   \epsfig{file=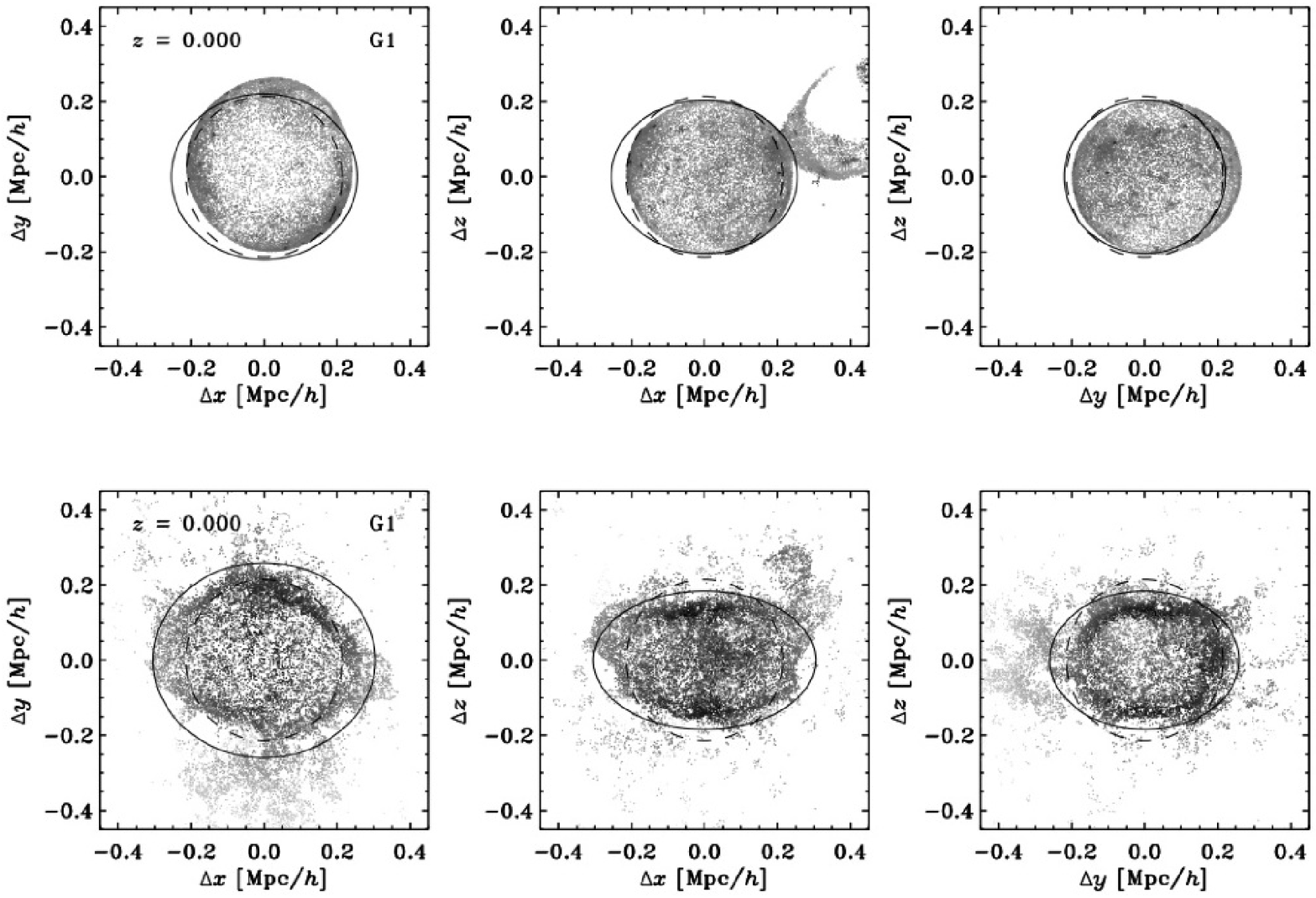, width=0.46\textwidth, angle=0}
\end{center}
\caption
{ Ellipsoid fit for the galaxy \simu{G1}{Box20b}.}
\label{f:hostshape.Box20b}
\end{figure}

We further note that the isodensity as well as the equipotential
surfaces in Figures~\ref{f:hostshape.01-01.NoSub.adweight} --
\ref{f:hostshape.Box20b} are in fact defining halo boundaries close to
the virial radius (dashed circles), just as intended. Or in other words, our method
ensures measuring the shapes of the haloes close to the pre-selected
\Riso\ value. 

\section{Finding the appropriate tube radius around the orbit}\label{app:tubemethod}

\subsection{Iteration}
For a given tube radius, we count all particles lying inside the tube
around the orbit, thus determining the fraction of 'tube
particles'. If more than 68.3\% (1 $\sigma$) of the particles are
lying inside, the radius is decreased, otherwise enlarged.  The
procedure is repeated, until we achieve a tube particle fraction of
$68.3\% \pm 0.5\%$. If no convergence is reached, we increase the
tolerance by a factor of 2, up to 2\%. In cases of deviations
exceeding 2\%, we skip the satellites. This usually is the case, if
the number of debris particles is so small, that a meaningful
statistic is not possible anymore.

\subsection{Particles inside/outside the tube}
We check for each particle, whether it lies inside a tube of a given radius or outside.

Instead of first searching for the orbit point closest to the considered particle, we go through the orbit points until one is found, where the particle lies in the tube next to it. Thus, we find \emph{one} region where the satellite particle lies inside the tube. In fact, there could be more than one suitable region, maybe at some other point the particle would even lie closer to the orbit path. Yet knowing that the particle lies inside the tube at some point is sufficient for our purposes.

\begin{figure}
\begin{center}
        \epsfig{file=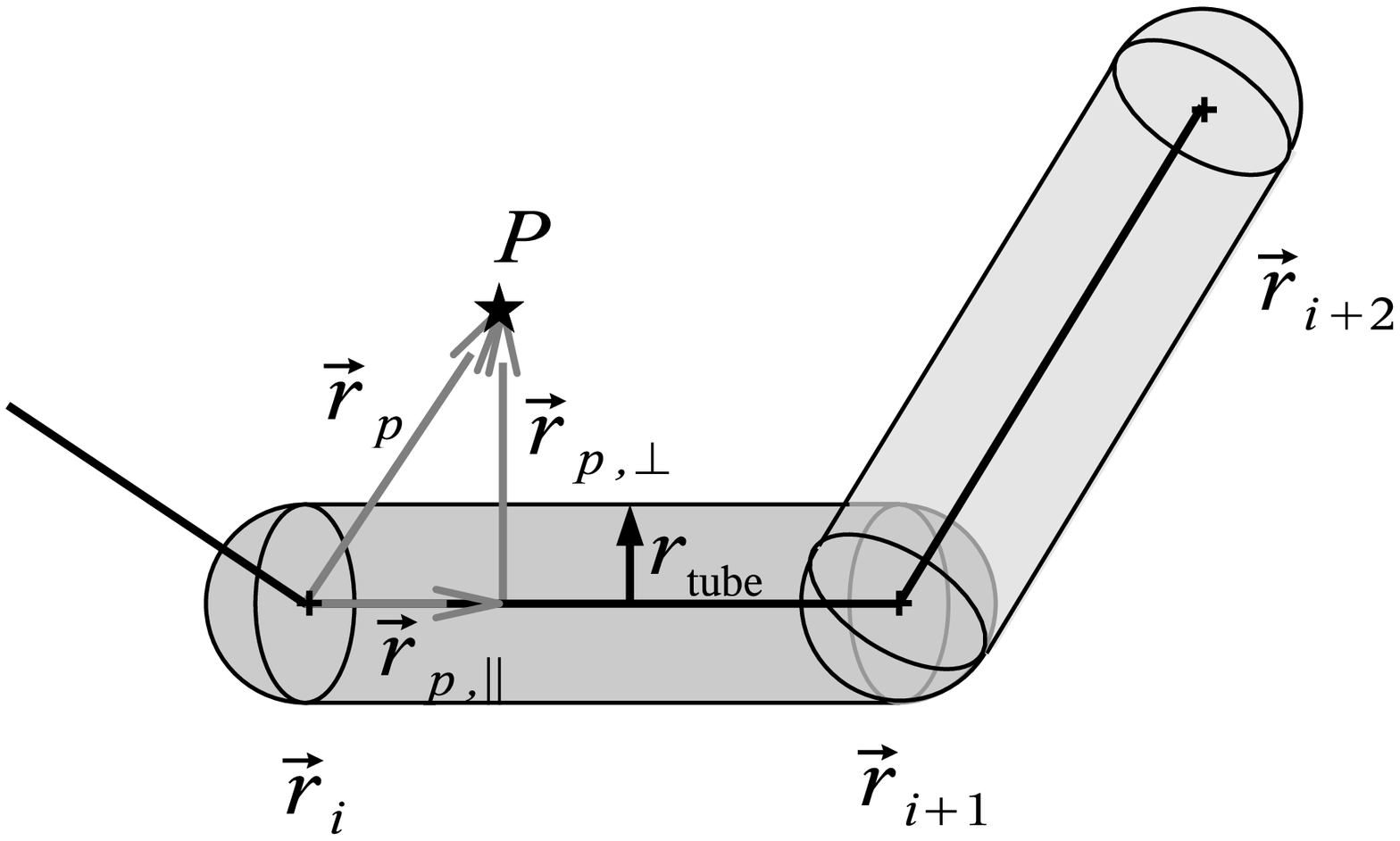, width=0.42\textwidth, angle=0}
\end{center}
\caption
{ Finding points inside/outside a tube of specified radius $r_t$ around the orbit path (thick black line). Not only a cylinder around the orbit path is used, but also the region in a half sphere at the end of each orbit piece is included. } 
\label{f:tube_method}
\end{figure}

\Fig{f:tube_method} illustrates the situation of particle $P$ lying outside the orbit tube. 
A particle at distance $\vec{r}_p$ from an orbit point lies inside the corresponding part of the orbit tube with radius $r_{\rm tube}$, if
\begin{itemize}
\item $r_{p,\perp} < r_{\rm tube}$ (within cylinder)
\item $r_{p, ||} \ge r_{\rm tube}$ (include cap of radius $r_{\rm tube}$ behind the considered orbit point)
\item $r_{p, ||} \le |\vec{r}_{i+1}-\vec{r}_i|+r_{\rm tube}$ (not too far beyond next orbit point)  
\end{itemize}
with the orbit points $\vec{r}_{i}$ and $\vec{r}_{i+1}$ and the components of $\vec{r}_{p}$ parallel ($\vec{r}_{p,||}$) and perpendicular ($\vec{r}_{p,\perp}$) to the orbit path. Particles are not only checked for lying in the cylinder of radius $r_{\rm tube}$ around the orbit path, but also a small half sphere ('cap') is allowed on front of and behind every orbit point. Otherwise, particles at the kink of the orbit might be wrongly classified as non-tube particles.

\label{lastpage}

\end{document}